\newcommand{\mass}[1]{\ensuremath{#1\, M_\odot}}

\newcommand{\Teff}{\ensuremath{T{_\mathrm{\!\!eff}}}}

\newcommand{\Z}{\ensuremath{Z}}
\newcommand{\Zinit}{\ensuremath{Z_\mathrm{init}}}
\newcommand{\Zsurf}{\ensuremath{Z_\mathrm{surf}}}
\newcommand{\FeH}{\ensuremath{\mathrm{[Fe/H]}}}
\newcommand{\deltaYdeltaZ}{\ensuremath{\frac{\Delta Y}{\Delta Z}}}
\newcommand{\chem}[2]{\ensuremath{^{#2}\kern-0.8pt\mathrm{#1}}}
\newcommand{\reac}[6]{\ensuremath{\,^{#2}\kern-0.8pt\mathrm{#1}\,({#3}\,,{#4})\,{}^{#6}\kern-0.8pt\mathrm{#5}\,}}

\documentclass[structabstract]{aa}  
\usepackage{graphicx}
\usepackage{enumerate}
\usepackage{txfonts}
\usepackage{natbib}

\usepackage{ulem}
\usepackage{color}

\begin{document}
   \title{Stellar mass and age determinations}

   \subtitle{I. Grids of stellar models from $Z$\,=\,0.006 to 0.04 and $M$\,=\,0.5 to \mass{3.5}}

   \author{N. Mowlavi\inst{1,2}
          \and
          P. Eggenberger\inst{1}
          \and
          G. Meynet\inst{1}
          \and
          S. Ekstr\"om\inst{1}
          \and
          C. Georgy\inst{3}
          \and
          A. Maeder\inst{1}
          \and
          C. Charbonnel\inst{1}
          \and
          L. Eyer\inst{1}
          }

   \institute{$^1$Observatoire de Gen\`eve, Universit\'e de Gen\`eve, chemin des Maillettes, 1290 Versoix, Switzerland\\
              $^2$ISDC, Observatoire de Gen\`eve, Universit\'e de Gen\`eve, chemin d'Ecogia, 1290 Versoix, Switzerland\\
              $^3$Centre de Recherche Astrophysique, Ecole Normale Sup\'erieure de Lyon, 46, all\'ee d'Italie, 69384 Lyon cedex 07, France\\
                \email{Nami.Mowlavi@unige.ch
                       }
             }

   \date{Received ...; accepted ...}

 
  \abstract
   {}
   {We present dense grids of stellar models suitable for comparison with observable quantities measured with great precision, such as those derived from binary systems or planet-hosting stars.
   }
   {We computed new Geneva models without rotation at metallicities $Z$\,=\,0.006, 0.01, 0.014, 0.02, 0.03 and 0.04 (i.e. [Fe/H] from $-0.33$ to $+0.54$) and with mass in small steps from 0.5 to \mass{3.5}.
   Great care was taken in the procedure for interpolating between tracks in order to compute isochrones.}
   {Several properties of our grids are presented as a function of stellar mass and metallicity.
Those include surface properties in the Hertzsprung-Russell diagram, internal properties including mean stellar density, sizes of the convective cores, and global asteroseismic properties.
   }
   {We checked our interpolation procedure and compared interpolated tracks with computed tracks.
   The deviations are less than 1\% in radius and effective temperatures for most of the cases considered.
   We also checked that the present isochrones provide nice fits to four couples of observed detached binaries and to the observed sequences of the open clusters NGC~3532 and M67.
   Including atomic diffusion in our models with $M$\,$<$\,\mass{1.1} leads to variations in the surface abundances that should be taken into account when comparing with observational data of stars with measured metallicities.
For that purpose, iso-$\Zsurf$ lines are computed.
These can be requested for download from a dedicated web page together with tracks at masses and metallicities within the limits covered by the grids.
The validity of the relations linking $Z$ and $\FeH$ is also re-assessed in light of the surface abundance variations in low-mass stars.
   }


   \maketitle

%
\section{Introduction}
\label{Sect:introduction}

Accurate grids of stellar models at various metallicities remain a basic tool in many areas of astrophysics, such as the study of stellar clusters, of stellar populations, of stellar nucleosynthesis, or of chemical evolution in galaxies.
Two observational developments have further strengthened the need for ever more accurate stellar model predictions in recent years.
The first concerns the improvements in stellar parameters determination from observations.
We can mention in this respect the determination to better than 3\% of stellar masses and radii of almost two hundred stars in eclipsing binary systems \citep{TorresAndersenGimenez10} or the precise measurement of the characteristics of planet-hosting stars.
The second development is related to the advent of large-scale surveys, initiated in the nineties with the EROS, MACHO, OGLE, Hipparcos, or SDSS experiments, to cite only a few.
Those surveys provide statistically significant data on millions to tens of millions of stars, and future ones are expected to increase the number of stars by one or two orders of magnitude.
We cite in this respect the future Gaia mission, planned for a launch in 2013, which will monitor 1 billion stars in our Galaxy, with distance determinations to better than 1\% for about 20 million of them allowing the building of the Hertzsprung-Russell (HR) diagram to an unprecedented degree of precision.
The Large Scale Synoptic Survey is another example of a future large scale-survey, which will complement Gaia by, among other goals, observing fainter objects.
The start of its operation is planned for the end of this decade.

Significant improvements have been made to the field of stellar modeling since the first extensive grids of Geneva stellar models published about two decades ago \citep[][and subsequent papers]{SchallerSchaererMeynet_etal92}.
They concern the equation of state, atomic diffusion, rotation or the solar abundances, to cite only a few.
The impact of their inclusion in our stellar models has been presented in a few papers \citep[e.g.][]{EggenbergerMeynetMaeder10,EggenbergerMiglioMontalban10}, justifying the need to update our previous grids of stellar models.

New grids of Geneva models covering stellar masses from 0.8 to \mass{120} at solar metallicity and computed with and without rotation are presented in a separate paper \citep{EkstroemGeorgyEggenberger11}.
In the present paper, we present dense grids of models covering stellar masses from 0.5 to \mass{3.5} and metallicities from $Z$\,=\, 0.006 to 0.04 with a distribution of points in the mass-metallicity plane dense enough to allow an accurate interpolation between computed models, but restricted to models without rotation and to phases from the zero-age main sequence (ZAMS) up to the base of the red giant branch (RGB).

The originality of the present study resides mainly in three aspects.
First, we use the solar abundances given by Asplund et al. (2005) as initial abundances.
Second, we provide tracks for a very dense sampling in initial masses for six metallicities.
Third, we construct an interpolation procedure between the tracks, allowing us to efficiently handle the change in the morphology of the tracks when the mass varies.

The paper is divided as follows.
Section~\ref{Sect:inputPhysics} presents the ingredients of the stellar evolution models
and Sect.~\ref{Sect:gridProperties} some general properties of the grids.
Section~\ref{Sect:tracks} describes the different types of tracks and isochrones constructed from the raw models.
A comparison of our model predictions with some observations and with some other tracks available in the literature is presented in Sect.~\ref{Sect:comparisons}.
The data made available for download are described in Sect.~\ref{Sect:dataDownload}.
Conclusions are then drawn in Sect.~\ref{Sect:conclusions}.

Three appendices complete the main body of the paper.
Appendix~\ref{Sect:FeH2Z} discusses the relation between the two standard representations of the metallicity, i.e. $\Z$, the mass fraction of all elements heavier than helium, and $\FeH$, the logarithm of the iron to hydrogen abundance ratio relative to the solar ratio.
Appendix~\ref{Sect:normalizationProcedure} describes the procedure used to construct basic tracks, which consists in a well defined series of stellar models chosen to insure easy interpolations in mass and metallicity.
Appendix~\ref{Sect:isochroneConstruction} describes the construction of isochrones.

\section{Ingredients of the stellar models}
\label{Sect:inputPhysics}

The input physics included in the Geneva evolution code is described in detail in \cite{EggenbergerMeynetMaeder08}, while the
chemistry and nucleosynthesis used for the grids is described in \cite{EkstroemGeorgyEggenberger11}.
We highlight here only a few points relevant to the present dense grids.

\paragraph{Chemical abundances}
\label{Sect:abundances}

\begin{table}
\caption{Initial chemical abundances of our grids.
}
\centering
\begin{tabular}{c c c c c}
  \hline
  $Z_\mathrm{init}$ & $\FeH_\mathrm{init}$ & $X_\mathrm{init}$ & $Y_\mathrm{init}$ & $Z_\mathrm{init}/X_\mathrm{init}$ \\[1mm]
  0.006 & $-0.331$ & 0.7383 & 0.2557 & 0.0081\\
  0.010 & $-0.103$ & 0.7291 & 0.2609 & 0.0137\\
  0.014 & $+0.048$ & 0.7200 & 0.2660 & 0.0194\\
  0.020 & $+0.212$ & 0.7063 & 0.2737 & 0.0283\\
  0.030 & $+0.402$ & 0.6834 & 0.2866 & 0.0439\\
  0.040 & $+0.542$ & 0.6606 & 0.2994 & 0.0606\\
  \hline
\end{tabular}
\label{Tab:gridAbundances}
\end{table}

The adopted initial chemical abundances are indicated in Table~\ref{Tab:gridAbundances} (see Appendix~\ref{Sect:FeH2Z} for more details).
For the mixture of the heavy elements, we adopt the ones from \cite{AsplundGrevesseSauval05}, except for Ne which we take from \cite{CunhaHubenyLanz06}.

\paragraph{Convection and overshooting}
\label{Sect:convection}

Convective zones are determined with the Schwarzschild criterion.
The convective core is extended with an overshoot parameter $\alpha_\mathrm{ov} \equiv d_\mathrm{ov}/H_\mathrm{P}$ of 0.10 for \mass{M\ge 1.7}.
Half of this overshooting, $\alpha_\mathrm{ov}$\,=\,0.05 is applied between 1.25 and \mass{1.7}, where the extent of the convective core, when it exists, is small.
If $d_\mathrm{ov}$ exceeds the dimension of the Schwarzschild core $R_\mathrm{Sch}$, then the total extension of the convective core is taken equal to $R_\mathrm{Sch} \times (1+d_\mathrm{over}/H_\mathrm{P})$.
This procedure avoids having an extension of the core superior to the radius of the initial core.
No overshooting is applied below \mass{1.25}.

The outer convective zone is treated according to the mixing length theory \citep{Bohm58}, with a solar calibrated value of the mixing-length parameter $\alpha_\mathrm{MLT} \equiv l/H_\mathrm{P}$ equal to 1.6467 based on models including atomic diffusion (see below).
This low value of the mixing-length parameter directly results from the use of the new solar abundances from 
\cite{AsplundGrevesseSauval05}.
It is, however, known that these abundances lead to solar models that reproduce helioseismic constraints in a less satisfactory manner than models using older solar abundances do \citep[see e.g.][and references therein]{BasuAntia08}.

\paragraph{Atomic diffusion}
\label{Sect:atomicDiffusion}

Atomic diffusion due to concentration and thermal gradients is included in the computation of models with initial masses below \mass{1.1}, following the Chapman-Enskog method \citep[][]{ChapmanCowling70}.
We refer to \cite{EggenbergerMeynetMaeder08} for more details about the modeling of atomic diffusion in the Geneva evolution code.
For masses above about \mass{1.1}, helium and heavy elements are rapidly drained out of the envelope when only atomic diffusion due to concentration and thermal gradients is included to model those stars (which have a thin convective envelope).
This results in surface abundances that are too low compared to observations.
Indeed, in those stars, the effects of atomic diffusion due to radiative forces and/or to macroscopic turbulent transport become significant.
These effects mainly counteract those of atomic diffusion due to concentration and thermal gradients.
Therefore, we assume here that above the (somewhat arbitrary) mass limit of \mass{1.1}, those models with no atomic diffusion probably better fit the observations.
We thus do not include atomic diffusion in stars with a mass equal to or greater than \mass{1.1}.

\paragraph{Equation of state}
\label{Sect:EOS}

Models with $M$\,$\ge$\,\mass{1.1} are computed with the general equation of state (EOS) used in the Geneva evolution code that is well suited to massive stars \citep{SchallerSchaererMeynet_etal92}.
The computation of reliable models of solar-type stars requires, however, a more specific equation of state, because non ideal effects such as Coulomb interactions become important for low-mass stars.
The OPAL EOS \citep{RogersNayfonov02} is then used for models with initial masses below \mass{1.1}.
We checked that the transition between models with diffusion and the OPAL EOS ($M$\,$<$\,\mass{1.1}) and models without diffusion and with the general EOS ($M$\,$\ge$\,\mass{1.1}) is smooth in the physical quantities at the center of our models.

\paragraph{Mass loss}
\label{Sect:massLoss}

No mass loss is applied to our models, which all have $M$\,$\le$\,\mass{3.5} and are computed only up to the base of the RGB.

\section{Grids properties}
\label{Sect:gridProperties}

\begin{figure}
  \centering
  \includegraphics[width=\columnwidth]{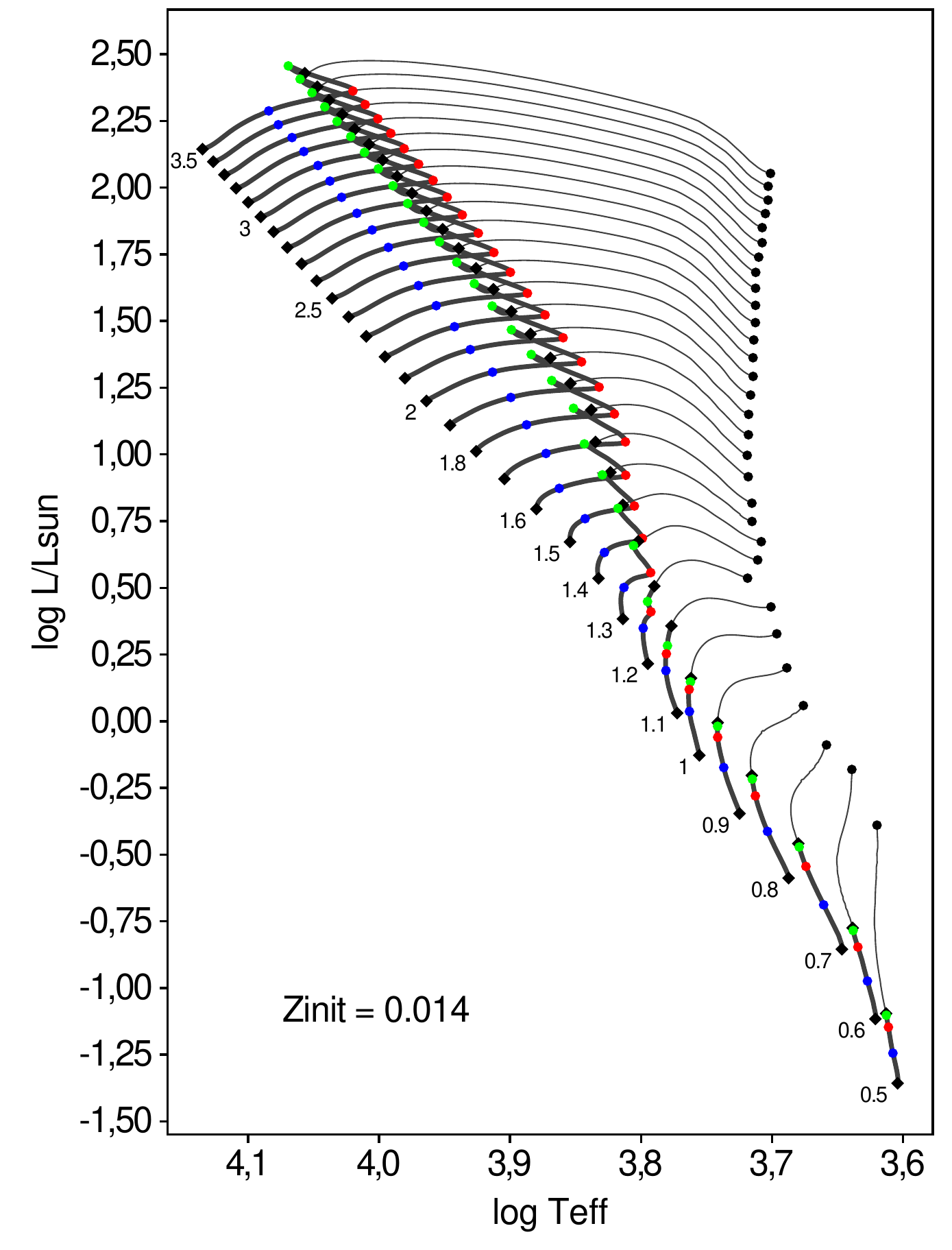}
  \caption{Evolutionary tracks in the HR diagram at $\Zinit = 0.014$.
  The main sequence is shown in thicker lines.
  Reference points defined in Sect.~\ref{Sect:basicTracks} are indicated by diamonds (ZAMS and TAMS) and filled circles (three reference points within the MS and one at the bottom of the RGB).
  For clarity, the tracks at stellar masses between 1.1 and \mass{1.2} and between 1.2 and \mass{1.3} are not shown.
  They are displayed in Fig.~\ref{Fig:HRfctZM} for $Z_\mathrm{init}$\,=\,0.006 and 0.04.
  }
\label{Fig:HRZ014}
\end{figure}

\begin{figure}
  \centering
  \includegraphics[width=\columnwidth]{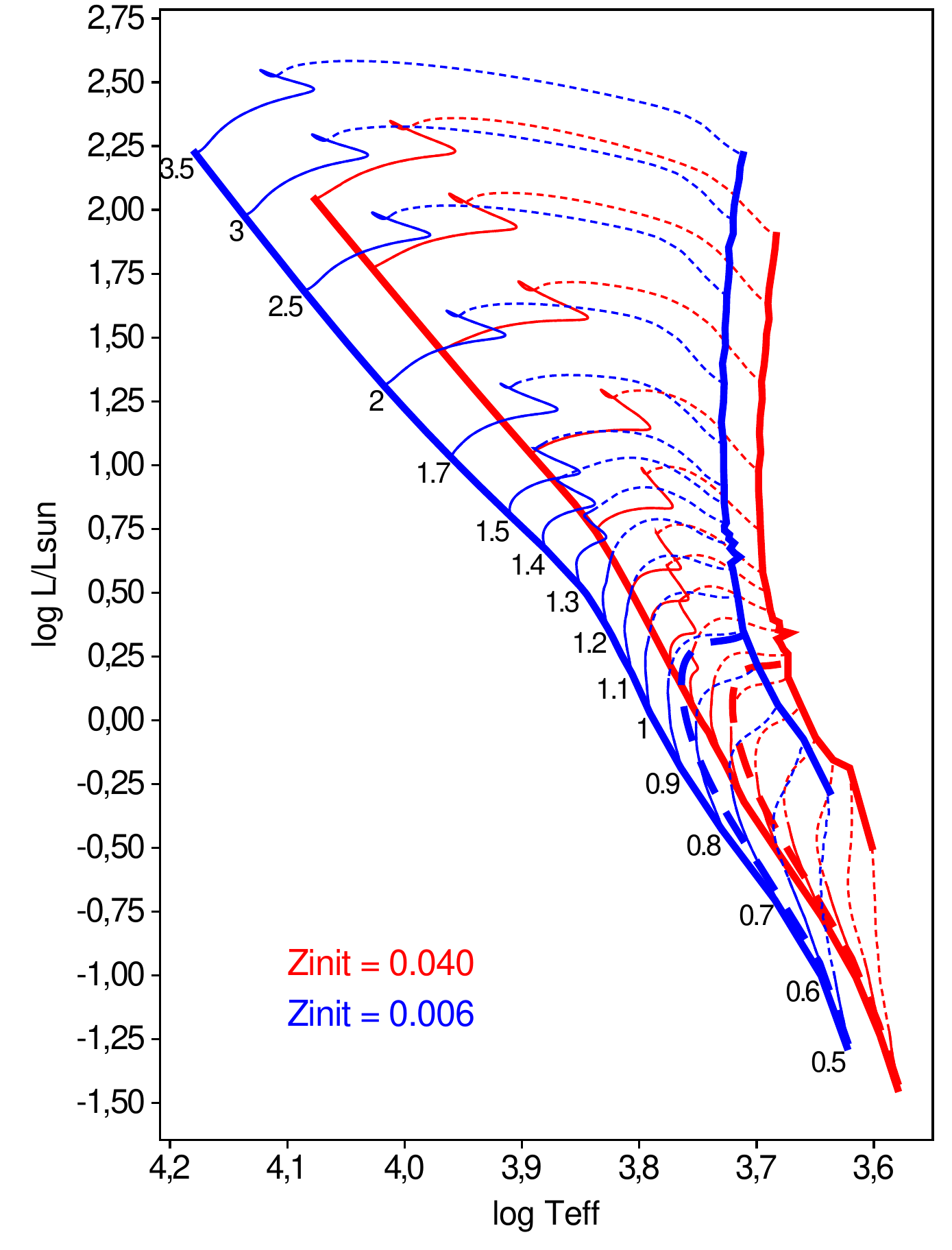}
  \caption{Evolutionary tracks at $\Zinit=0.006$ (in blue, hot tracks) and 0.040 (in red, cool tracks).
  Only selected stellar masses are plotted for clarity, as labeled next to the ZAMS location of the low-metallicity grid.
  The main sequences are drawn in thin continuous lines and the post-main sequences in thin dotted lines.
  Thick continuous lines denote the location of the ZAMS (hot sides) and of the onset of the RGB (cool sides).
  Also shown for each of the two metallicity grids are the isochrones at 13.5~Gyr (thick dashed lines on the low-luminosity side).
  }
\label{Fig:HRSummary}
\end{figure}

Our dense grids of stellar models are computed for initial metallicities $\Zinit$ and masses $M$ equal to
\begin{equation}
\left\{
\begin{array}{cllcl}
  \Zinit & = & \multicolumn{3}{l}{0.006, ~0.010, ~0.014, ~0.020, ~0.030, ~0.040}\\
  \\
  M & = & 0.5 & [0.1]  & 1.1~\mass{},\\
    &   & 1.1 & [0.02] & 1.3~\mass{},\\
    &   & 1.3 & [0.1]  & 3.5~\mass{}
\end{array}
\right.
\label{Eq:gridDefinition}
\end{equation}
where the notation $M_\mathrm{a}$ [d$M$] $M_\mathrm{b}$ stands for a range of masses between $M_\mathrm{a}$ and $M_\mathrm{b}$ in steps of d$M$.
Table~\ref{Tab:gridAbundances} gives some information on the initial abundances at each metallicity.

Section~\ref{Sect:diagrams} summarizes the model predictions in two diagrams relating quantities derivable from observations, namely the classical HR diagram and the $\log \Teff$ -- $\log \overline{\rho}/\overline{\rho}_\odot$ diagram plotting the mean stellar density against effective temperature.
Section~\ref{Sect:convectiveCores} then addresses the onset of convection in the core of intermediate-mass stars, Sect.~\ref{Sect:ages} discusses the ages of our models, and Sect.~\ref{Sect:surfaceAbundances} briefly presents the effect of atomic diffusion in low-mass models.

\subsection{Summary diagrams}
\label{Sect:diagrams}

\paragraph{Hertzsprung-Russell diagram}

\begin{figure}
  \centering
  \includegraphics[width=\columnwidth]{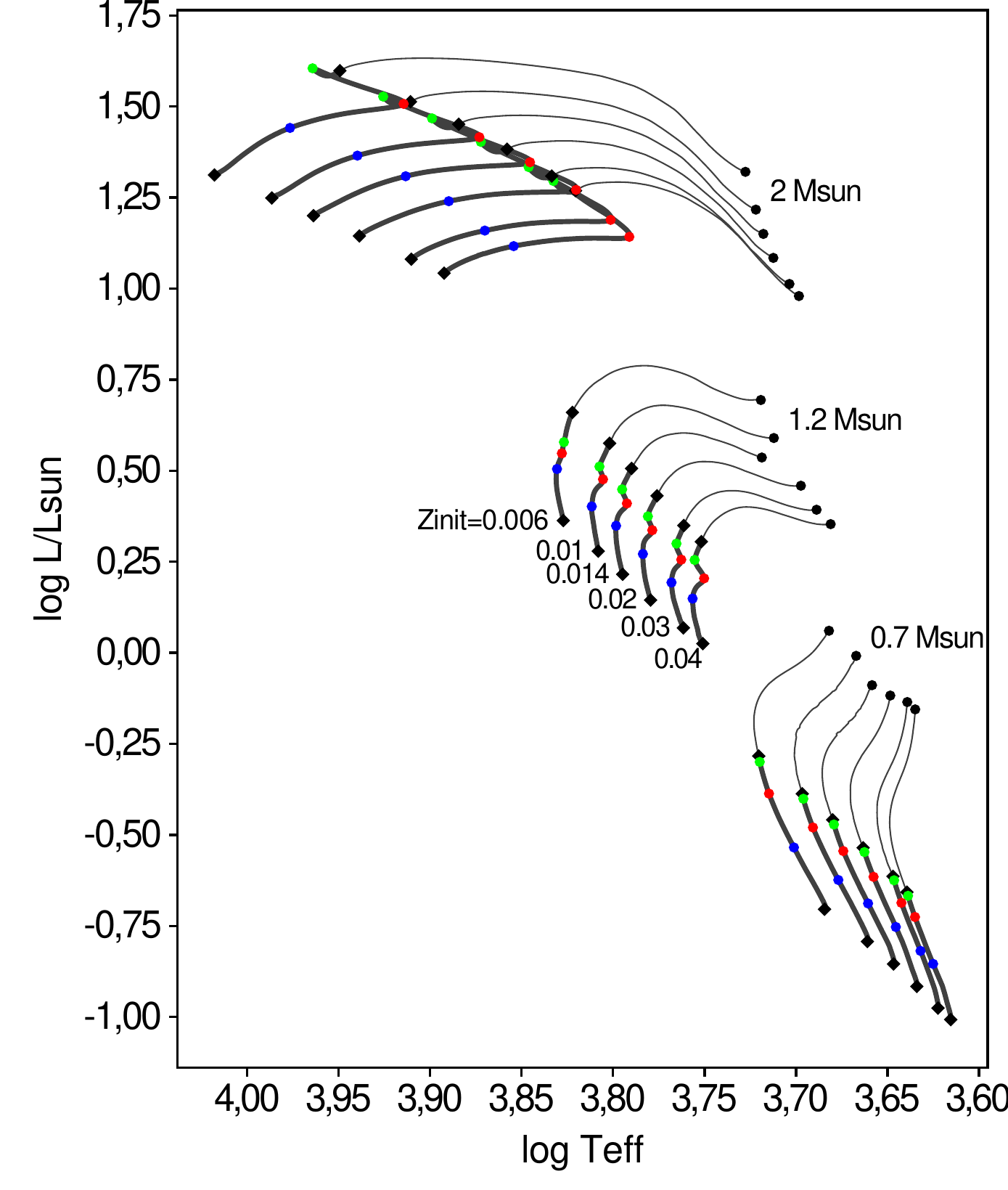}
  \caption{Evolutionary tracks of the 0.7, 1.2, and \mass{2} models at the different metallicities of our grids as labeled in the figure.
  Reference points are indicated on the tracks by filled diamonds and circles in a similar way to Fig.~\ref{Fig:HRZ014}.
  }
\label{Fig:HRfctZ}
\end{figure}

\begin{figure}
  \centering
  \includegraphics[width=\columnwidth]{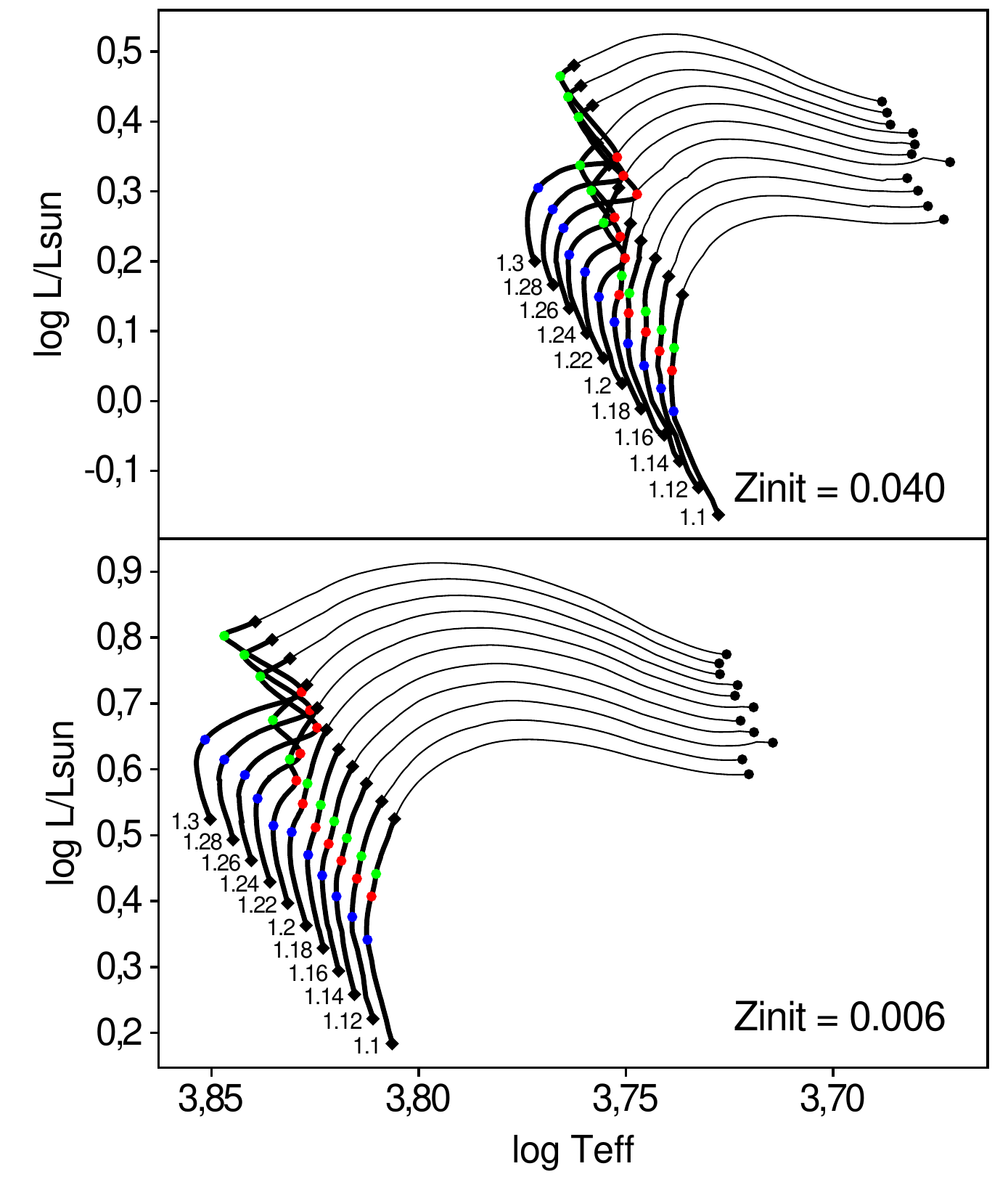}
  \caption{Evolutionary tracks at $Z_\mathrm{init}$\,=\,0.04 (upper panel) and 0.006 (lower panel) for the intermediate-mass stars from 1.1 to \mass{1.3}, as labeled in the panels.
  Reference points are indicated on the tracks by filled diamonds and circles in a similar way as in Fig.~\ref{Fig:HRZ014}.
  }
\label{Fig:HRfctZM}
\end{figure}

The HR diagram of our \Zinit\,=\,0.014 tracks is shown in Fig.~\ref{Fig:HRZ014}, with the ZAMS, the terminal age main sequence (TAMS), and the base of the RGB.
The base of the RGB is defined as the point on the post-MS track where the slope of the line connecting the TAMS and the running point on the post-MS tracks starts to increase (see Sect.~\ref{Sect:PostMSConstruction}).
Three other reference points chosen on the MS are defined in Sect.~\ref{Sect:basicTracks} and used for constructing the basic tracks.
Two features are worth noticing in the HR diagram.
First, a hook appears at the end of the MS phase in the morphology of the MS tracks at stellar masses above \mass{1.1}.
The filled red and green points locate respectively the red and the subsequent blue turn off points.
The hooks lead to a degeneracy in the HR diagram, whereby a given set ($L$, $\Teff$) at a given metallicity can correspond to up to three stellar models, each with a different stellar mass.
Second, from ZAMS to the red turn-off point (or to the TAMS in the absence of hook), $\Teff$ strictly increases with time for masses below \mass{1}, strictly decreases with time for stellar masses above \mass{1.4}, and has a decreasing-then-increasing behavior for masses in between those limits. 

The area covered by all our models in the HR diagram is shown in Fig.~\ref{Fig:HRSummary}, from $\Zinit$\,=\,0.006 (hotter tracks) to 0.04 (cooler tracks).
The isochrones at 13.5~Gyr plotted in the figure start on the \mass{0.5} tracks and remain very close to the ZAMS up to a mass around \mass{0.75} because very low-mass stars do not evolve much in the time span of 13~Gyr (the MS lifetime of the \mass{0.5} star ranges between 69~Gyr at $Z_\mathrm{init}$\,=\,0.006 and 76~Gyr at $Z_\mathrm{init}$\,=\,0.04, much greater than the age of the Universe).
 
The metallicity dependence of the tracks is shown in Fig.~\ref{Fig:HRfctZ} for three stellar masses.
The morphology of the tracks at \mass{1.2} is particularly interesting as it shows a metallicity dependence of the MS hook in the HR diagram due to the sensitivity of the onset of convection to metallicity (see Sect.~\ref{Sect:convectiveCores}).
The tracks in the critical stellar mass range between 1.1 and \mass{1.3}, where convection sets in in the core of MS stars, are shown in more detail in Fig.~\ref{Fig:HRfctZM} for the two extreme metallicities considered in our grids (see discussion in Sect.~\ref{Sect:convectiveCores}).

\paragraph{Effective temperature - mean density diagram}

\begin{figure}
  \centering
  \includegraphics[width=\columnwidth]{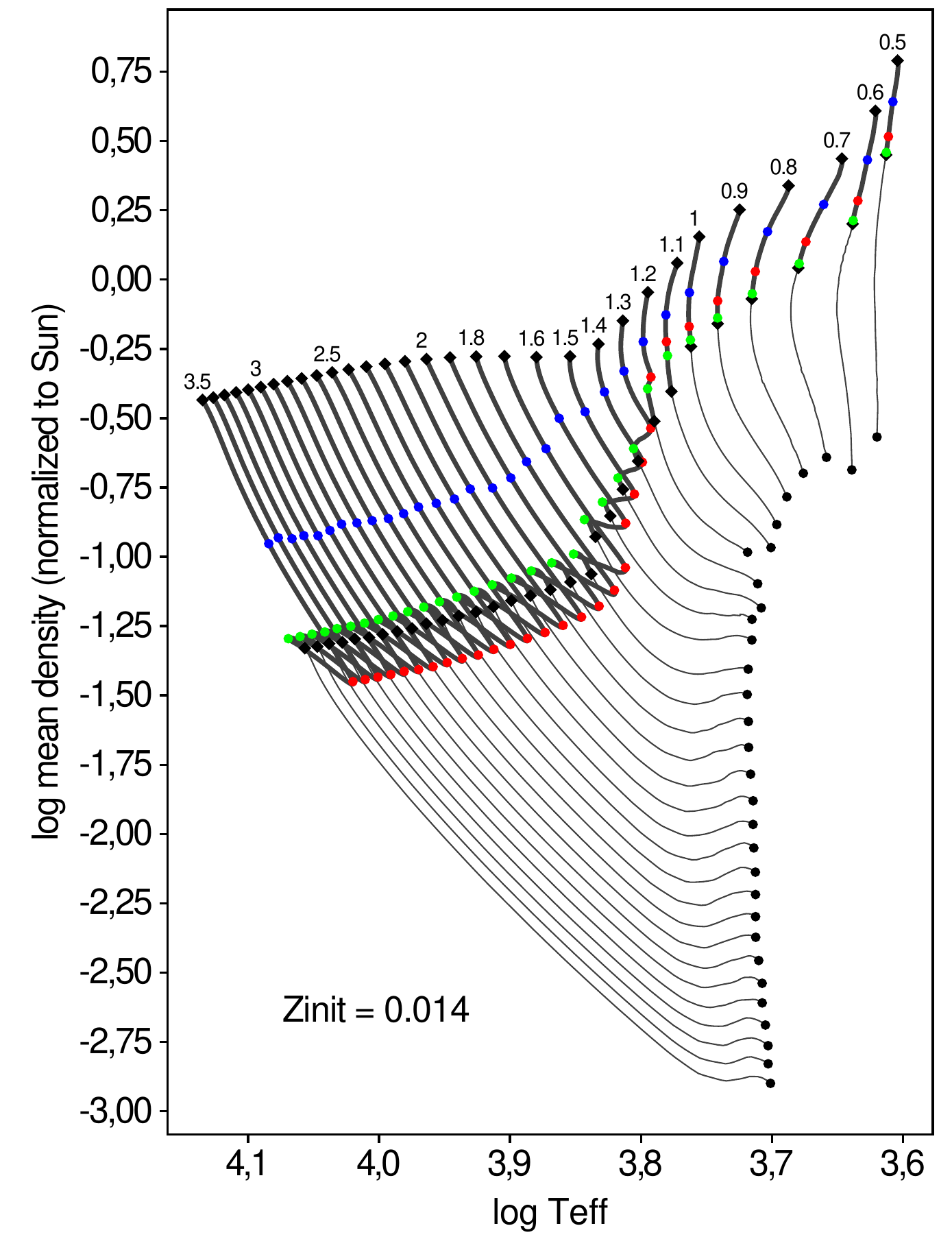}
  \caption{Evolution in the $\log \Teff - \log \overline{\rho}/\overline{\rho}_\odot$ diagram at $\Zinit$\,=\,0.014.
  See Fig.~\ref{Fig:HRZ014} for the explanation of the symbols.
  }
\label{Fig:TeffMeanRhoZ014}
\end{figure}

\begin{figure}
  \centering
  \includegraphics[width=\columnwidth]{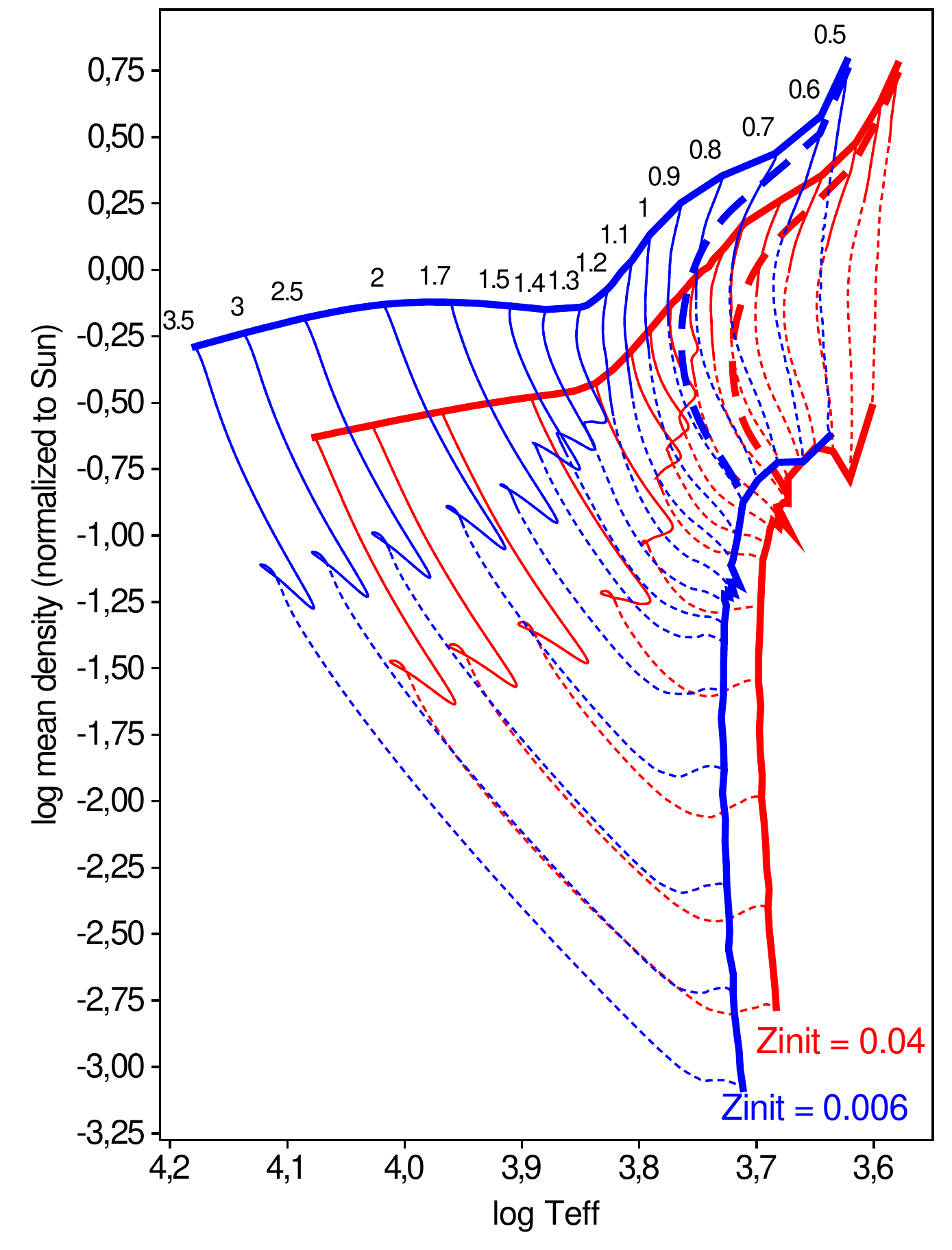}
  \caption{Evolution in the $\log \Teff - \log \overline{\rho}/\overline{\rho}_\odot$ diagram at $\Zinit$\,=\,0.006 and 0.04.
  See Fig.~\ref{Fig:HRSummary} for the explanation of the symbols.
  }
\label{Fig:TeffMeanRhoSummary}
\end{figure}

The mean density $\overline{\rho}= M / (\frac{4}{3} \pi R^3)$ of a star can be determined with relatively good accuracy in systems hosting exoplanets \citep[e.g.][]{SozzettiTorresCharbonneau_etal07,GillonSmalleyHebb_etal09}.
This quantity generally decreases with age for a given star, as shown in Fig.~\ref{Fig:TeffMeanRhoZ014}, which displays the tracks followed by the \Zinit\,=\,0.014 models in the $\log \Teff - \log \overline{\rho}/\overline{\rho}_\odot$ plane.

The range of stellar mean densities and effective temperatures covered by our grids is shown in Fig.~\ref{Fig:TeffMeanRhoSummary} for the two extreme metallicities considered here.
For the very low-mass stars ($M$\,$\lesssim$\,0.8), however, as in the HR diagram, a portion of the $\log \Teff - \log \overline{\rho}/\overline{\rho}_\odot$ plane is unreachable within the age of the Universe.
The frontier of the realistic region is indicated in Fig.~\ref{Fig:TeffMeanRhoSummary} by the 13.5~Gyr isochrones, starting on the \mass{0.5} tracks close to the ZAMS and ending at the base of the RGB on the 0.9 (at \Zinit\,=\,0.006) and $\sim$\mass{1} (at \Zinit\,=\,0.04) tracks.

\subsection{Onset of convection in the core}
\label{Sect:convectiveCores}

\begin{figure}
  \centering
  \includegraphics[width=\columnwidth]{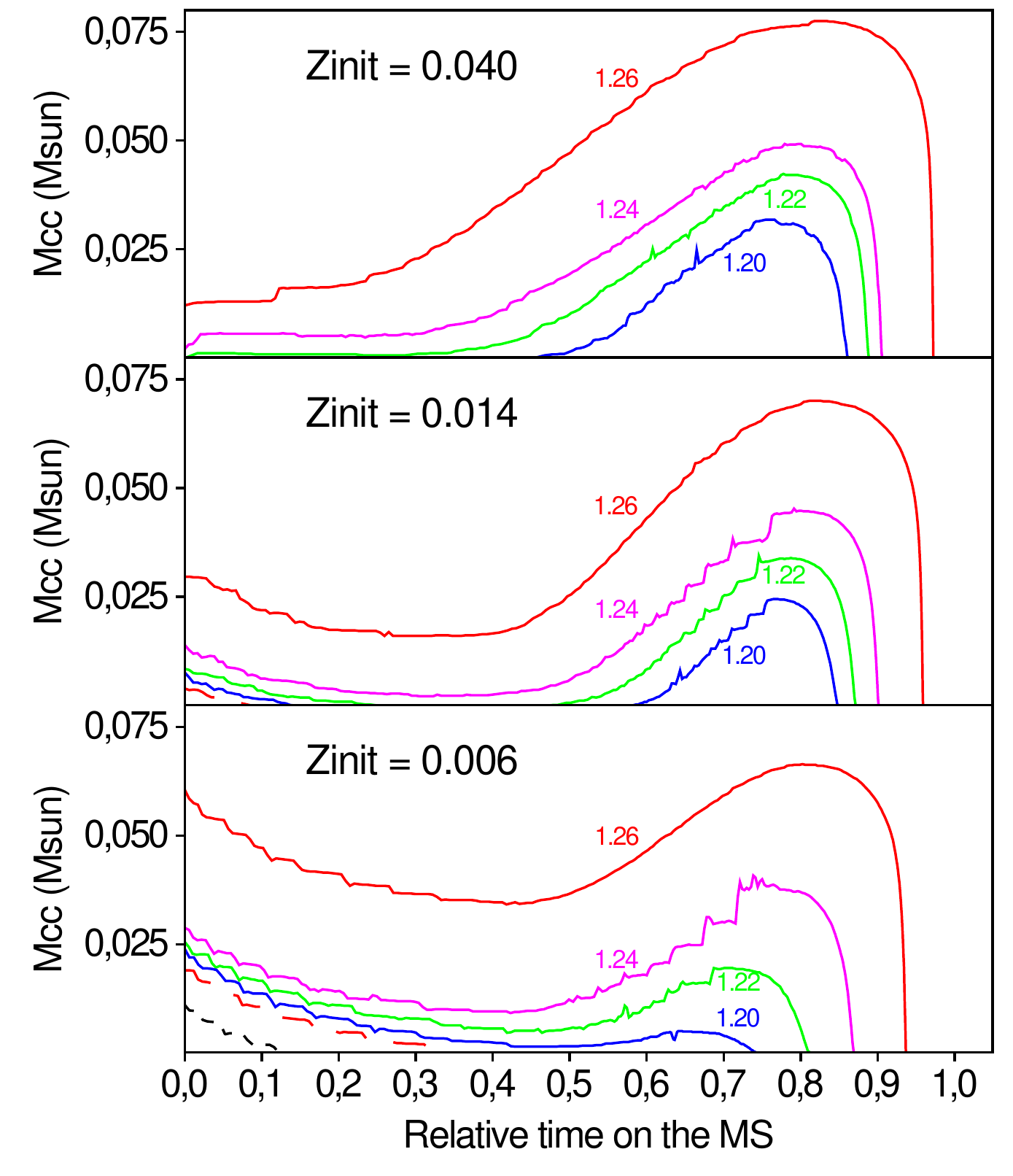}
  \caption{Evolution of the mass of the convective core as a function of the relative time spent on the main sequence (0 is ZAMS and 1 is TAMS) for models at $Z_\mathrm{init}$\,=\,0.040 (upper panel), 0.014 (middle panel), and 0.006 (lower panel).
  The models are for stellar masses of, from top to bottom, 1.26, 1.24, 1.22, and \mass{1.20} (as labeled on the continuous lines in all panels), \mass{1.18} (long dashed lines in the middle and lower panels), and \mass{1.10} (short dashed line in the lower panel).
  }
\label{Fig:convCore}
\end{figure}

\begin{figure}
  \centering
  \includegraphics[width=\columnwidth]{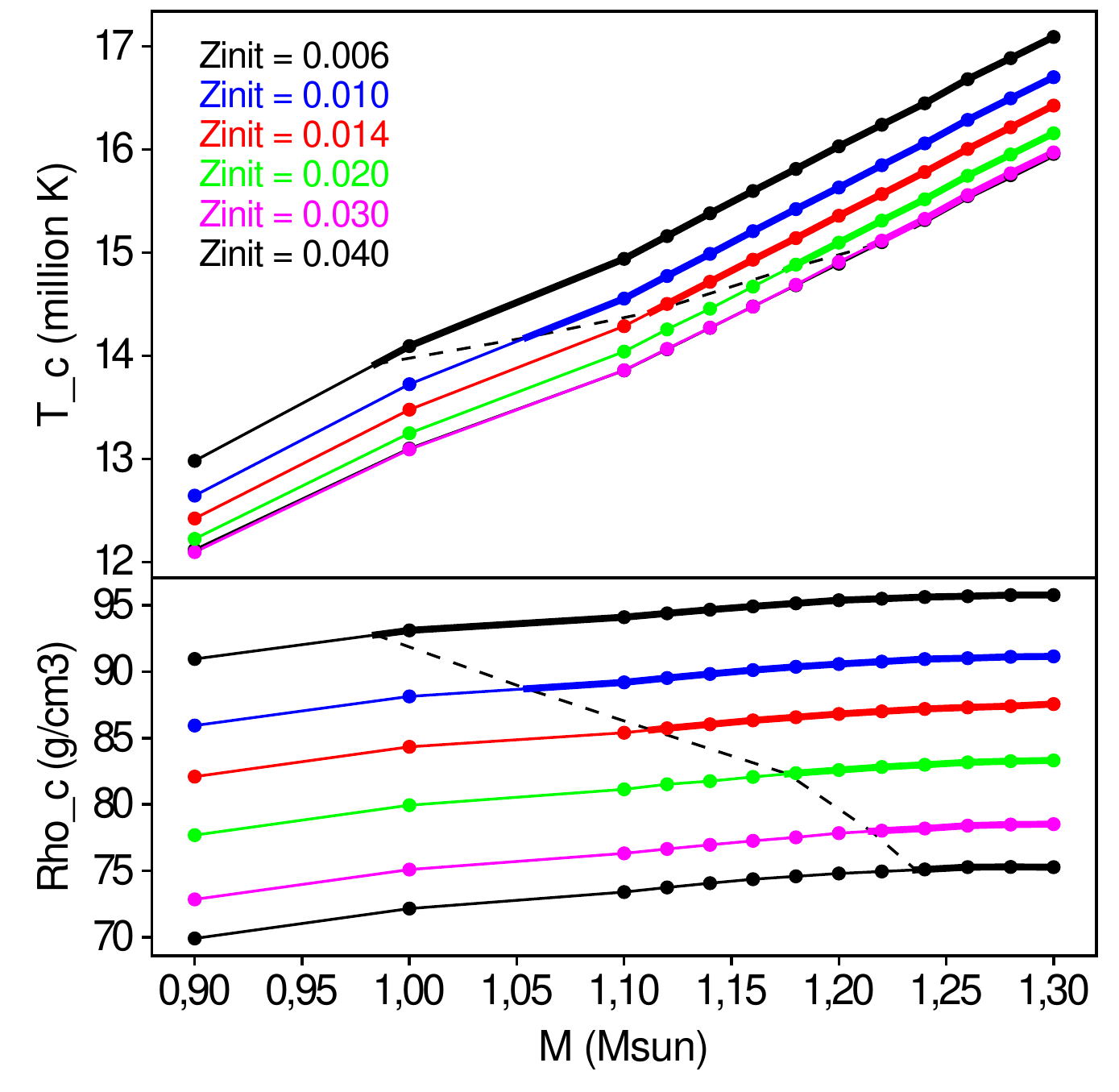}
  \caption{\textbf{Upper panel:} Central temperature, in million Kelvin, at ZAMS of stellar models at different metallicities as a function of the stellar mass of the models.
  Each dot represents a computed model.
  The thick lines indicate the range of stellar masses for each metallicity, where a convective core is present on the  ZAMS.
  The dotted line connects the core temperature at the lowest stellar mass having a convective core, for the different metallicities.
  \textbf{Lower panel:} Same as upper panel, but for the central density.
  }
\label{Fig:TcZAMS}
\end{figure}

\begin{figure}
  \centering
  \includegraphics[width=\columnwidth]{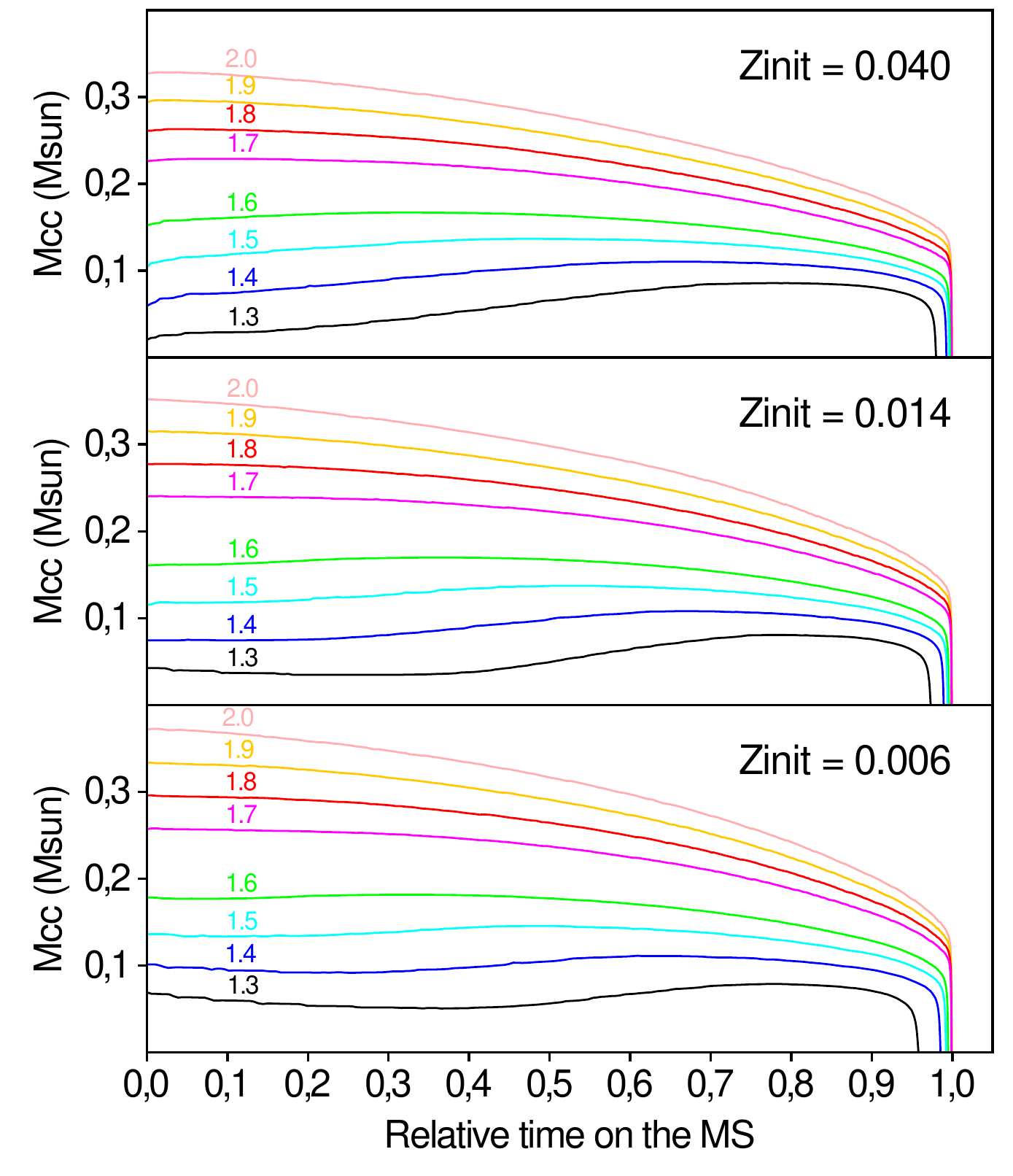}
  \caption{Same as Fig.~\ref{Fig:convCore}, but for stellar masses of, from bottom to top, 1.3 to \mass{2} in steps of \mass{0.1} as labeled on the lines.
  }
\label{Fig:convCoreHigherMasses}
\end{figure}

\begin{figure}
  \centering
  \includegraphics[width=\columnwidth]{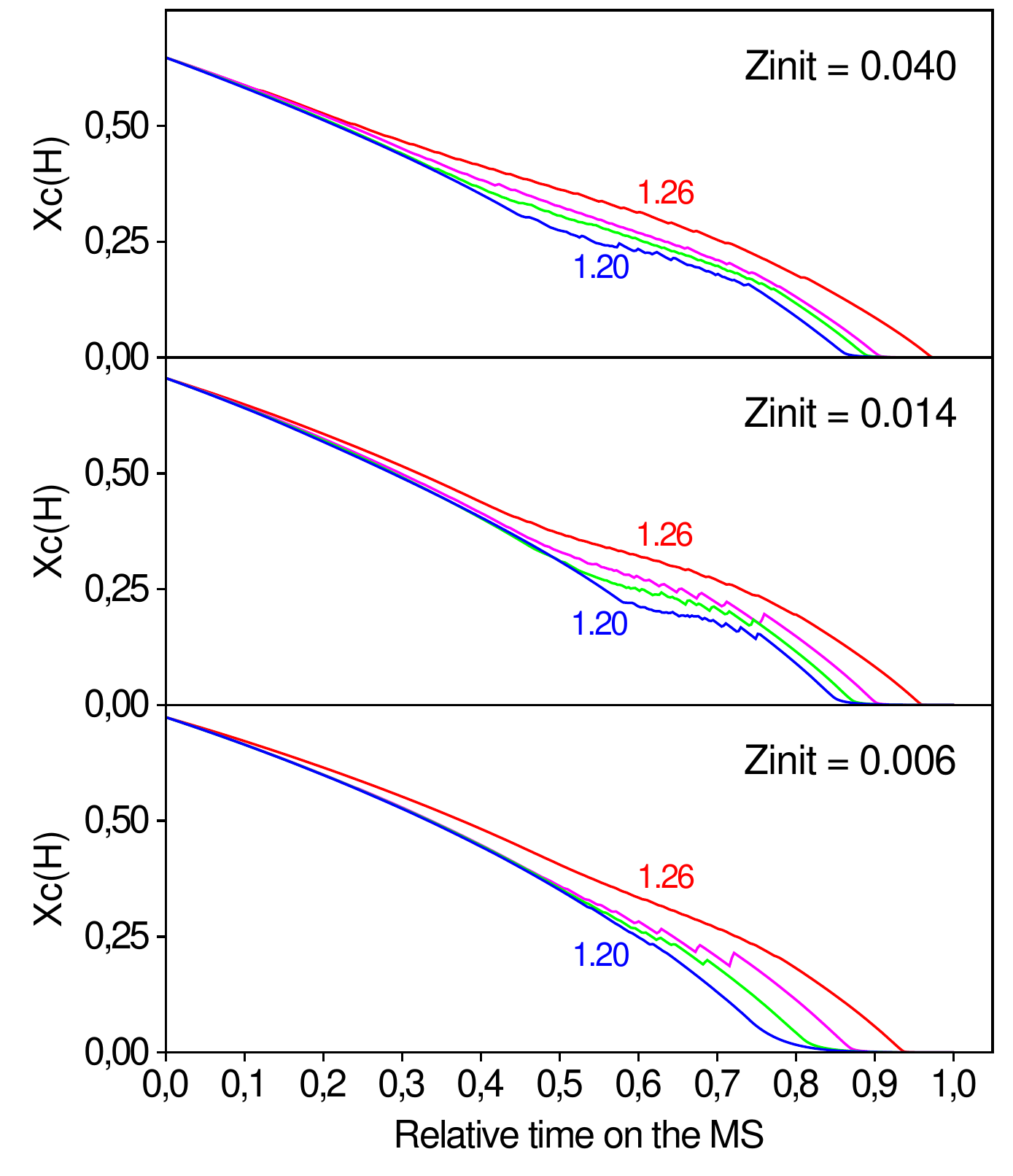}
  \caption{Same as Fig.~\ref{Fig:convCore}, but for the evolution of the hydrogen mass fraction at the center of the star in the models from 1.20 to \mass{1.26}.
  }
\label{Fig:XcH}
\end{figure}

There is a mass limit below which stellar models burn hydrogen radiatively  throughout the MS phase, and another mass limit above which the core is convective all through the MS phase.
Those mass limits depend on metallicity.
For \Zinit\,=\,0.014 stars, they are $\sim$\mass{1.20} and $\sim$\mass{1.24}, respectively.
Between those two mass limits, a star may have a convective core at ZAMS that disappears with age and/or a convective core that appears in the middle of the MS, grows thereafter and disappears at the end of the MS.
This is seen in Fig.~\ref{Fig:convCore}, which displays the evolution of the mass of the convective core in the 1.1 to \mass{1.26} models as a function of relative time on the MS at three different metallicities.
In the \mass{1.20} model at \Zinit\,=\,0.014 (middle panel of the figure), for example, a convective core exists at ZAMS and disappears after 0.74~Gyr (when the hydrogen mass fraction at the center of the star, $X_\mathrm{c}$, equals 0.61).
A convective core then reappears at 1.89~Gyr ($X_\mathrm{c}$\,=\,0.22), reaches its maximum growth in mass at 3.79~Gyr ($X_\mathrm{c}$\,=\,0.13), and disappears at 4.18~Gyr ($X_\mathrm{c}$\,=\,0.015).
The MS ends at 4.93~Gyr ($X_\mathrm{c}$\,=\,$10^{-5}$).
In the \mass{1.22} stellar model (MS duration of 4.62~Gyr), the initial convective core disappears at an age of 1.15~Gyr ($X_\mathrm{c}$\,=\,0.53), and reappears at 2.25~Gyr ($X_\mathrm{c}$\,=\,0.32).

There are two processes at work in making the core unstable against convection.
The first process is due to the fusion of \chem{He}{3} through the reaction \chem{He}{3} + \chem{He}{3} $\rightarrow$ \chem{He}{4} + 2~p.
It is responsible for the occurrence of a convective core right from the ZAMS in stars with $M$\,$\gtrsim$\mass{1.1} (at \Zinit\,=\,0.014).
That the \chem{He}{3} + \chem{He}{3} reaction is very energetic (it depends on the 18$^\mathrm{th}$ power of temperature at $12\times 10^6$~K) leads in those stars to the appearance of a convective core at ZAMS, with a mass that decreases with time as \chem{He}{3} burns.
The core temperatures at ZAMS are shown in Fig.~\ref{Fig:TcZAMS}.
The lower stellar mass at which the core is convective on the ZAMS corresponds to the beginning of the thick line at the left.
It is computed by extrapolation from the mass of the convective core as a function of stellar mass, and is equal to 0.984, 1.055, 1.113, 1.176, 1.215 and \mass{1.236} at \Zinit\,=\,0.006, 0.010, 0.014, 0.020, 0.030, and 0.040, respectively.

Thus one sees that the minimum mass for having a convective core on the ZAMS increases when the metallicity increases.
This comes from the fact that, for a given mass, hydrogen burning occurs at higher temperature when the metallicity decreases (since the luminosity of the star is higher).
This will allow the \chem{He}{3} + \chem{He}{3} reaction to release a greater amount of energy since the initial abundance of \chem{He}{3} does not depend \citep[or in a very marginal way, see e.g.][]{LagardeCharbonnelDecressin11} on the initial metallicity.
This favors the appearance of a convective core at lower initial mass at low metallicity.

The second process that makes the central regions of a main sequence star unstable against convection is the high temperature dependence of the CNO burning.
It comes into play when the temperature reaches a level high enough for the CNO cycles to outweigh  the p-p chains energetically.
This occurs for stellar masses higher than about \mass{1.2}, and is at the origin of the growth of the convective core in the second part of the MS as shown in Fig.~\ref{Fig:convCore}.
The evolution of the convective core in the models up to \mass{2} is shown in Fig.~\ref{Fig:convCoreHigherMasses}.
It is only for stellar masses above $\sim$\mass{1.7} that the mass of the convective core is a strictly decreasing function of time during the whole MS, since the CNO chains are highly efficient from the ZAMS in those stars.

The mass of the convective core is therefore not always a strictly decreasing function of time during the MS.
The growth of the convective core with time, especially at stellar masses between 1.2 and \mass{1.25}, implies mixing of fresh hydrogen from the outer layers into the H-depleting core.
The discrete time steps needed to compute evolutionary models result in sudden increases in the hydrogen abundance in the core, leading to discontinuous abundance profiles at its border.
These sudden additions of a discrete amount of hydrogen to the core may lead to `spikes' in the hydrogen mass fraction with time.
Figure~\ref{Fig:XcH}, which displays the evolution of $X_\mathrm{c}$ with time, shows that such spikes occur for stellar models in the range \mass{1.20-1.24}.
The hydrogen mass fraction in the core of MS stars is thus not necessarily a strictly decreasing function of time either.

We add two remarks concerning convective cores in MS models.
The first remark concerns the sensitivity of their extent at ZAMS to pre-MS modeling.
\cite{MorelProvostBerthomieu00} performed a study of solar models and compared the evolution of models when the pre-MS phase is accounted for and when it is not.
The authors conclude that both sets of models become nearly identical as soon as the nuclear reactions start to work at equilibrium.
With the ZAMS defined when $X_\mathrm{c}(\mathrm{H})$ has decreased by 0.25\% in our models (see Sect.~\ref{Sect:basicTracks}), a value chosen to ensure that nuclear equilibrium is reached for all masses and metallicities in our grids, our MS models are independent of the pre-MS evolution.
We also note that with our definition of the ZAMS, our solar track does not have a convective core at ZAMS.
The convective core that develops at the early stage of the solar track \citep[see][]{MorelProvostBerthomieu00} has exhausted by the time of the ZAMS.

The second remark concerns the sensitivity of the extent of the convective cores on the constitutive physics of the models.
\cite{MagicSerenelliWeiss_etal10} show, for example, how the constitutive physics of the models, such as the extension of the convective core, can reproduce a hook in the critical mass range \mass{1.2-1.3}.
The study was done in relation to the open cluster M67, which has a turn-off mass precisely in that mass range and which displays a clear hook in the color-magnitude diagram (CMD).
We discuss this case further in Sect.~\ref{Sect:comparisons}

\subsection{Main sequence lifetimes}
\label{Sect:ages}

\begin{figure}
  \centering
  \includegraphics[width=\columnwidth]{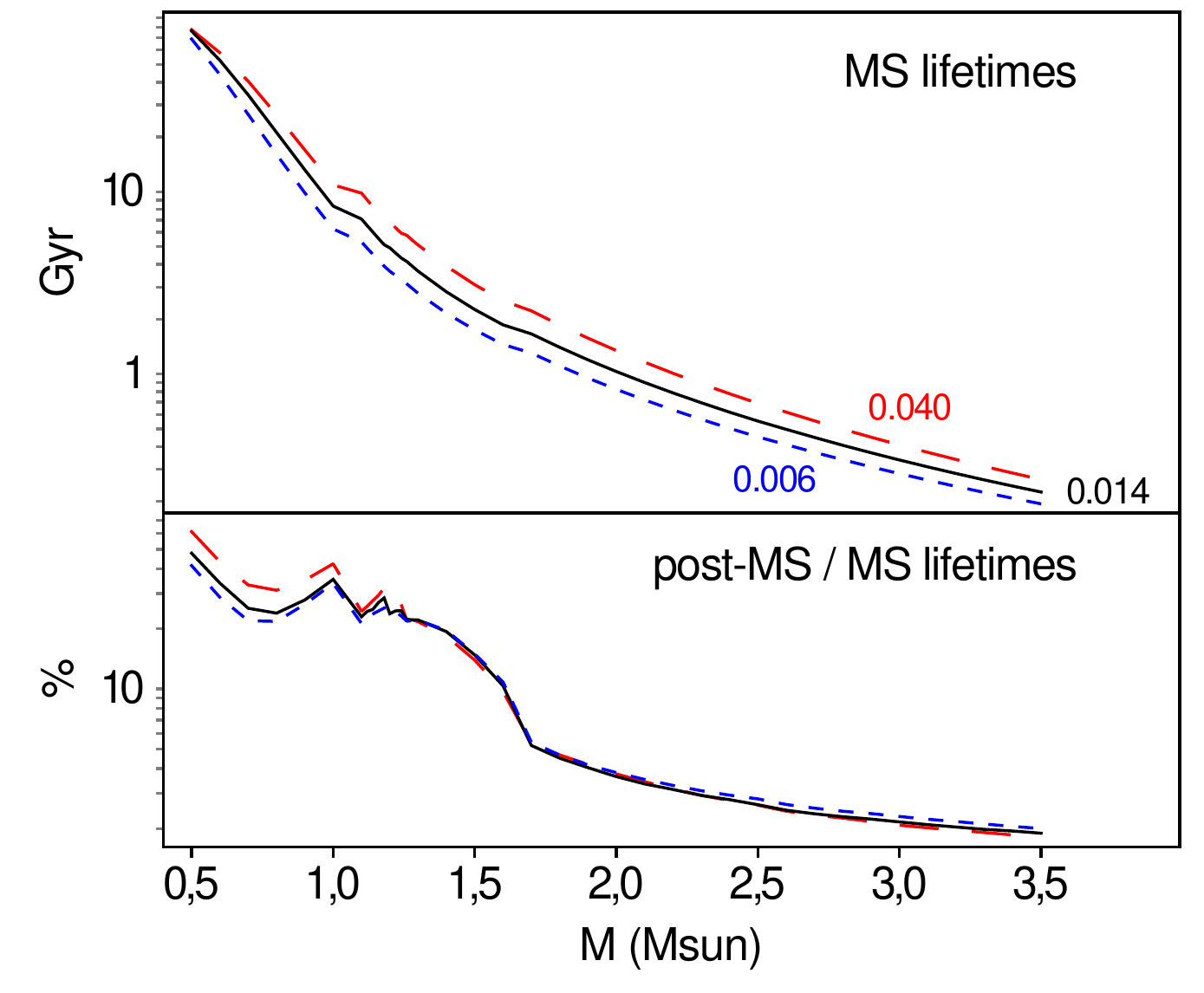}
  \caption{\textbf{Upper panel:} Main sequence lifetimes in Gyr as a function of initial stellar mass at three different metallicities: $Z_\mathrm{init}$\,=\,0.006 (short-dashed blue line), 0.014 (continuous black line), and 0.040 (long-dashed red line).
  \textbf{Lower panel:} Same as upper panel, but for the time, relative to the MS lifetime, needed to the models to cross the HR diagram from the end of the MS to the bottom of the RGB.
  }
\label{Fig:MsAges}
\end{figure}

\begin{figure}
  \centering
  \includegraphics[width=\columnwidth]{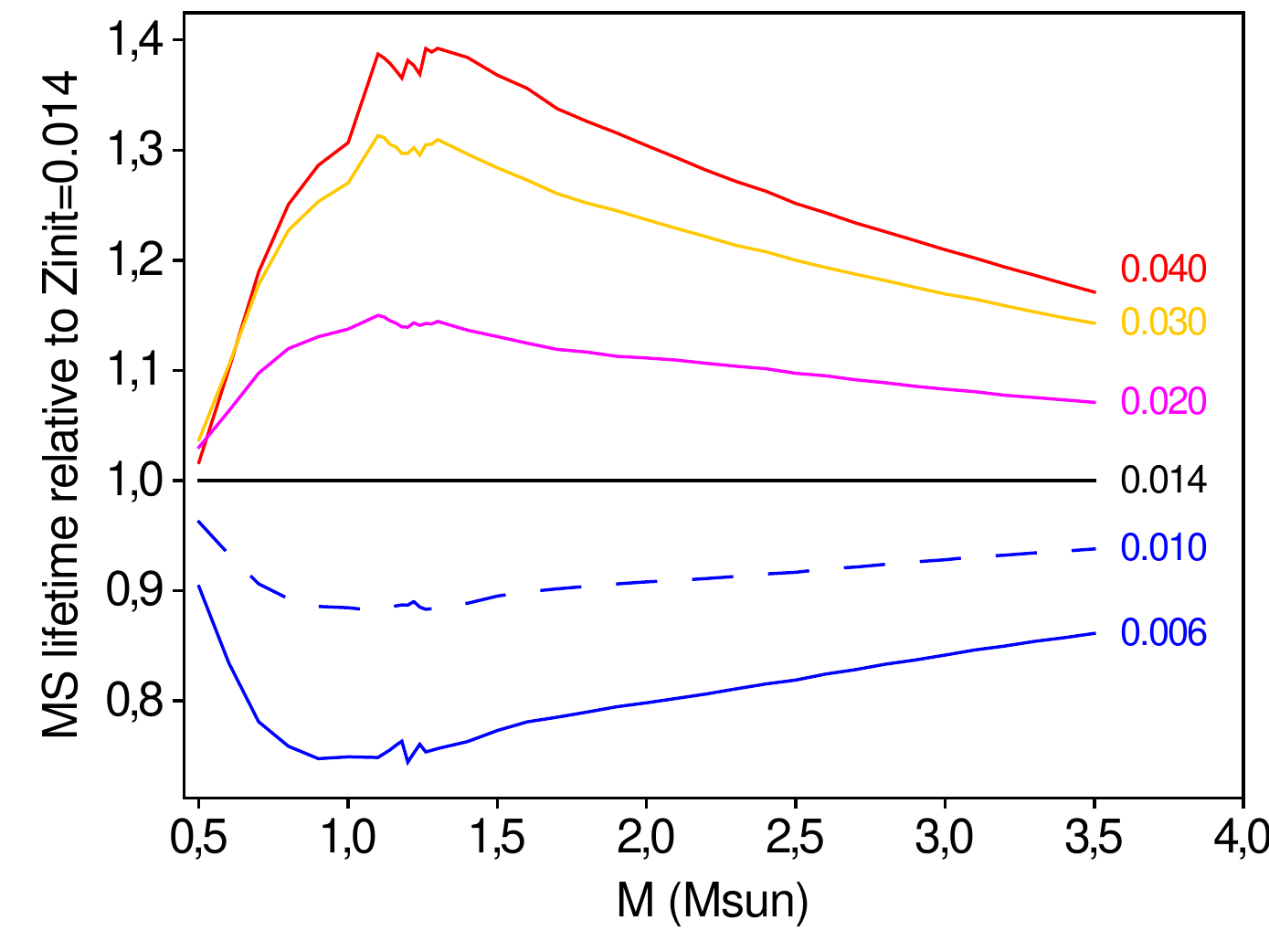}
  \caption{Main sequence lifetimes, as a function of initial stellar mass, of models at different metallicities as labeled next to the curves.
  The lifetimes are displayed relative to the lifetime of the considered stellar mass at \Zinit\,=\,0.014.
  }
\label{Fig:MsAgesM}
\end{figure}

\begin{figure}
  \centering
  \includegraphics[width=\columnwidth]{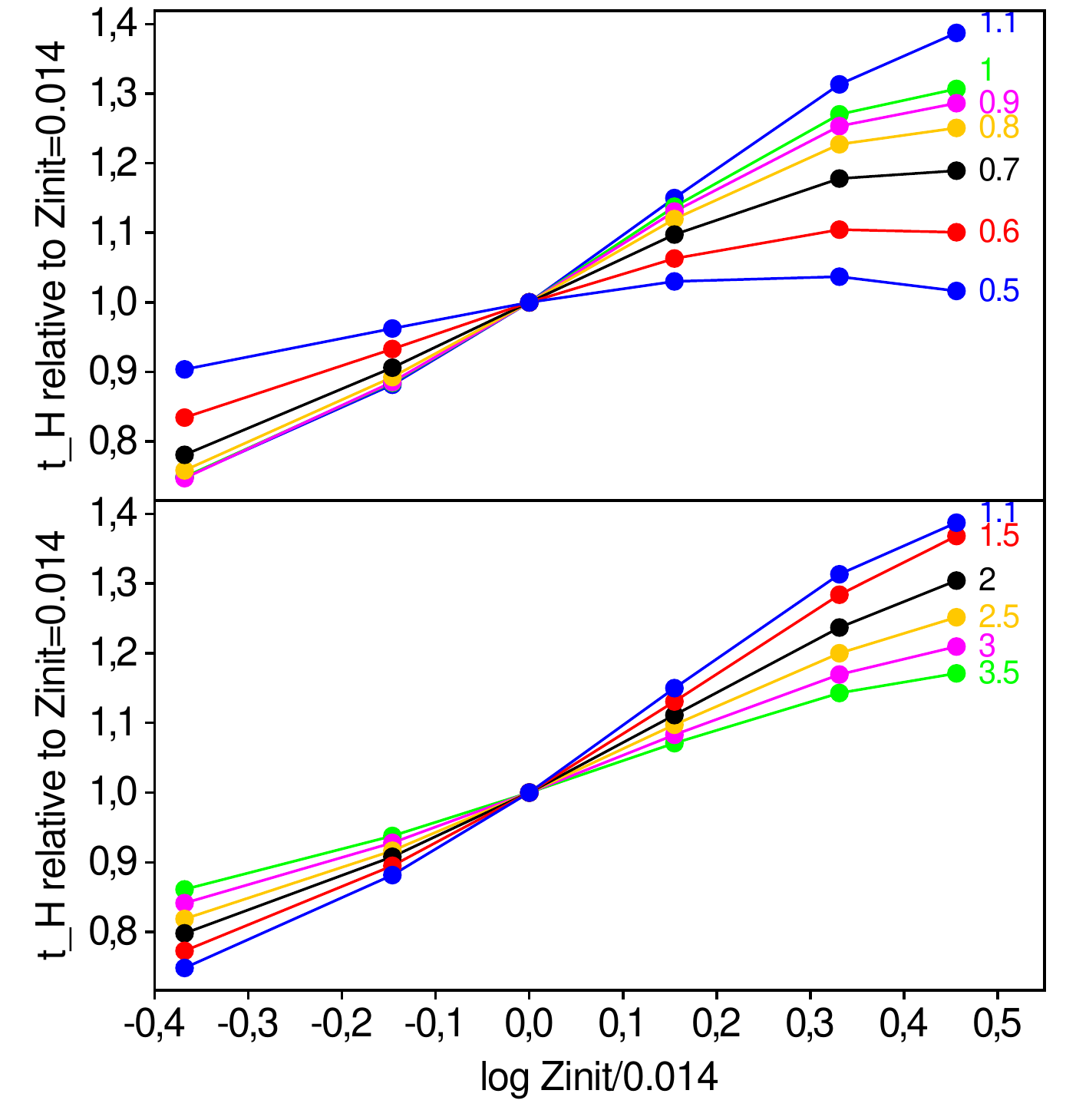}
  \caption{\textbf{Upper panel:} Main sequence lifetimes as a function of the logarithm of the metallicity relative to solar, for models from 0.5 to \mass{1.1} (bottom to top curves) as labeled on the right of the curves.
  The lifetimes are normalized for each stellar mass to its value at $Z_\mathrm{init}$\,=\,0.014.
  \textbf{Lower panel:} Same as upper panel, but for selected models from 1.1 to \mass{3.5} (top to bottom curves).
  }
\label{Fig:MsAgesZ}
\end{figure}

Figure~\ref{Fig:MsAges} displays the MS lifetimes (upper panel), $t_\mathrm{H}$, as a function of stellar mass at three metallicities.
It is a well known decreasing function of stellar mass, mainly due to the increase in stellar luminosity with stellar mass.
The time needed to reach the base of the RGB from the TAMS is plotted in the lower panel of Fig.~\ref{Fig:MsAges}.
It amounts to about 30\% of $t_\mathrm{H}$ for models with masses below $\sim$\mass{1.2}, decreases rapidly above that mass to $\sim$3\% at \mass{1.8}, and then does so more slowly for more massive stars, reaching $\sim$2\% at \mass{3.5}.

The sensitivity of $t_\mathrm{H}$ with stellar mass can be analyzed, at any given metallicity, by comparison with the lifetimes at \Zinit\,=\,0.014.
This comparison, shown in Fig.~\ref{Fig:MsAgesM}, reveals a maximum sensitivity of $t_\mathrm{H}$ to $Z$ at stellar masses between 0.8 and \mass{1.5}.
At lower masses, $t_\mathrm{H}$ becomes less sensitive to $Z$, since it is almost equal in all \mass{0.5} models with \Zinit\,$\geqslant$\,$Z_{\odot,\mathrm{init}}$.
This is due to the independence of the p-p chain to metals and to the lower sensitivity of the p-p reaction rates to temperature.

The sensitivity of $t_\mathrm{H}$ with $Z$ is shown in Fig.~\ref{Fig:MsAgesZ}.
The plot confirms the results highlighted above from Fig.~\ref{Fig:MsAgesM}, according to which the highest sensitivity is around \mass{1.2} and the lowest at the smallest mass \mass{0.5} considered in our grids.
The figure reveals, however, a pattern of $t_\mathrm{H}$ as a function of $Z$ that differs from the pattern found in \cite{MowlaviMeynetMaeder_etal98} (hereafter MMM98) with the old Geneva models.
While the \mass{3} was found by MMM98 to indeed be less sensitive to metallicity than the \mass{1} models (see their Fig.~5), in agreement with the results of this paper, the models at metallicities higher than solar showed a trend of decreasing $t_\mathrm{H}$ with $Z$, which is not found in this paper.
This results first from the adoption in the present work of the solar metallicity of Asplund et al. (2005), implying that a given value of $\Zinit/Z_{\odot,\mathrm{init}}$ corresponds now to a lower value of \Zinit.
In the present work, we therefore never reach the very extreme metallicities explored in MMM98.
Second, the slope $\Delta Y/\Delta Z$\,=\,1.2857 adopted in this paper is a factor of almost two lower than in MMM98.
This prevents to have the significant drop of the initial H abundance recorded at high metallicities in MMM98 and which contributed to their decrease of $t_\mathrm{H}$ with metallicity.
Moreover, many physical ingredients have been updated, such as the opacities, some nuclear reaction rates, and the overshooting parameter, which may of course contribute to explaining this difference as well.

In the cases where the mass of the convective core grows with time, the addition of fresh hydrogen naturally increases the main sequence lifetime.
The question then arises whether numerical simulations do adequately model this increase.
The evolution of the mass of the convective core with time (Fig.~\ref{Fig:convCore}) suggests that the global evolution of the convective core during the MS is not affected much by the numerical spikes.
It must also be noted that the effective temperature and the surface luminosity of the models
are evolving smoothly with time irrespective of the occurrence of spikes.
We thus assume that the MS lifetimes are not affected by those spikes.
The situation is quite different from what happens during the core helium-burning phase where spikes in the core helium abundance appear in the growing core at a late stage of helium burning, when the mass fraction of helium has decreased below 10\%.
The spikes at that stage lead to a significant increase in the core helium-burning lifetime, the later becoming sensitive to the numerical parameters adopted in the simulations.
We do not expect that difficulty during the MS.

\subsection{Surface abundances}
\label{Sect:surfaceAbundances}

\begin{figure}
  \centering
  \includegraphics[width=\columnwidth]{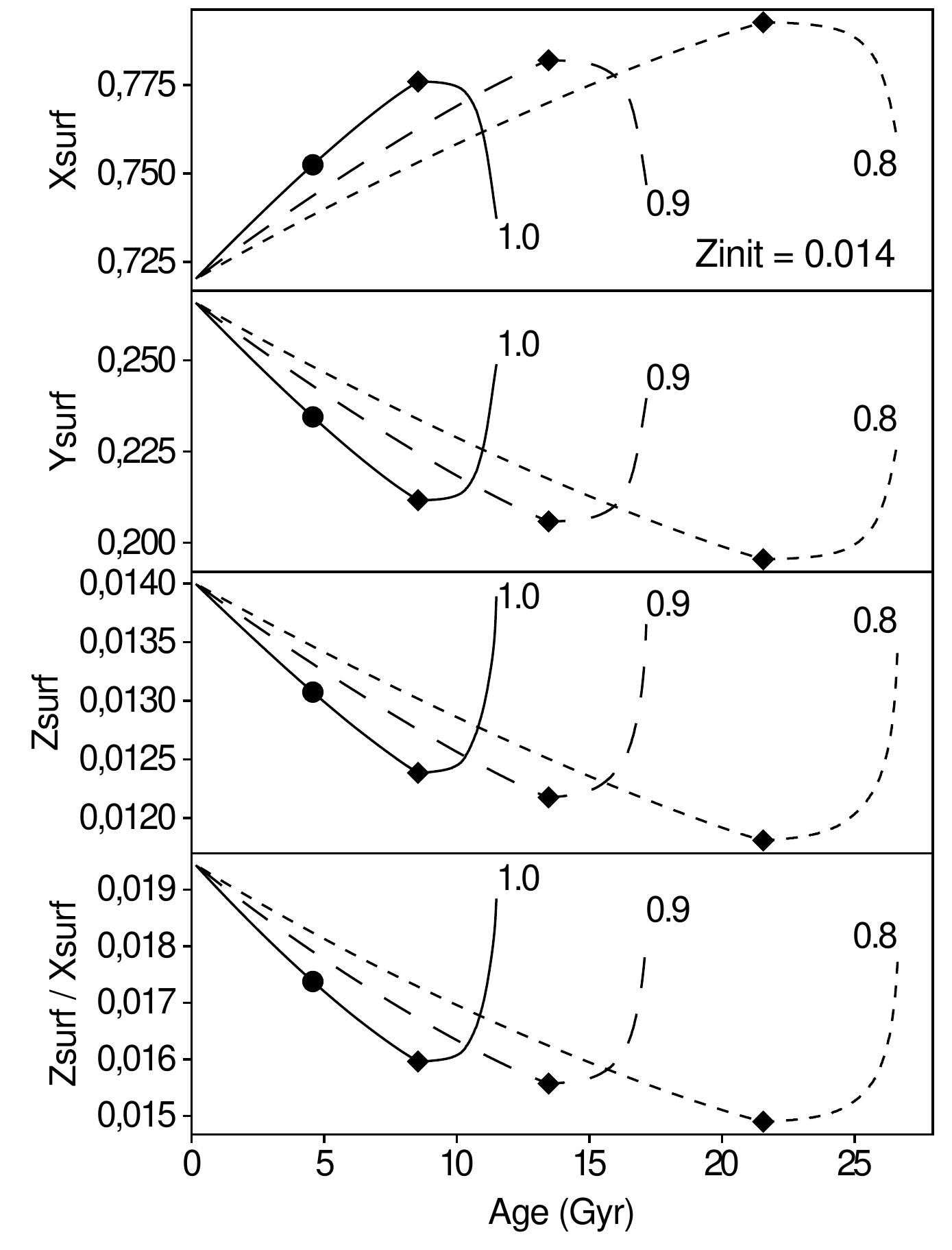}
  \caption{Hydrogen mass fraction (upper panel), helium mass fraction (second panel from top) and metallicity (two lower panels) at the surface of the 0.8 (dashed-short), 0.9 (dashed-long) and \mass{1.0} (continuous) \Zinit\,=\,0.014 models as a function of age.
  Filled diamonds indicate the end of the main sequence phase.
  Filled circles on the \mass{1} tracks locate the solar model at an age of 4.57~Gyr.
  }
\label{Fig:XsDiffusion}
\end{figure}

\begin{figure}
  \centering
  \includegraphics[width=\columnwidth]{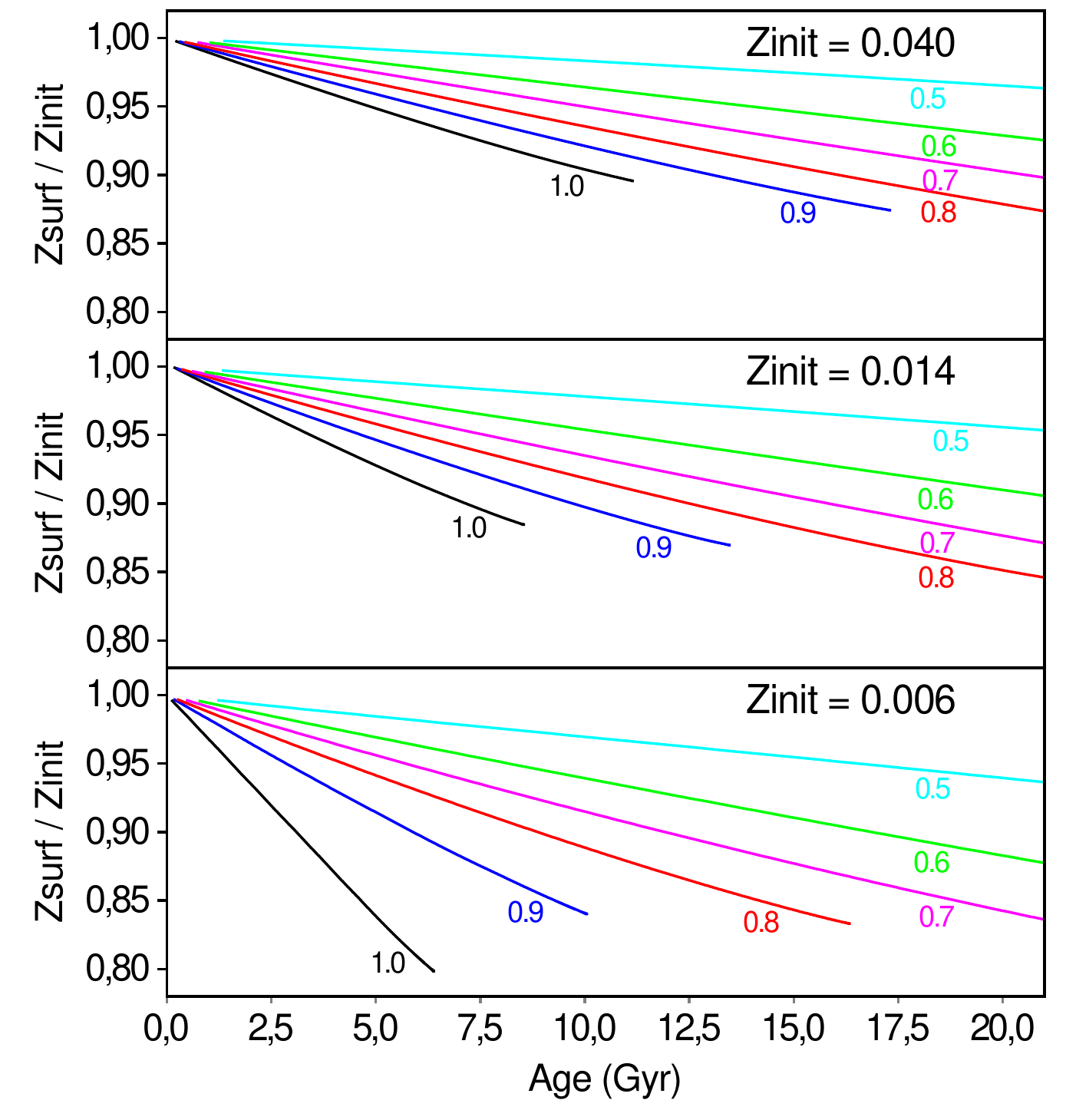}
  \caption{Evolution of the surface metallicity during the MS in all our models with $M$\,$\leqslant$\,\mass{1} at $Z_\mathrm{init}$\,=\,0.04 (upper panel), 0.014 (middle panel), and 0.006 (lowest panel).
  The X-axis has been restricted to the age interval [0-20]~Gyr.
  }
\label{Fig:XsDiffusionZ}
\end{figure}

Atomic diffusion leads to nonconstant surface abundances with time.
In the \mass{1.0}, \Zinit\,=\,0.014 model, for example, the surface hydrogen mass fraction $X_\mathrm{surf}$ increases by about 7\% during the main sequence (MS), while that of helium, $Y_\mathrm{surf}$,  decreases by almost 20\%.
This is displayed in, respectively, the upper and second panels from the top of Fig.~\ref{Fig:XsDiffusion}.
As a result, the surface metallicity decreases by 11\% during the MS of that star, reaching $Z_\mathrm{surf}$\,=\,0.0124 at the end of that phase (third panel from top of Fig.~\ref{Fig:XsDiffusion}).
It evolves back to its initial value during the post-MS phase as the convective envelope deepens into the star and homogenizes back the diffused elements.
Similar behaviors are seen in the evolution of the surface abundances at stellar masses below \mass{1.0}, as confirmed in Fig.~\ref{Fig:XsDiffusion} for the 0.9 and \mass{0.8} models at \Zinit\,=\,0.014.
The change of the surface metallicity due to diffusion is thus not negligible during the MS phase.

Figure~\ref{Fig:XsDiffusionZ} further displays the evolution of the surface metallicity during the MS for all our models that include diffusion at \Zinit\,=\,0.04 (upper panel), 0.014 (middle panel) and 0.006 (lowest panel).
The fraction of metal that depletes from the surface owing to atomic diffusion into the deeper layers of the star decreases with increasing metallicity for a given stellar mass, due to the different densities at the base of the convective envelope.
As a result, 10~\% of the metal mass fraction is depleted by the end of the MS in the \mass{1} model at $\Zinit$\,=\,0.04, and 20~\% at 0.006.

\subsection{Global asteroseismic quantities}
\label{Sect:asteroseismology}

\begin{figure}
  \centering
  \includegraphics[width=\columnwidth]{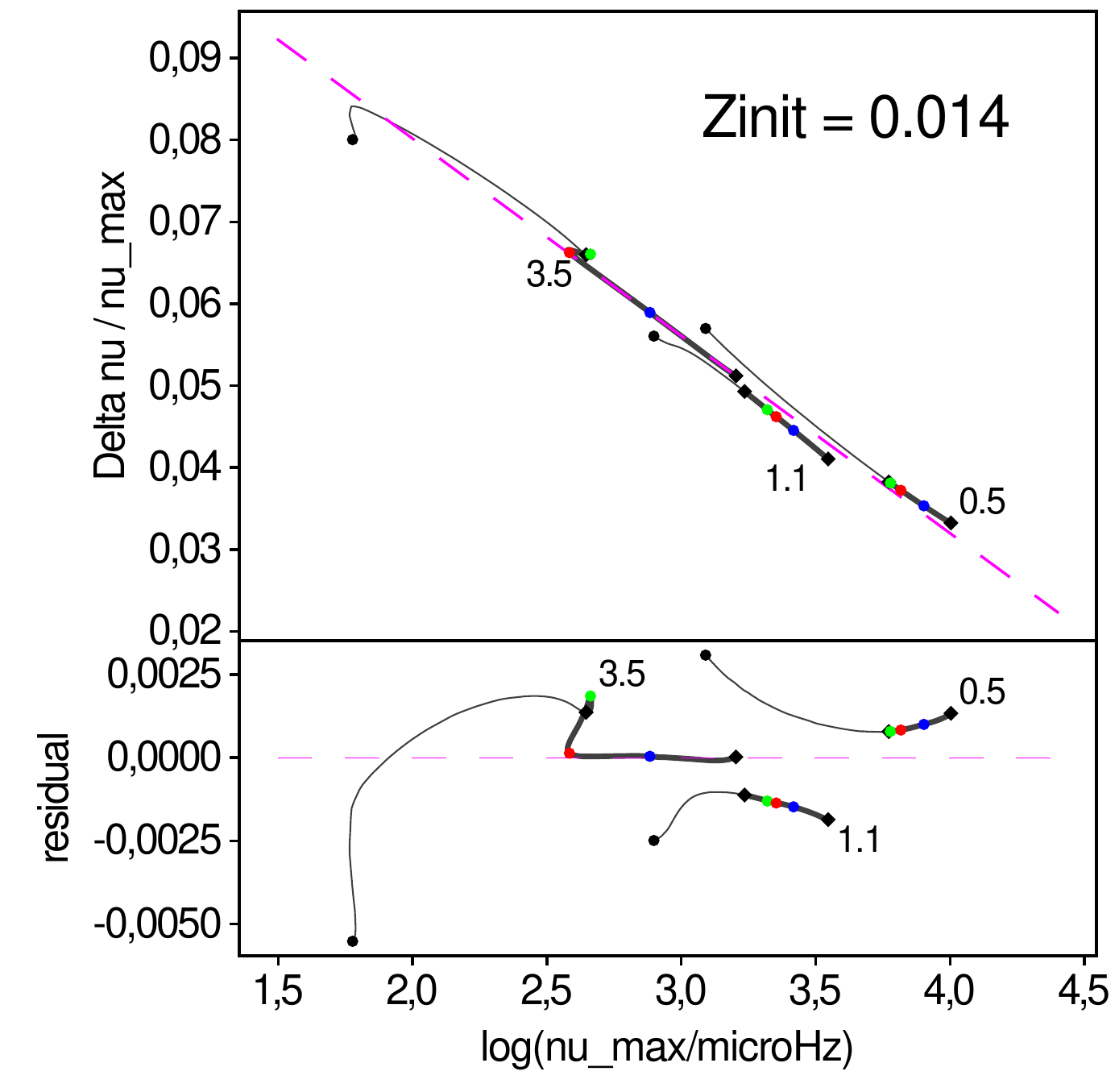}
  \caption{\textbf{Top panel:} Evolutionary tracks at $\Zinit$\,=\,0.014 in the $\Delta \nu/\nu_\mathrm{max}$ versus $\log(\nu_\mathrm{max})$ diagram.
  For clarity, only the 0.5, 1.1, and \mass{3.5} tracks are plotted, as labeled next to the tracks.
  The main sequence is shown in thicker lines.
  The ZAMS is located on the right of each track.
  Reference points are indicated with symbols as in Fig.~\ref{Fig:HRZ014}.
  The dashed line is a straight line passing through the ZAMS and the first MS hook of the \mass{3.5} star, given by Eq.~\ref{Eq:DeltaNuOverNuMax}.
  The frequencies are expressed in $\mu$Hz.
  \textbf{Bottom panel}: same as top panel, but for the residual of $\Delta \nu/\nu_\mathrm{max}$ with respect to the straight line in the top panel.
  \label{Fig:asteroDeltaNuZ}
  }
\end{figure}

By providing a wealth of information on the internal structure of the Sun, the solar five-minute oscillations stimulated various attempts to obtain similar observations for other stars.
These past years, a growing number of detections of solar-like oscillations have been obtained for different kinds of stars by ground-based spectrographs and space missions.
Scaling relations relating asteroseismic parameters to global stellar properties are largely used to analyze these data.
These global asteroseismic quantitites are indeed of great help to constrain stellar parameters and to obtain thereby new information about stellar evolution without the need to perform a full asteroseismic analysis \cite[see e.g.][]{StelloChaplinBruntt_etal09,MiglioMontalbanBaudin_etal09,ChaplinKjeldsenChristensenDalsgaard_etal11}.
In this section, we provide some asteroseismic properties of our models that can be directly computed from their global properties using scaling relations:
\\
\begin{itemize}
\item the large frequency separation $\Delta \nu$ that is proportional to the square root of the mean density of the star \cite[e.g.][]{Ulrich86}:
  \begin{equation}
    \Delta \nu = \Delta \nu_\odot \times \frac{(M/M_\odot)^{0.5}}{(R/R_\odot)^{1.5}} \;,
  \label{Eq:astro_DeltaNu}
  \end{equation}
with $\Delta \nu_\odot$\,=\,134.9~$\mu$Hz \citep{ChaplinKjeldsenBedding_etal11};
\vskip 2mm
\item the frequency $\nu_\mathrm{max}$ at the peak of the power envelope of the oscillations where the modes exhibit their strongest amplitudes, given by \citep{BrownGillilandNoyesRamsey91,KjeldsenBedding95}:
  \begin{equation}
    \nu_\mathrm{max} = \nu_{\mathrm{max},\odot} \times \frac{M/M_\odot}{(R/R_\odot)^2 \; (\Teff/\Teff_{,\odot})^{0.5}} \; ,
  \label{Eq:astro_numax}
  \end{equation}
with $\nu_{\mathrm{max},\odot}$\,=\,3150~$\mu$Hz \citep{ChaplinKjeldsenBedding_etal11};
\vskip 2mm
\item and the maximum oscillation amplitude $A_\mathrm{max}$ \cite[e.g.][]{ChaplinKjeldsenBedding_etal11}:
  \begin{equation}
   A_\mathrm{max} = A_{\mathrm{max},\odot} \times \left(\frac{L}{L_\odot}\right)^s \left(\frac{M}{M_\odot}\right)^{-t} \left(\frac{\Teff}{\Teff_{,\odot}}\right)^{-r}
  \label{Eq:astro_Amax}
  \end{equation}
  where the solar value $A_{\mathrm{max},\odot}$ of the maximum oscillation amplitude depends on the instrument used to measure the oscillations.
For the Kepler mission, for example, $A_{\mathrm{max},\odot} \approx 2.5$~ppm \citep{ChaplinKjeldsenBedding_etal11}.
  Following \cite{KjeldsenBedding95} we adopt $r$\,=\,2 and take $s$\,=\,0.838 and $t$\,=\,1.32 from \cite{HuberBeddingStello_etal11}.
\end{itemize}

\begin{figure}
  \centering
  \includegraphics[width=\columnwidth]{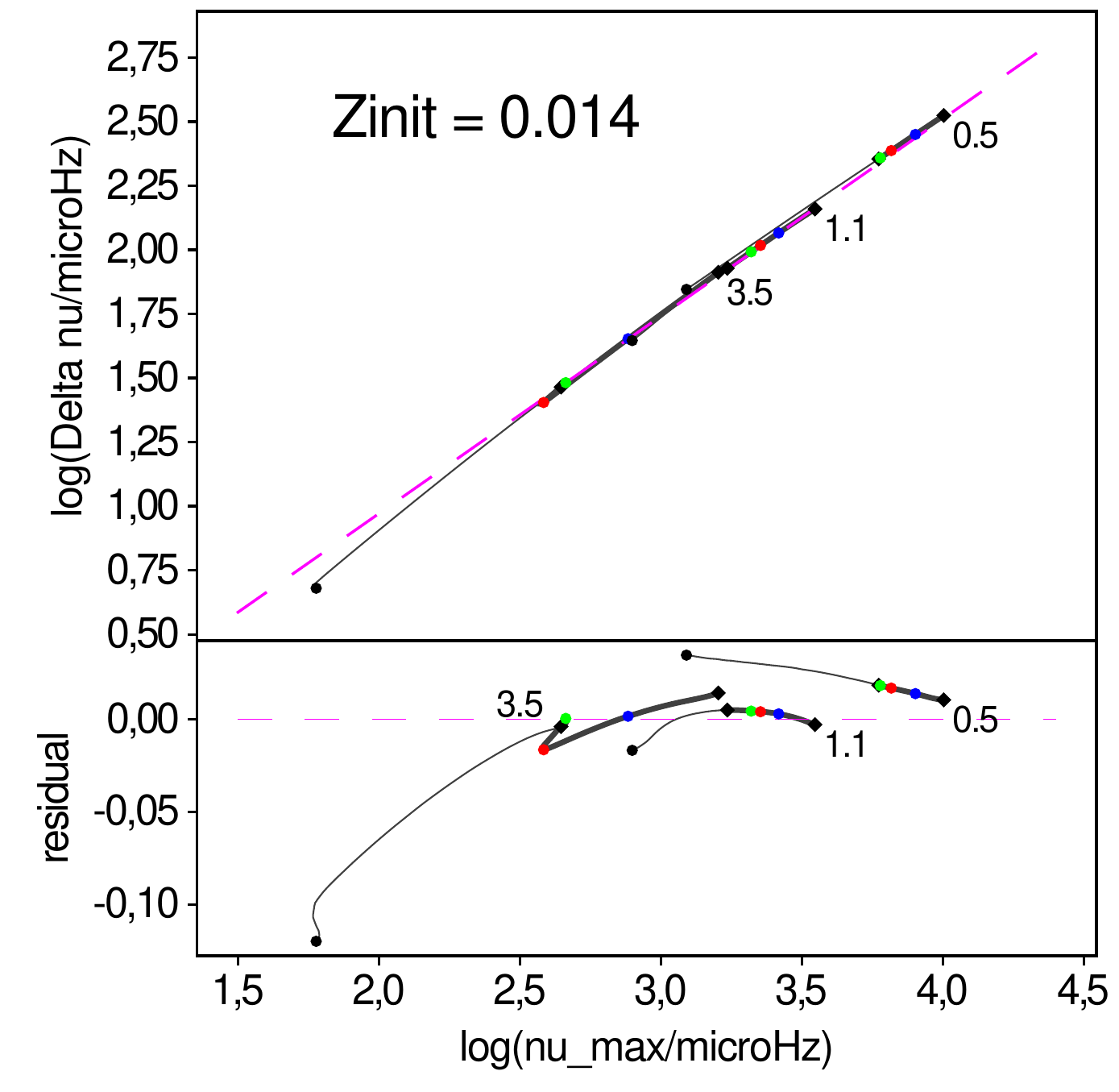}
  \caption{Same as Fig.~\ref{Fig:asteroDeltaNuZ}, but for $\log(\Delta \nu)$ versus $\log(\nu_\mathrm{max})$.
  The dashed line in the top panel is a straight line with slope 0.77 passing through the ZAMS of the \mass{1.1} model.
  }
\label{Fig:asteroLogDeltaNuZ}
\end{figure}

\begin{figure}
  \centering
  \includegraphics[width=\columnwidth]{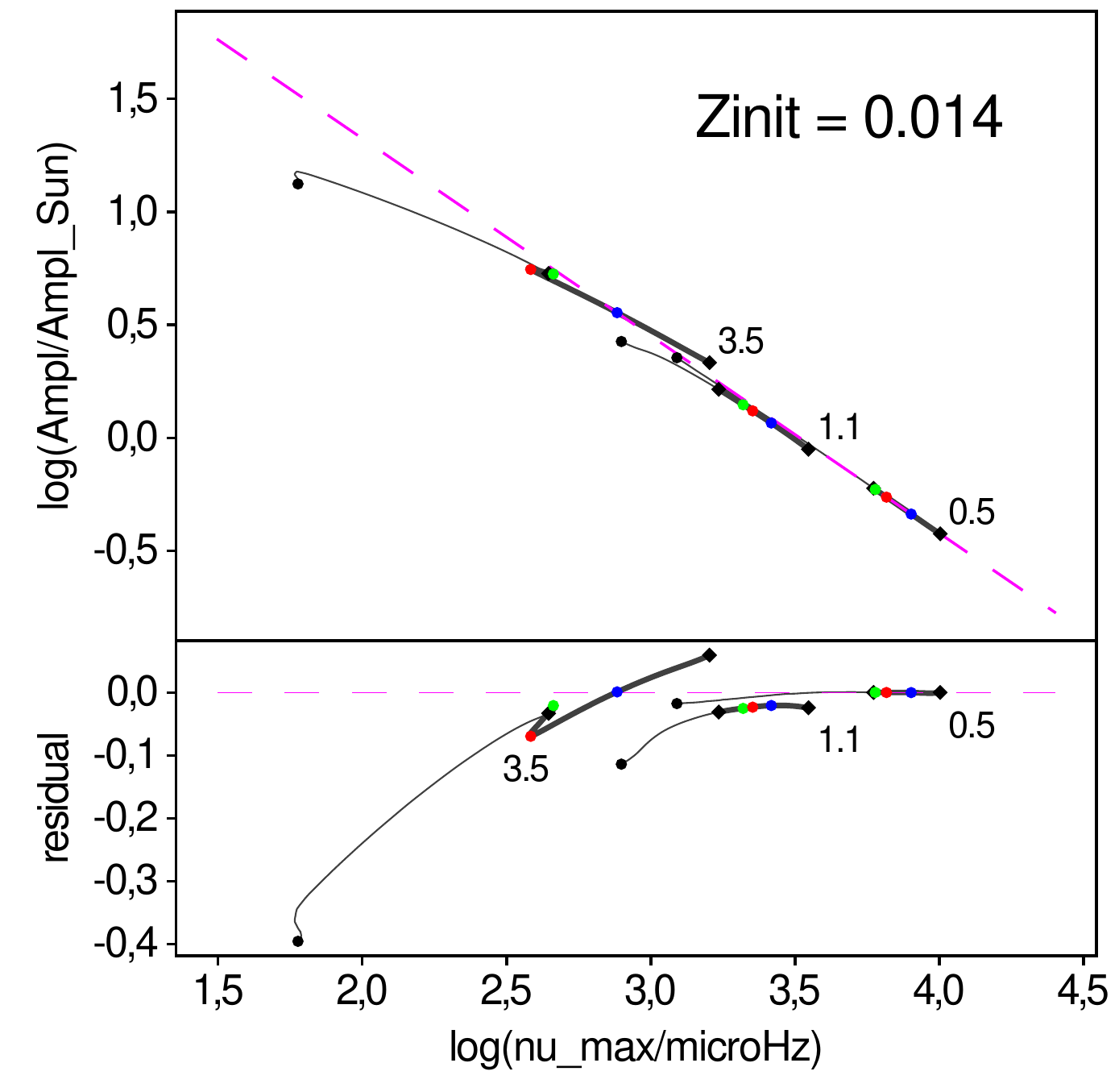}
  \caption{Same as Fig.~\ref{Fig:asteroDeltaNuZ}, but for $A_\mathrm{max}/A_{\mathrm{max},\odot}$ versus $\nu_\mathrm{max}$.
   The dashed line in the top panel is a straight line passing through the ZAMS and TAMS of the \mass{0.5} star, given by Eq.~\ref{Eq:AmaxOverAmaxSun}.
  }
\label{Fig:asteroAmaxZ014}
\end{figure}

\noindent We do not test here the validity of the scaling relations \ref{Eq:astro_DeltaNu} to \ref{Eq:astro_Amax}, but simply assume that these relations hold for the different masses and metallicities computed for these grids.

The asteroseismic data of our grids are summarized in the top panels of Figs.~\ref{Fig:asteroDeltaNuZ} to \ref{Fig:asteroAmaxZ014}.
Figure~\ref{Fig:asteroDeltaNuZ} displays $\Delta \nu/\nu_\mathrm{max}$, Fig.~\ref{Fig:asteroLogDeltaNuZ} $\log \Delta \nu$
and Fig.~\ref{Fig:asteroAmaxZ014} $A_\mathrm{max}/A_{\mathrm{max},\odot}$, all three versus $\log(\nu_\mathrm{max})$.
The oscillation frequencies and the large frequency separation decrease during the evolution of the stars, as expected from the decrease inof their mean density (cf. Fig.~\ref{Fig:TeffMeanRhoZ014}), while the amplitude of their oscillation increases.

The MS tracks are seen to obey an almost linear relation in each of the three figures.
Suggested linear relations are shown by the dashed lines in the top panels of the figures.
The residuals of the tracks with respect to that linear relation are shown in the bottom panels.
In the [$\log(\nu_\mathrm{max})$, $\Delta \nu/\nu_\mathrm{max}$] plane (Fig.~\ref{Fig:asteroDeltaNuZ}), the linear relation is constructed  based on the ZAMS and the point of first hook of the \mass{3.5} star.
It is given by
\begin{equation}
  \Delta \nu \, / \, \nu_\mathrm{max} = 0.128 - 0.0241 \; \log(\nu_\mathrm{max}) \;.
\label{Eq:DeltaNuOverNuMax}
\end{equation}
The coefficients of this relation vary only slightly with metallicity, the relation being $\Delta \nu \, / \, \nu_\mathrm{max} = 0.133-0.0249\;\log(\nu_\mathrm{max})$ at \Zinit\,=\,0.006 and $\Delta \nu \, / \, \nu_\mathrm{max} = 0.124-0.0235\;\log(\nu_\mathrm{max})$ at \Zinit\,=\,0.04.

In the [$\log(\nu_\mathrm{max})$, $\log(\Delta \nu)$] plane (Fig.~\ref{Fig:asteroLogDeltaNuZ}), we adopt a linear relation with a slope of 0.77 as derived observationally \citep{StelloChaplinBasu09} and passing through the ZAMS of the \mass{1.1} track.
The residuals displayed in the bottom panel of Fig.~\ref{Fig:asteroLogDeltaNuZ} show very good agreement with our predictions.
Our models actually predict a slope of 0.74 for the MS track of the 0.5 and \mass{1.1}, and a larger slope of 0.82 for the MS track of the \mass{3.5}.
 
In the [$\log(\nu_\mathrm{max})$, $A_\mathrm{max}/A_{\mathrm{max},\odot}$] plane, the linear relation is constructed  based on the ZAMS and the TAMS of the \mass{0.5}.
It is given by
\begin{equation}
  \log (A_\mathrm{max}/A_{\mathrm{max},\odot}) = 3.072 - 0.8734 \; \log(\nu_\mathrm{max}) \;.
\label{Eq:AmaxOverAmaxSun}
\end{equation}
The coefficients of this relation vary only slightly with metallicity as well.
The relation is $\log (A_\mathrm{max}/A_{\mathrm{max},\odot}) = 3.111 - 0.8789 \; \log(\nu_\mathrm{max})$ at \Zinit\,=\,0.006 and $\log (A_\mathrm{max}/A_{\mathrm{max},\odot}) = 3.037 - 0.8704 \; \log(\nu_\mathrm{max})$ at \Zinit\,=\,0.04.

\section{From tracks to isochrones}
\label{Sect:tracks}

We computed 234 \textsl{raw tracks} of evolutionary model, one for each \{$\Zinit, M$\} couple taken from the list (\ref{Eq:gridDefinition}).
The raw tracks start with an initial model close to the ZAMS and stop when the stellar model reaches the base of the RGB.
The time steps between two successive models in each sequence are chosen to ensure good modeling of both the chemical (nucleosynthesis and atomic diffusion) and physical evolution.
They vary with mass, metallicity and stage of evolution of the stellar models. 

In practice, for computing isochrones, we use tracks, hereafter called \textsl{basic tracks}, described by fewer models than those defining the raw tracks.
The models in the basic tracks are chosen in order to ensure reliable interpolation in mass or in metallicity between them.
Interpolations are necessitated because, in order to compute isochrones, we need to construct tracks for initial masses (and metallicities) that are not among the 234 raw tracks.
Besides this, these interpolations have to be made between models that are at equivalent evolutionary stages.
By this, we mean models that are either at the same evolutionary stage (for instance, the same mass fraction of hydrogen in the center or at the same relative time) or at the same position in the HR diagram (for instance, at the red turn off).
We construct in this way 234 basic tracks, each consisting of 601 well chosen stellar models.
In each basic track, the models with the same numbering are all equivalent, in the sense defined above.
Each basic track is identified by a combination $\{\Zinit, M\}$ taken from (\ref{Eq:gridDefinition}).

As explained above, evolutionary tracks at a \{\Zinit, $M$\} couple not available in the set of basic tracks can be computed by interpolating among basic tracks.
They are thus not stored in files, but can be computed on demand through the web page (see below).
They are called \textsl{interpolated tracks}.

In addition to the evolutionary tracks, we also construct lines in the HR diagram linking models of equal initial mass, having different initial metallicities and ages, but showing equal surface abundances.
We call these lines \textsl{iso-$\Zsurf$ lines}.
Each iso-$\Zsurf$ line is identified by a pair \{\Zsurf, $M$\}.

Finally, to be complete, we recall that \textsl{isochrones} are lines linking models having the same age and initial metallicity.
They are identified by a couple \{$\Zinit$, age\}.
The basic tracks, interpolated tracks, iso-$\Zsurf$ lines and isochrones are respectively described in Sects.~\ref{Sect:basicTracks}, \ref{Sect:interpolatedTracks}, \ref{Sect:isoZsurfTracks}, and \ref{Sect:isochrones}.

\subsection{Basic tracks}
\label{Sect:basicTracks}

\begin{figure}
  \centering
  \includegraphics[width=\columnwidth]{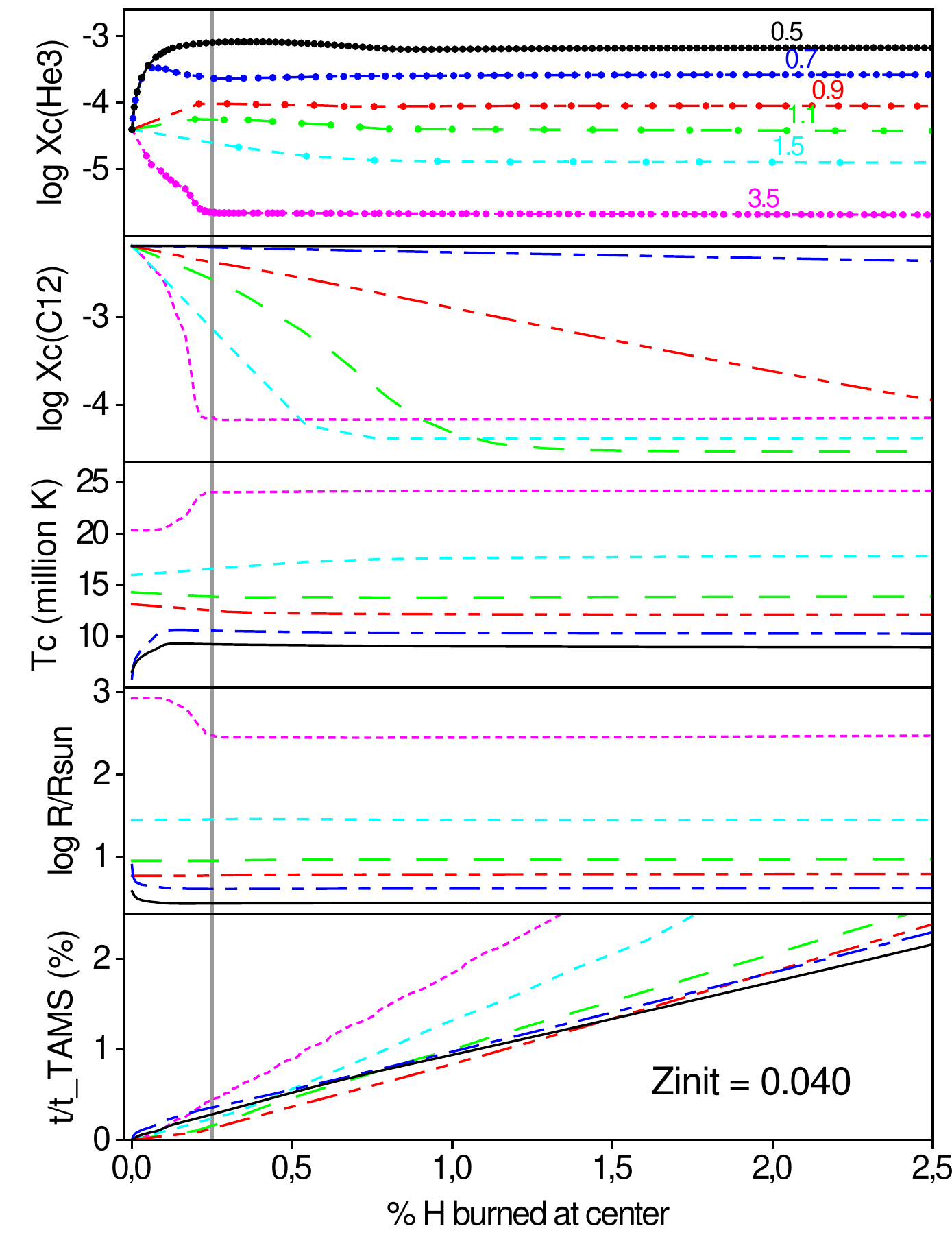}
  \caption{Evolution of several quantities in raw models at $Z_\mathrm{init}$\,=\,0.04 as a function of the percentage of hydrogen burned at the center.
  {\bf From top panel to bottom:}
  logarithm of the \chem{He}{3} mass fraction at the center of the star $\log X_\mathrm{c}(\chem{He}{3})$,
  $\log X_\mathrm{c}(\chem{C}{12})$,
  core temperature in $10^6$~K,
  logarithm of the stellar radius relative to the solar radius
  and percentage of time elapsed since the first raw model relative to the age at TAMS.
  Models of 0.5 (continuous lines), 0.7 (long-short dashed lines), 0.9 (long-short-short dashed lines), 1.1 (long dashed lines), 1.5 (short dashed lines), and \mass{3.5} (dotted lines) are displayed, as labeled in the top panel.
  Each marker in the top panel represents a computed model.
  The vertical line indicates the percentage of burned hydrogen taken to define the ZAMS in all our grids.
  }
\label{Fig:ZamsDef}
\end{figure}

Basic evolutionary tracks are constructed from the raw evolutionary tracks by first identifying reference points on each track in the $\log R-\log L$ diagram, and then defining `equivalent' models by regularly distributing them between the reference points.
We refer to Appendix~\ref{Sect:normalizationProcedure} for a detailed description of the reference points.

The ZAMS is defined at the time when $X_\mathrm{c}(\mathrm{H})$ has decreased by 0.25\% at the center of the star.
It ensures that the models have reached a gravitationally stable configuration.
This is shown in Fig.~\ref{Fig:ZamsDef}, which displays, among other quantities, the evolution of the stellar radius (second panel from bottom) and of the central temperature (third panel from bottom) as a function of the fraction of central H burned in the case of \Zinit\,=\,0.04 models.
The equilibrium state is reached at an even earlier stage of H burning in the grids of lower metallicities.
 
The time taken by our initial models to reach the ZAMS amounts to 0.1 to 0.6\% of the MS durations, depending on the mass of the star\footnote{
This `pre-MS' time has no physical meaning since the pre-MS phase is not followed in our simulations.
}.
By that time, the mass fractions of some nuclides have already been substantially modified in their core.
Most noticeable is \chem{He}{3}, the evolution of which, at the center of the stars, is shown in Fig.~\ref{Fig:ZamsDef} (top panel) for selected stellar masses.
\chem{He}{3} is produced by deuterium burning during the pre-MS phase in all our stars, but is destroyed in more massive, hotter stars through \chem{He}{3}\,+\,\chem{He}{3}.
The net result, at the time of ZAMS, is an enrichment of \chem{He}{3}, with respect to its initial abundance, in the core of stars less massive than \mass{1.1} and a depletion in stars more massive than this mass.
During the MS phase, the evolution of \chem{He}{3} is dictated by the pp chain.

The ages of our models are taken relative to the initial models.
The ages at ZAMS are thus not zero, but reflect the time necessary for the initial models to get to thermal equilibrium and burn 0.25\% of hydrogen in their core (see Sect.~\ref{Sect:convectiveCores} for further discussion).
The ZAMS model of our solar track, for example, has an age of 190 million years, and it takes an extra 4.36 billion years for the ZAMS model to reach solar luminosity, resulting in an age of the solar model of $4.56\times10^9$~yr.
The TAMS is set at the time when $X_\mathrm{c}(\mathrm{H})$ has decreased to $10^{-5}$.

Three reference points are further identified along the MS tracks to locate `homologous points' that have comparable evolutionary states.
Those reference points are important to ensure easy interpolation between homologous points at different masses and metallicities.
Their definition is arbitrary to some extent, and varies from one author to another depending on the characteristics of the grids and their expected usage.
The central abundances of H and He and the morphology of the tracks in the HR diagram tend to be used for the identification of those reference points \citep[e.g.][]{SchallerSchaererMeynet_etal92,BergbuschVandenberg92,EkstroemGeorgyEggenberger11}.
In this work, we adopt a slightly modified procedure that takes the morphology of the tracks in the $\log R-\log L$ diagram into account.
This is made necessary by the diversity of morphologies of the tracks in the HR diagram in the mass range considered in our grids.
The procedure is detailed in Appendix~\ref{Sect:normalizationProcedure}.
Two reference points are related to the hooks observed in this diagram (as well as in the HR diagram) for $ M$\,$\gtrsim$\,\mass{2.5}, and one reference point is set between the ZAMS and the first turn off point (see Appendix~\ref{Sect:normalizationProcedure}).

Finally, the last model in our tracks is located at the base of the RGB, after the star has crossed the HR diagram.
Six reference points are thus defined: the ZAMS, the three reference points in the MS, the TAMS, and the base of the RGB.
Their locations in the HR diagram are illustrated in Fig.~\ref{Fig:HRZ014} for the \Zinit\,=\,0.014 grid.
They define five intervals of time, four in the MS and one in the post-MS.

The basic tracks are then constructed by choosing \{100, 100, 100, 80, 220\} models in the five respective time intervals\footnote{When a model required in the basic track does not correspond to a computed model in the raw track, it is interpolated from the raw data.} defined by the reference points, plus one model at the base of the RGB.
The models are distributed linearly in age in the first two intervals on the MS, in $\log X_\mathrm{c}(\mathrm{H})$ in the last two intervals on the MS, and linearly in age in the post-MS interval.
Each basic track thus contains 601 points.
The reason $X_\mathrm{c}(\mathrm{H})$ cannot be used in all MS intervals is that the hydrogen abundance in the core is not a monotonically decreasing function of time in all models (see Appendix~\ref{Sect:normalizationProcedure}).

\subsection{Interpolated tracks}
\label{Sect:interpolatedTracks}

\begin{figure}
  \centering
  \includegraphics[width=\columnwidth]{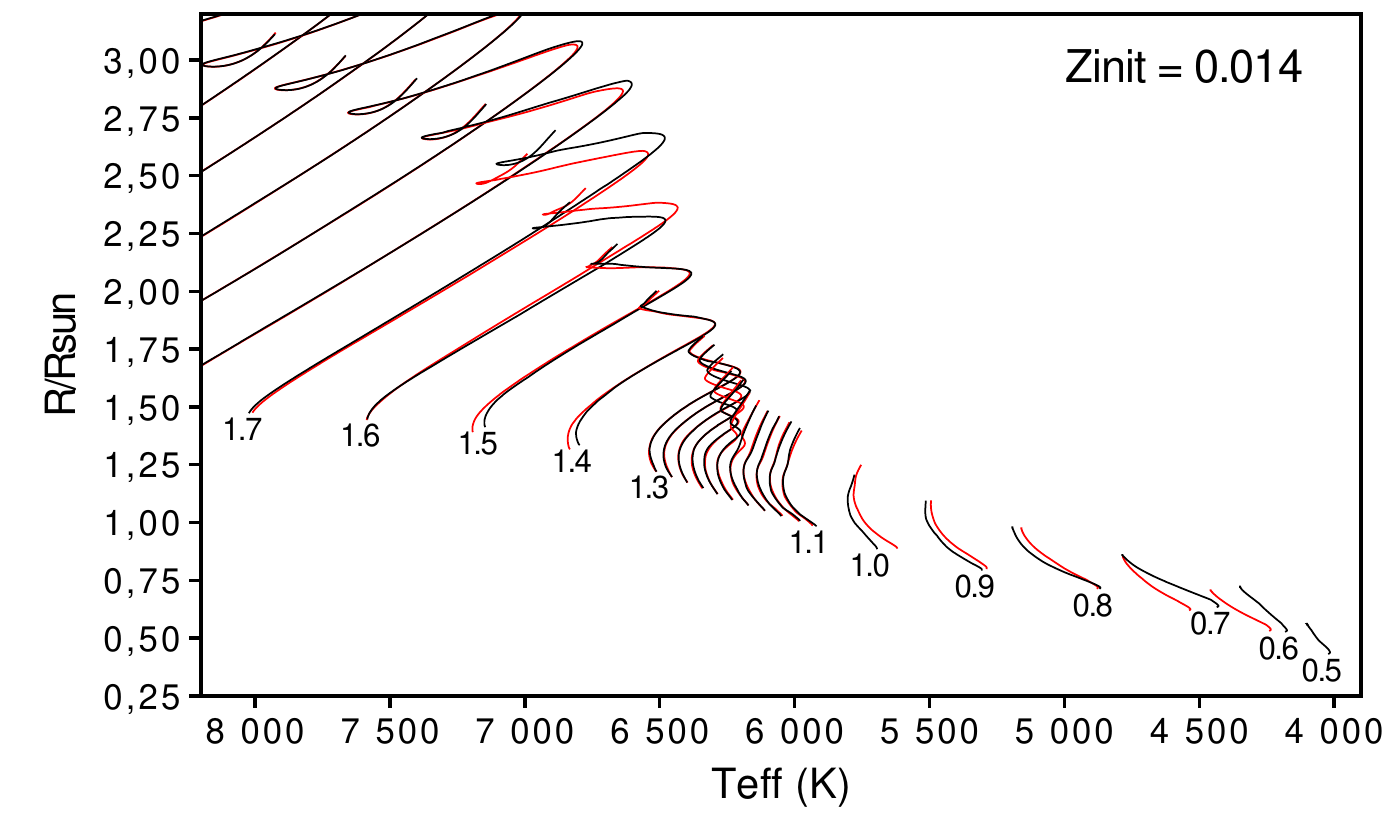}
  \caption{Basic (black) and interpolated (red) main sequence tracks at $Z_\mathrm{init}$~=~0.014, in the mass range between 0.5 and \mass{2} where the morphology of the tracks differs most from one track to the next.
           The interpolated tracks are obtained by interpolating in mass the two encompassing basic tracks.
           The masses of the tracks are indicated on the figure except for tracks between 1.1 and \mass{1.3} where tracks are equally spaced with mass steps of \mass{0.02}.
  }
\label{Fig:InterpTeffRTracks}
\end{figure}

\begin{figure*}
  \centering
  \includegraphics[width=0.675\columnwidth]{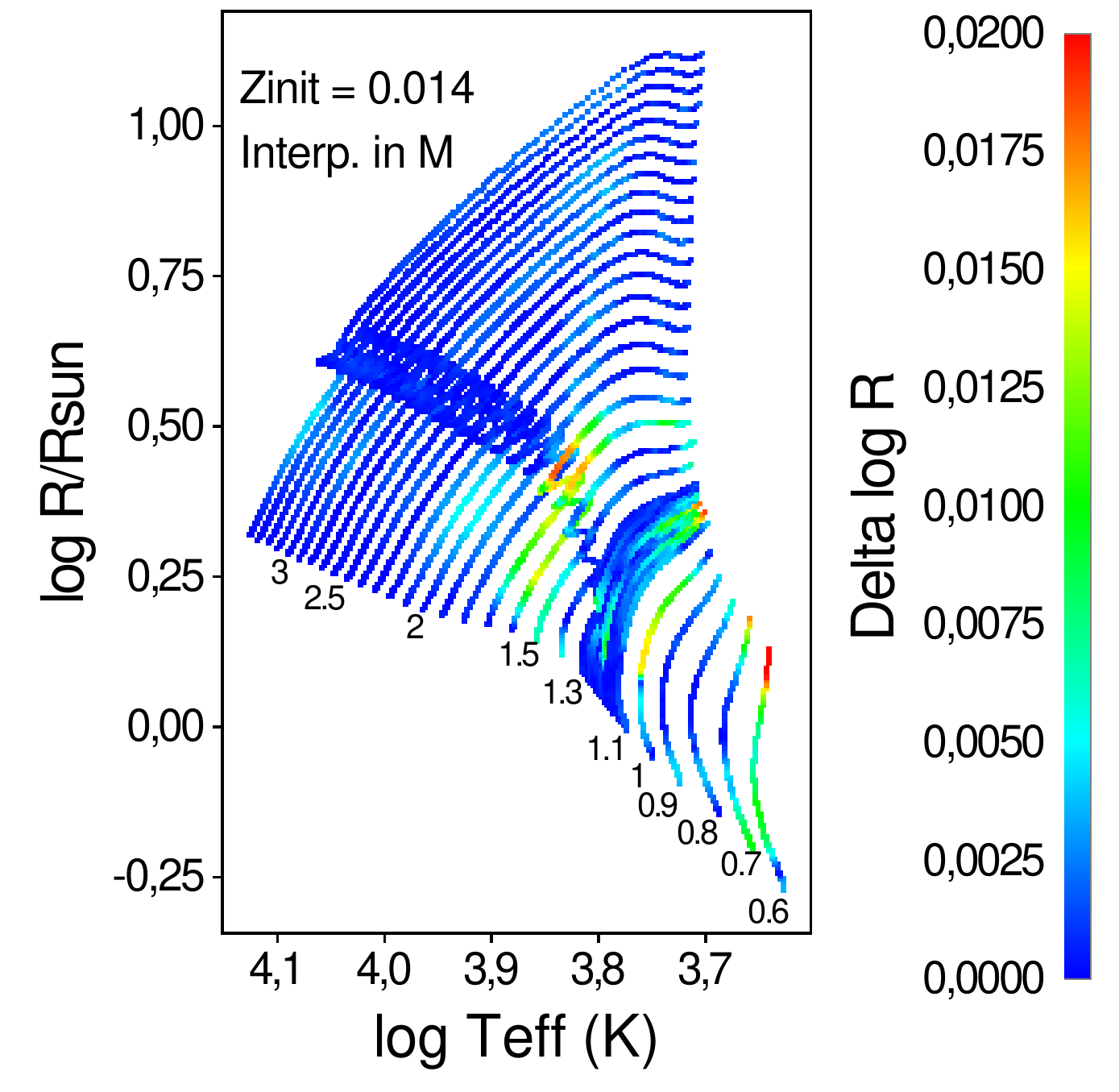}
  \includegraphics[width=0.675\columnwidth]{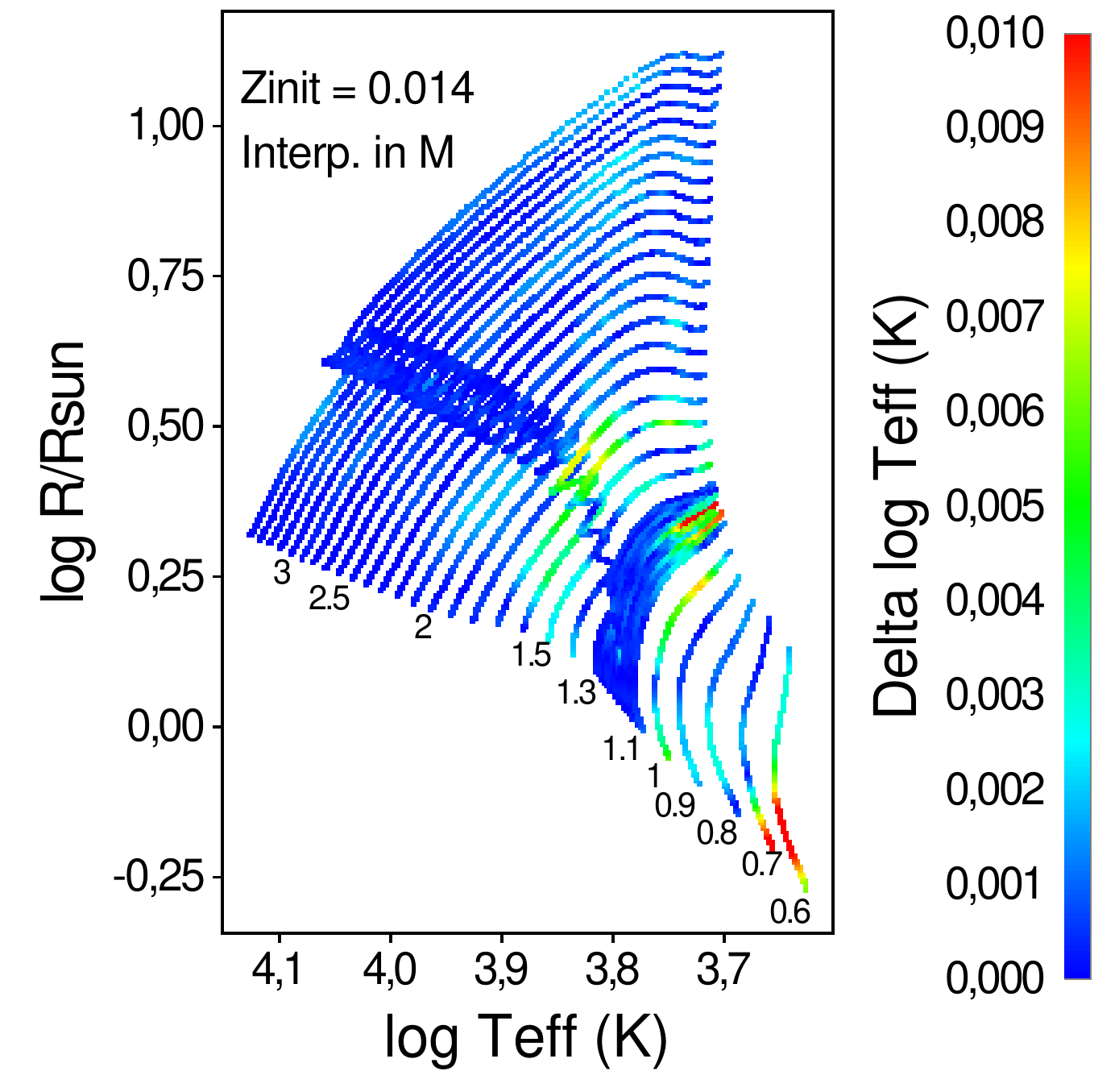}
  \includegraphics[width=0.675\columnwidth]{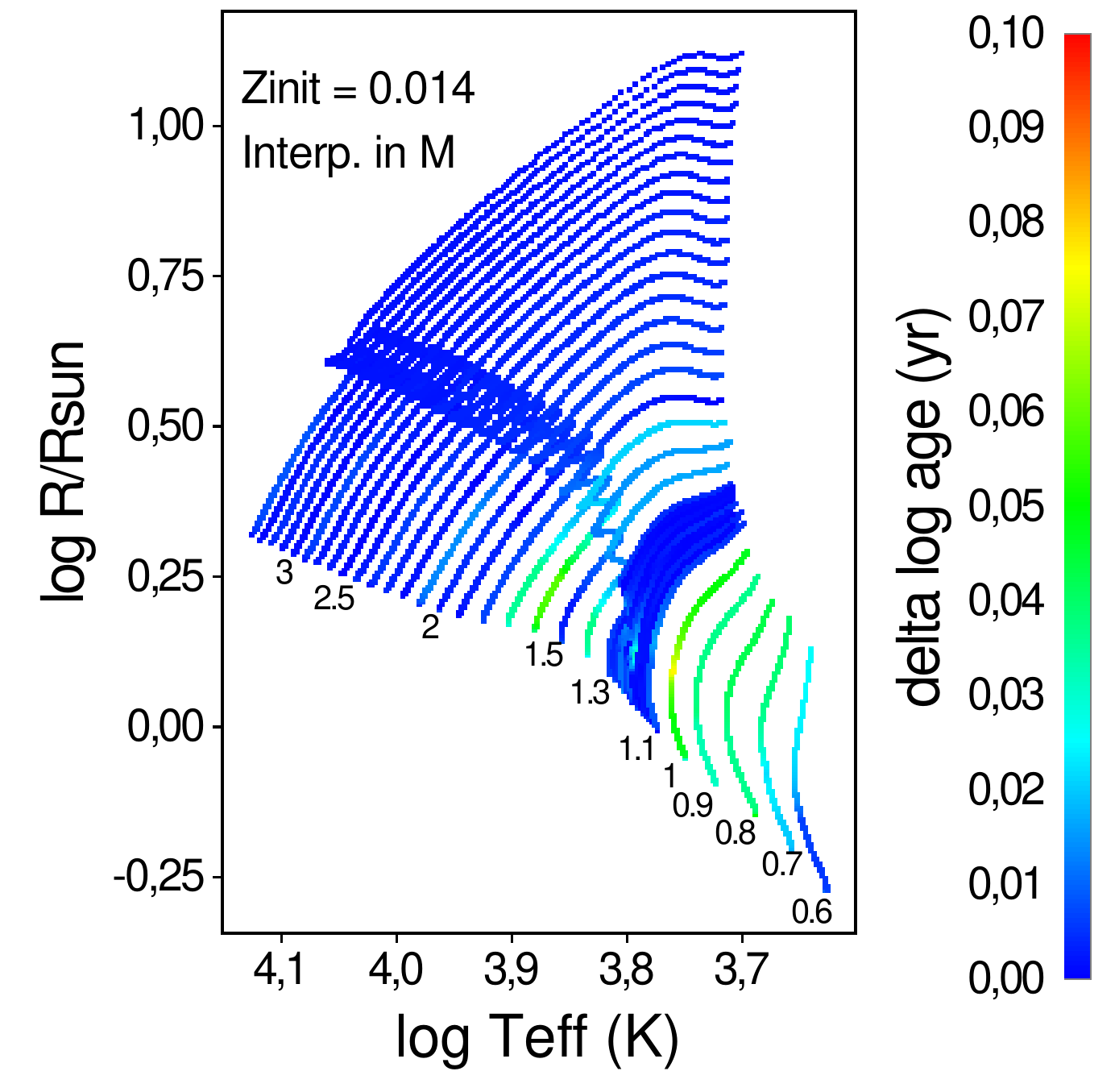}
  \caption{Accuracy of the interpolation procedure in mass.
  Original tracks at a given mass are compared to tracks interpolated between the two basic masses encompassing the given mass.
  The color-coded data represents the difference of the logarithm of the radius (left panel), the logarithm of the effective temperature (middle panel), and the logarithm of the age (right panel) of equivalent models between the basic and interpolated tracks.
  All tracks are at \Zinit\,=\,0.014.
  }
\label{Fig:InterpAccM}
\end{figure*}

\begin{figure*}
  \centering
  \includegraphics[width=0.675\columnwidth]{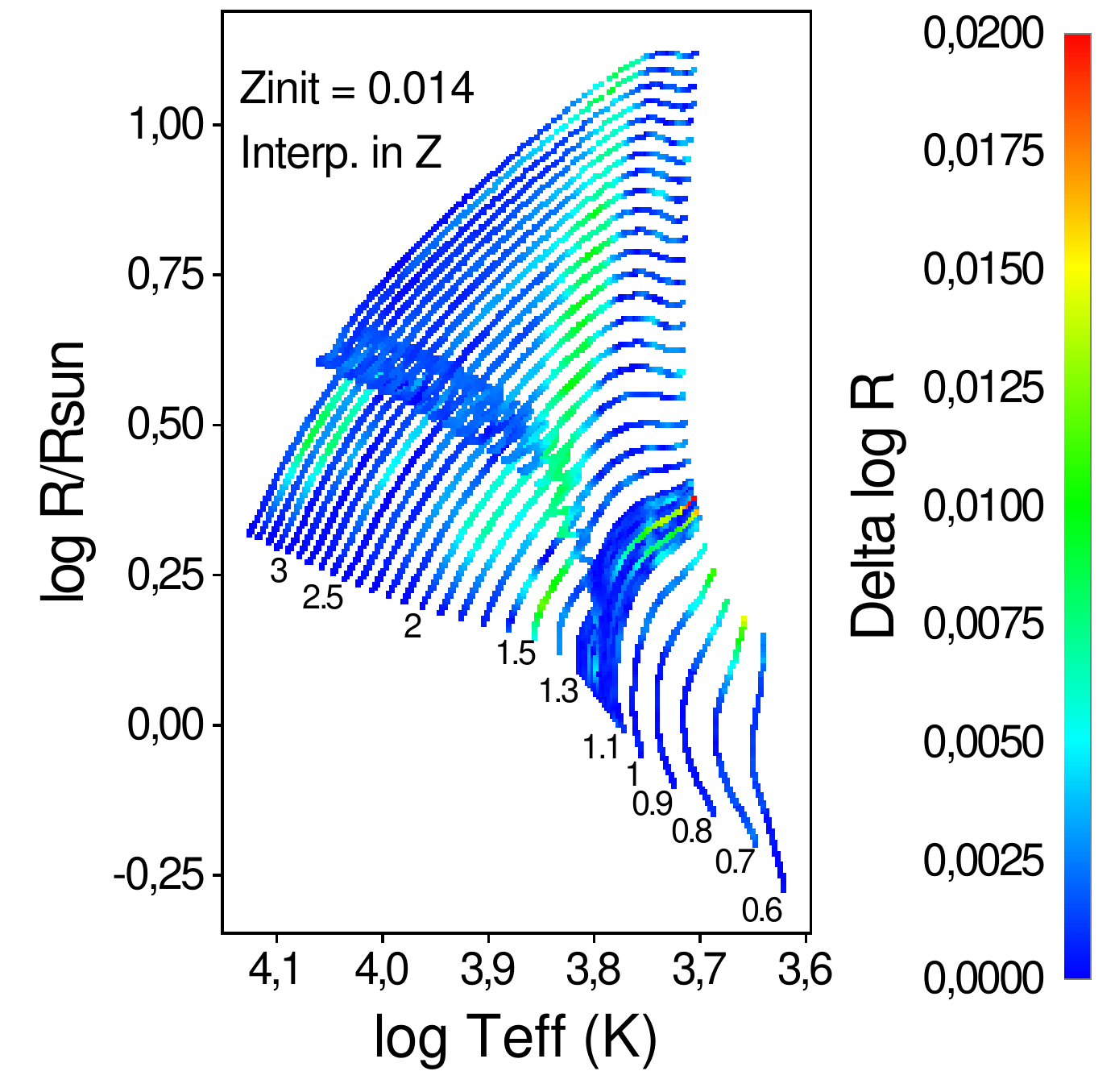}
  \includegraphics[width=0.675\columnwidth]{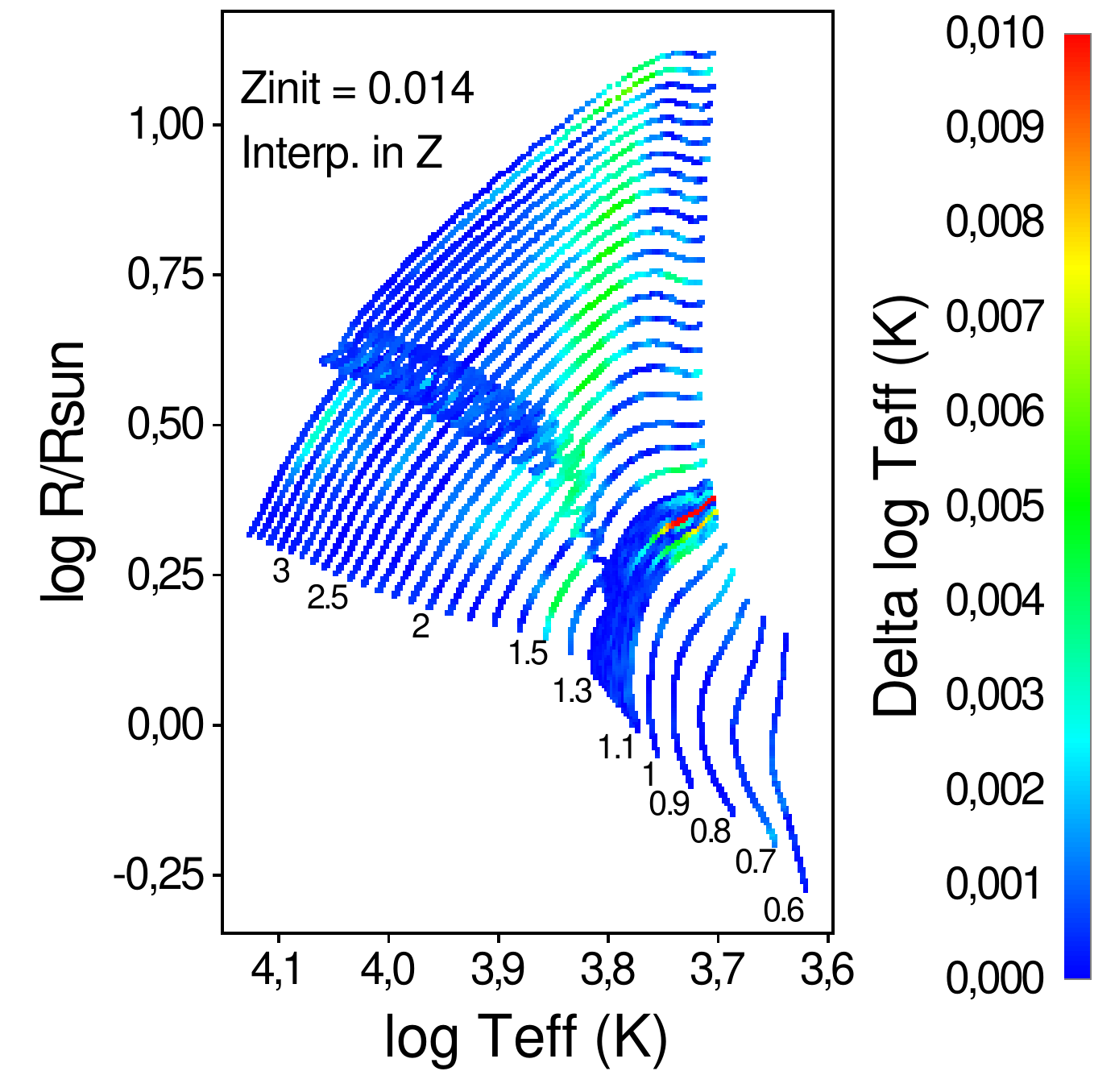}
  \includegraphics[width=0.675\columnwidth]{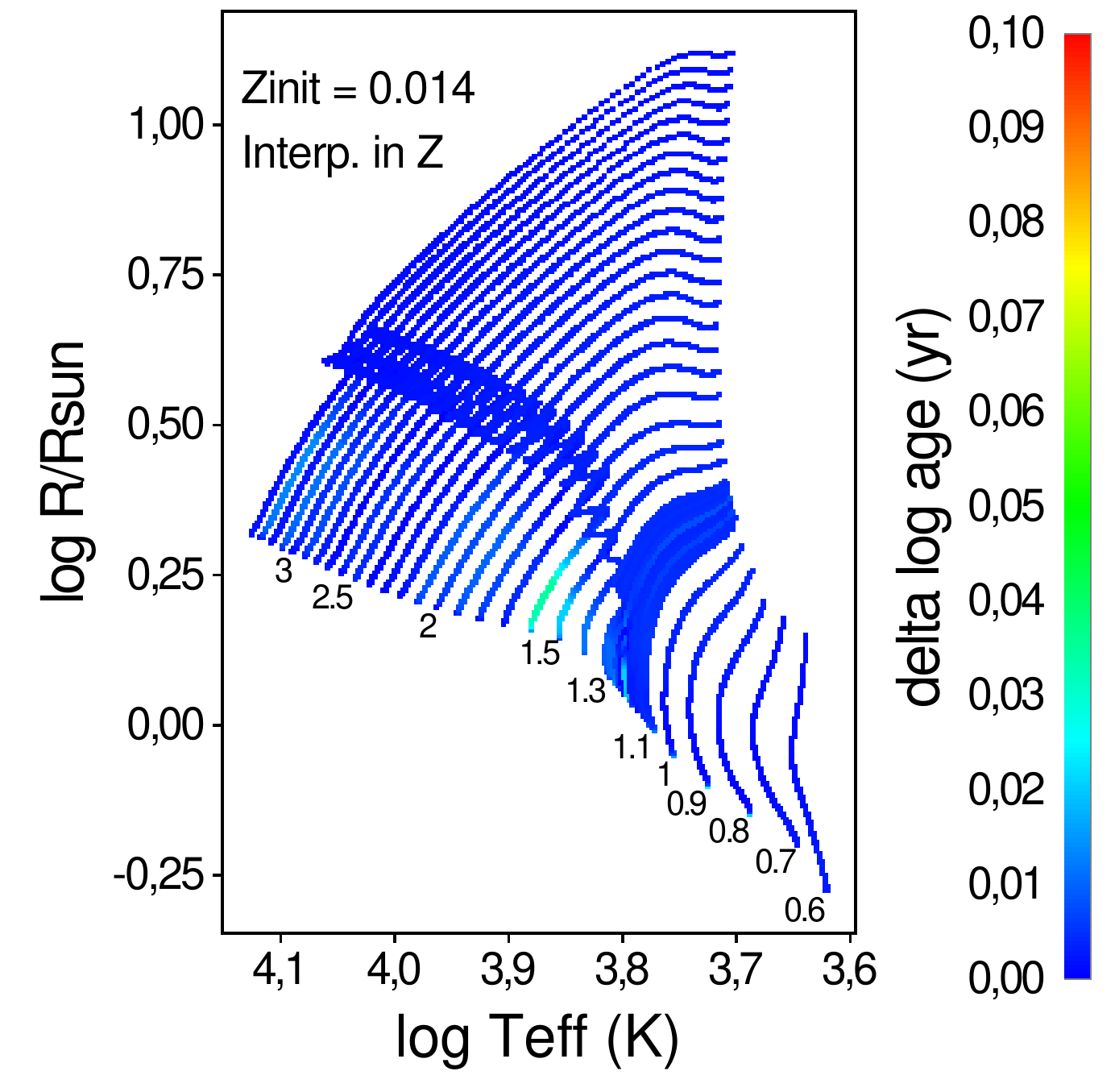}
  \caption{Same as Fig.~\ref{Fig:InterpAccM}, but for the interpolation accuracy in metallicity.
  Original tracks at $Z_\mathrm{init}$\,=\,0.014 are compared to tracks interpolated between $Z_\mathrm{init}$\,=\,0.01 and 0.02.
  The color scales of the three panels are kept identical to those in Fig.~\ref{Fig:InterpAccM}.
  }
\label{Fig:InterpAccZ}
\end{figure*}

Evolutionary tracks at \{\Zinit,$M$\} couples not available in the basic tracks are interpolated, on demand, from the basic tracks.
The interpolation is done in two steps.
Tracks are first interpolated in metallicity:
the two consecutive metallicities $Z_{\mathrm{init,a}}$ and $Z_{\mathrm{init,b}}$ of our grids that encompass the requested $Z_\mathrm{init}$ (i.e. $Z_{\mathrm{init,a}}\!\le\!Z_\mathrm{init}\!\le\!Z_{\mathrm{init,b}}$) and the two consecutive masses $M_\mathrm{a}$ and $M_\mathrm{b}$ in the grids that encompass the requested $M$ (i.e. $M_\mathrm{a}\!\le\!M\!\le\!M_\mathrm{b}$) are identified and interpolated in $Z_\mathrm{init}$ to construct tracks at ($Z_\mathrm{init}$, $M_\mathrm{a}$) and ($Z_\mathrm{init}$, $M_\mathrm{b}$).
These tracks are then interpolated in mass to obtain the requested evolutionary track at ($Z_\mathrm{init}$, $M$).
The interpolation procedure ultimately reduces to interpolating between equivalent models from basic tracks,  where the interpolation quantity is either mass or metallicity.
Interpolation of stellar quantities between two models is performed in logarithm for the metallicity, mass, effective temperature, surface luminosity and abundance mass fractions, and linearly for the age.

To evaluate the accuracy of the interpolation procedure, we construct, for each basic track $i$ of mass $M_i$, a track $i'$ interpolated at mass $M_i$ from the adjacent tracks $i-1$ and $i+1$, and compare the interpolated track to the basic track.
Figure~\ref{Fig:InterpTeffRTracks} compares the basic and interpolated tracks for the MS phase of \Zinit\,=\,0.014 models in the mass range where the morphology of the tracks varies the most from one track to the next.
Three regions are seen to be more sensitive to interpolation.
The first is for masses between 1.6 and \mass{1.7}, where the amplitude of overshooting is increased (see Sect.~\ref{Sect:inputPhysics}), leading to longer MS tracks.
The second region concerns masses between 1.18 and \mass{1.26}, where the interpolation inaccuracy results from the transition between purely radiative MS core models at low masses to fully convective core MS models at high masses.
This difficulty is partially palliated by the higher density of the tracks between 1.1 and \mass{1.3} in our grids.
And the third region concerns lower mass stars, especially between 0.6 and \mass{0.8}.

Figure~\ref{Fig:InterpAccM} shows the deviation of the stellar radius, effective temperature and age (all quantities in log) of the models between the basic and interpolated tracks.
The deviations are seen to be less than 1\% in radius and 0.5\% in effective temperature for most of the interpolated models, and 5\% in age, with a maximum deviation of 2.6, 1.3, and 7.8~\% encountered for  $R$, $\Teff$, and $t$, respectively, in any of the \Zinit\,=\,0.014 interpolated models.
In practice, a track interpolated in our grids has an even better accuracy than those results since the interpolation is done between two successive tracks of a grid rather than between one every two tracks of the grid.

A similar analysis, but in metallicity, is displayed in Fig.~\ref{Fig:InterpAccZ}.
It shows that our dense grids lead to a very good interpolation in metallicity.
This good interpolation accuracy holds true at all metallicities considered in our grids.

Finally, we tested how accurately the radius, effective temperature, and age can be obtained at any given point in the $\log \Teff$ - $\log R$ diagram \textit{without} interpolation, but by considering the closest basic track in that diagram.
The accuracies reach values as high as 11, 4.3, and 23\% for $R$, $\Teff$, and the age, respectively, which are at least three to four times worse than with interpolation.
Using interpolated tracks is thus strongly recommended.

\subsection{Iso-\Zsurf\ lines}
\label{Sect:isoZsurfTracks}

The change in surface abundances with time in models including atomic diffusion (see Sect.~\ref{Sect:surfaceAbundances}) implies that the measured metallicity at the surface of a star may be lower than its initial metallicity.
The distinction between surface and initial metallicities in models including diffusion has already been stressed in the literature, for example in relation with MS fitting of subdwarfs \citep[e.g.][]{MorelBaglin99,LebretonPerrinCayrel_etal99,SalarisGroenewegenWeiss00}, binary stars fitting \citep[e.g.][]{LebretonPerrinCayrel_etal99}, and globular cluster distance and $\Delta Y/\Delta Z$ determinations \citep[e.g.][]{SalarisGroenewegenWeiss00}.
We stress in this respect that our solar track at \Zinit\,=\,0.014 corresponds to $\FeH_\mathrm{init}$\,=\,+0.048 and not $\FeH_\mathrm{init}$\,=\,0.00.

In this case, to determine the age and the mass of a star from its observed position in the HR diagram, it may be erroneous to use tracks computed with an initial metallicity corresponding to that measured at its surface. A way to circumvent this difficulty is to use lines in the HR diagram that link models of equal initial mass, having different initial metallicities and ages, and showing equal surface abundances. We call these lines \textsl{iso-$\Zsurf$ lines}.
By searching the values of the initial mass passing through the observed positions, it is possible to determine the initial mass, the initial metallicity, and the age of the star.

\begin{figure}
  \centering
  \includegraphics[width=\columnwidth]{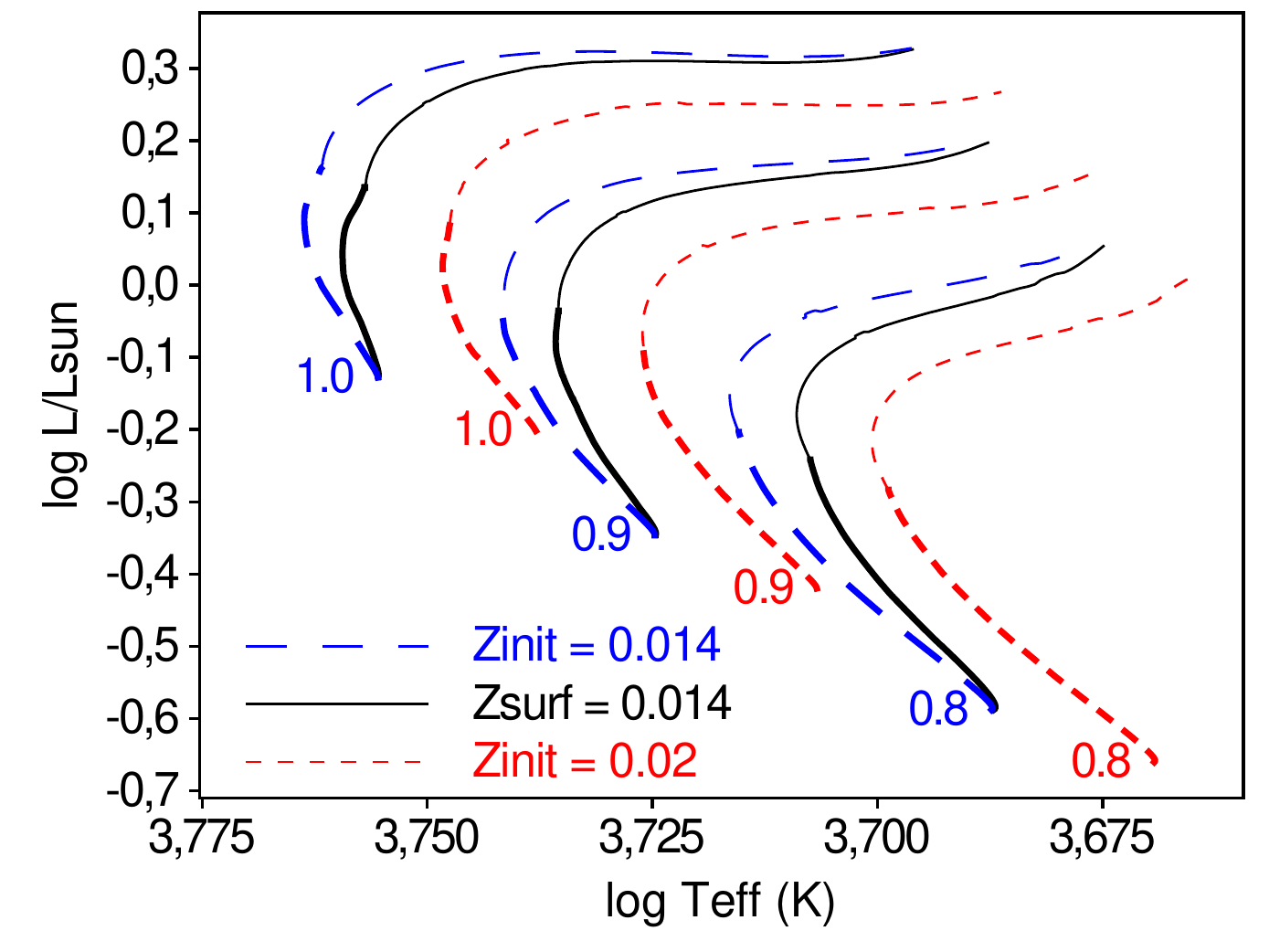}
  \caption{Iso-$\Zsurf$ lines (continuous lines) in the HR diagram at 0.8, 0.9, and \mass{1.0} as labeled in the figure.
   Also shown are the evolutionary tracks at $\Zinit=0.014$ (long-dashed lines) and 0.020 (short-dashed lines), from which the iso-$\Zsurf$ tracks are constructed by interpolation between equivalent models.
  }
\label{Fig:figIsoZsurfHR}
\end{figure}

\begin{figure}
  \centering
  \includegraphics[width=\columnwidth]{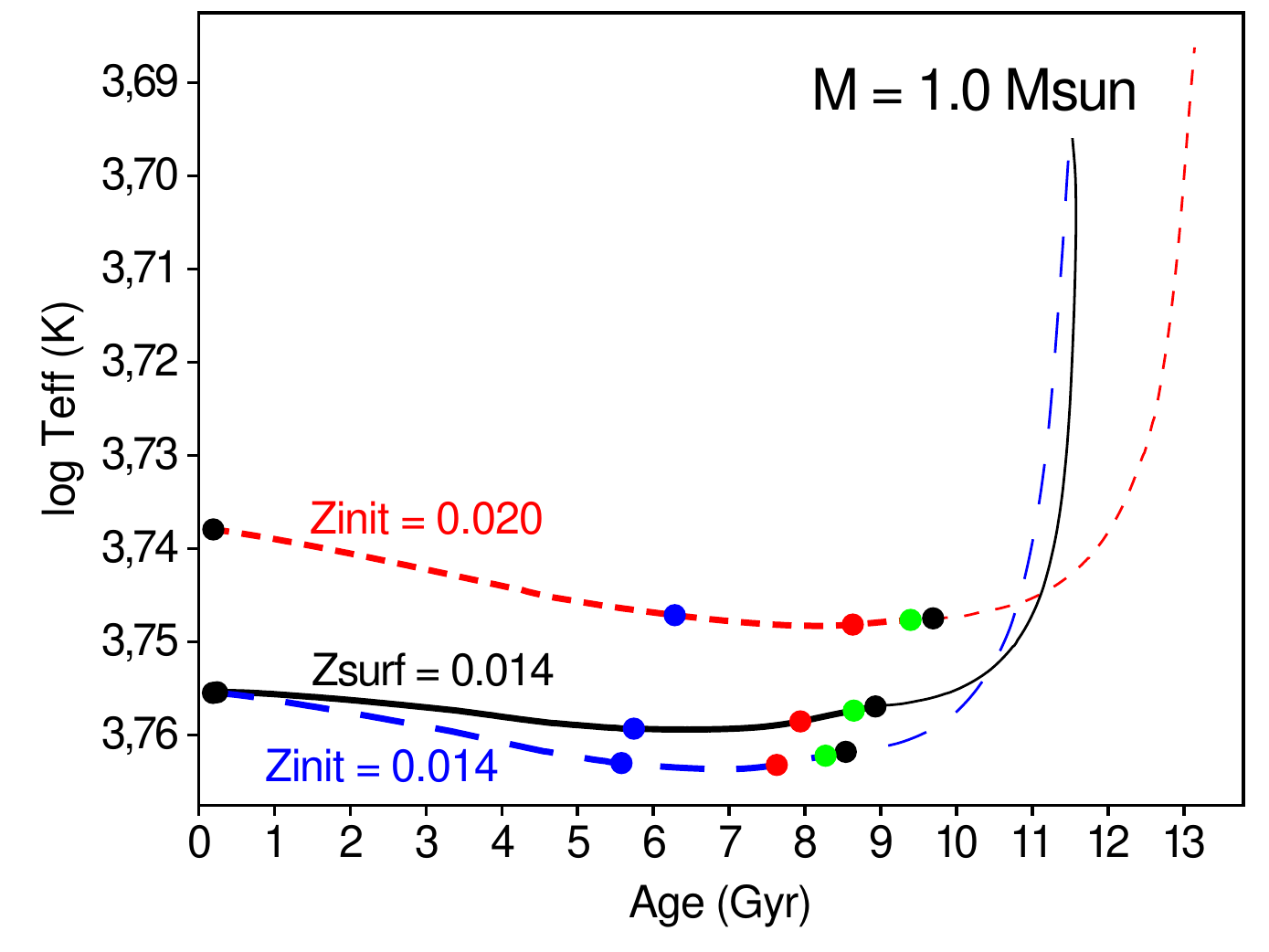}
  \caption{Continuous line: effective temperature of the models of the iso-$\Zsurf$ lines at \mass{1} versus the age of the models.
  For comparison, the evolutionary tracks at $\Zinit=0.014$ (long-dashed lines) and 0.020 (short-dashed lines) are also shown.
  The filled circles indicate the positions of the reference points on the MS.
  }
\label{Fig:figIsoZsurfTeff_M1p0}
\end{figure}

Figure~\ref{Fig:figIsoZsurfHR} illustrates iso-$\Zsurf$ lines at \Zsurf\,=\,0.014 at three stellar masses, as well as the corresponding evolutionary tracks at $\Zinit=0.014$ and 0.020 from which they are constructed.
The ZAMS models of the iso-$\Zsurf$ lines are identical to the evolutionary tracks at $\Zinit=\Zsurf$.
Evolved models on the MS of the iso-$\Zsurf$ lines then deviate from the evolutionary tracks.
In the post-MS phase, the iso-$\Zsurf$ models come back close to the evolutionary models at $\Zinit$\,=\,$\Zsurf$ as the surface chemical elements are mixed again with the deep layers due to the deepening of the convective envelope.

Iso-$\Zsurf$ lines are, of course, \textit{not} evolutionary tracks.
In particular, the ages of the models in an iso-$\Zsurf$ line do not necessarily increase with model number.
This is illustrated in Fig.~\ref{Fig:figIsoZsurfTeff_M1p0}, which displays the iso-$\Zsurf$ track at \mass{1.0} and $\Zsurf$\,=\,$0.014$.
The models in the post-MS phase are seen to have an age that decreases towards the end of the track, getting back close to the ages of the evolutionary track at $\Zinit=0.014$.

\subsection{Isochrones}
\label{Sect:isochrones}

\begin{figure}
  \centering
  \includegraphics[width=\columnwidth]{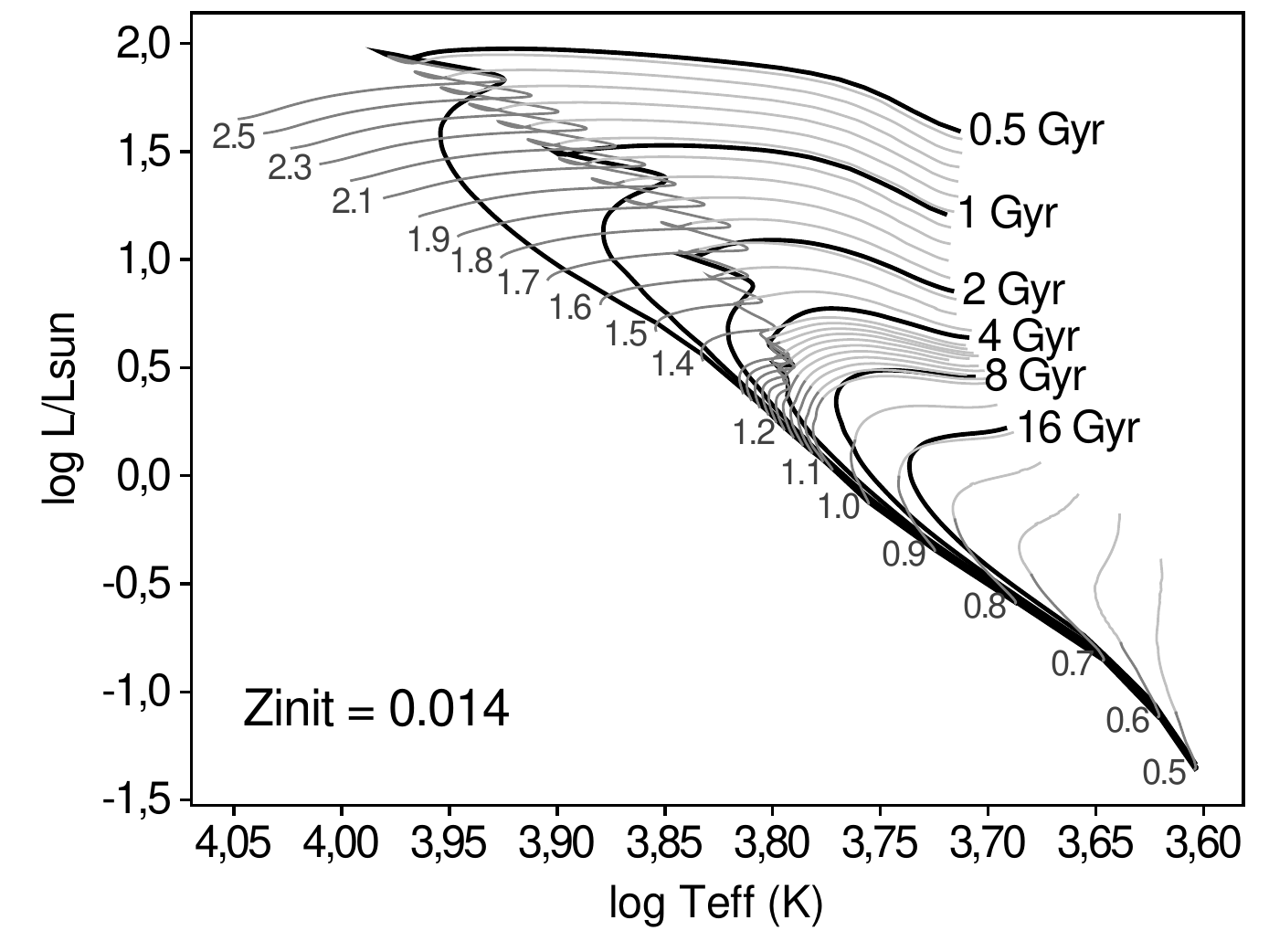}
  \caption{Isochrones in the HR diagram for five different ages as labeled in Gyr next to the last model of the isochrones (displayed in thick black lines).
  Also shown in gray are the evolutionary tracks for initial masses from 0.8 to \mass{2.6} as labeled next to the ZAMS models of the tracks.
  The main sequences of the evolutionary tracks are shown in dark gray and the post-MS phases in light gray.
  Isochrones and tracks are at $\Zinit=0.014$.
  }
\label{Fig:isochronesHR}
\end{figure}

Isochrones consist of a succession of models all having the same age and initial metallicity, but with increasing initial stellar masses from one model to the next in the track.
The track displayed by an isochrone in the HR diagram corresponds to the position of single-aged stellar populations, such as those obtained in single-aged stellar clusters or starburts.

Examples of \Zinit\,=\,0.014 isochrones between 0.5 and 8~Gyr are shown in Fig.~\ref{Fig:isochronesHR}.
They are computed from the basic tracks by taking care to adequately reproduce the hooks at the end of the MS.
The details of an isochrone construction procedure are given in Appendix~\ref{Sect:isochroneConstruction}.
We note that the 4.0~Gyr isochrone presents more than two hooks in the HR diagram.
This is related to the development of the convective core, as explained in the appendix.


\section{Comparison with observations and other grids}
\label{Sect:comparisons}

In this section, we compare our models with observations of several binary systems (Sect.~\ref{Sect:binarySystems}), with two open clusters (Sect.~\ref{Sect:openClusters}) and with some stellar grids from the literature (Sect.~\ref{Sect:otherGrids}).

\subsection{Comparison with binary systems}
\label{Sect:binarySystems}

\begin{figure}
  \centering
  \includegraphics[width=\columnwidth]{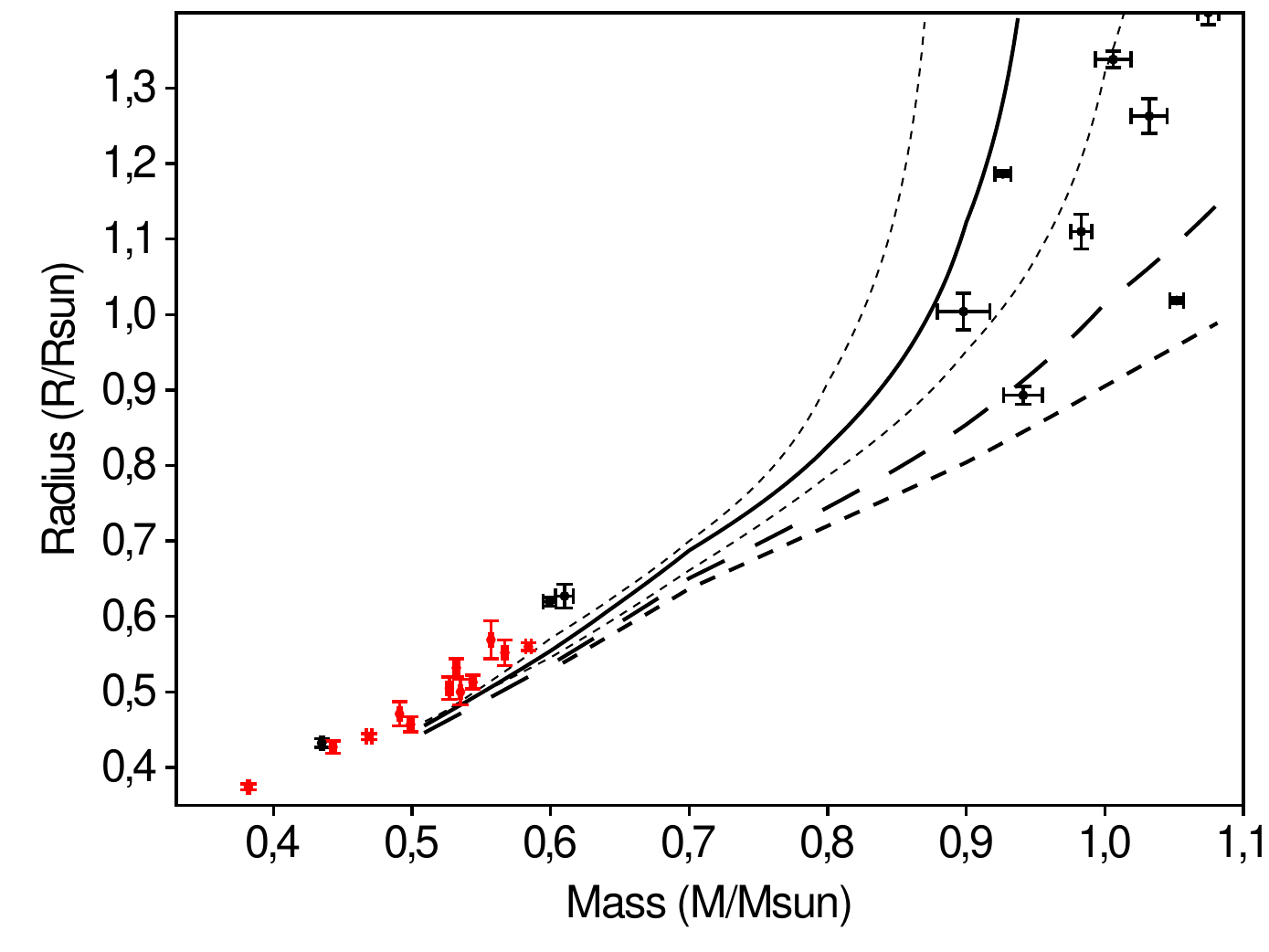}
  \caption{Masses and radii of detached binary stars from \cite{KrausTuckerThompson_etal11} (red points) and \cite{TorresAndersenGimenez10} (black points) in the mass range below \mass{1.1}.
  Isochrones from our grids are shown at 14 (continuous line), 5 (long dashed line) and 1 Gyr (short dashed line) for $\Zinit$\,=\,0.014.
  Isochrones at 14 Gyr are also shown for $\Zinit$\,=\,0.006 (upper dotted line) and 0.04 (lower dotted line).
  }
\label{Fig:massRadius}
\end{figure}

\begin{figure}
  \centering
  \includegraphics[width=0.8\columnwidth]{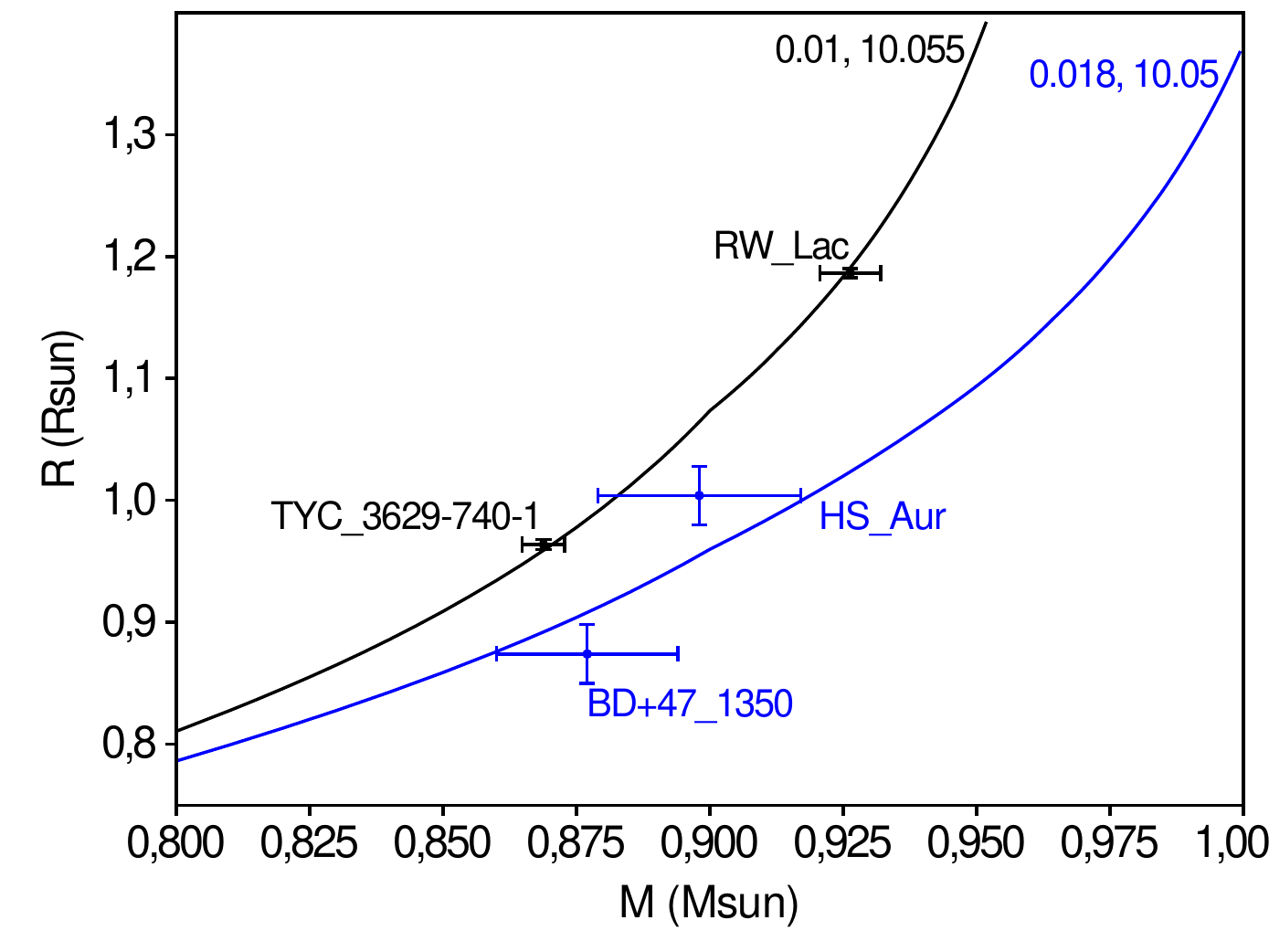}
  \includegraphics[width=0.8\columnwidth]{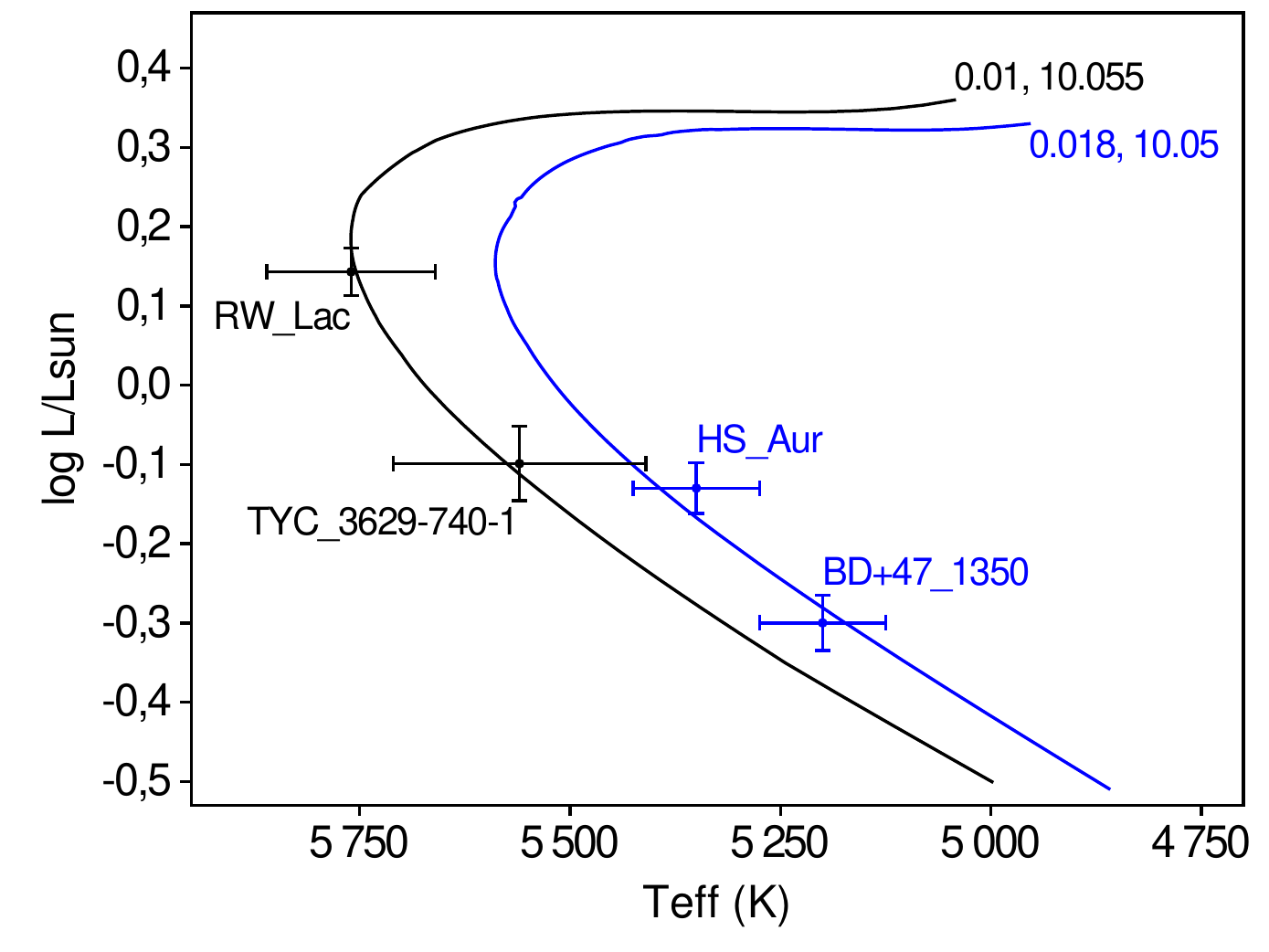}
  \caption{
  Isochrone fits to binaries 90 (RW Lac + TYC 3629-740-1) and 91 (HS Aur + BD+47 1350) of \cite{TorresAndersenGimenez10} in the mass-radius (upper panel) and HR (lower panel) diagrams.
  The positions of the stars in those diagrams are shown with the error bars quoted by those authors.
  Isochrone fitting is performed visually.
  The numbers next to the isochrones are the initial metallicity $\Zinit$ and the logarithm of the age in years.
  }
\label{Fig:binaries1}
\end{figure}

\begin{figure}
  \centering
  \includegraphics[width=0.8\columnwidth]{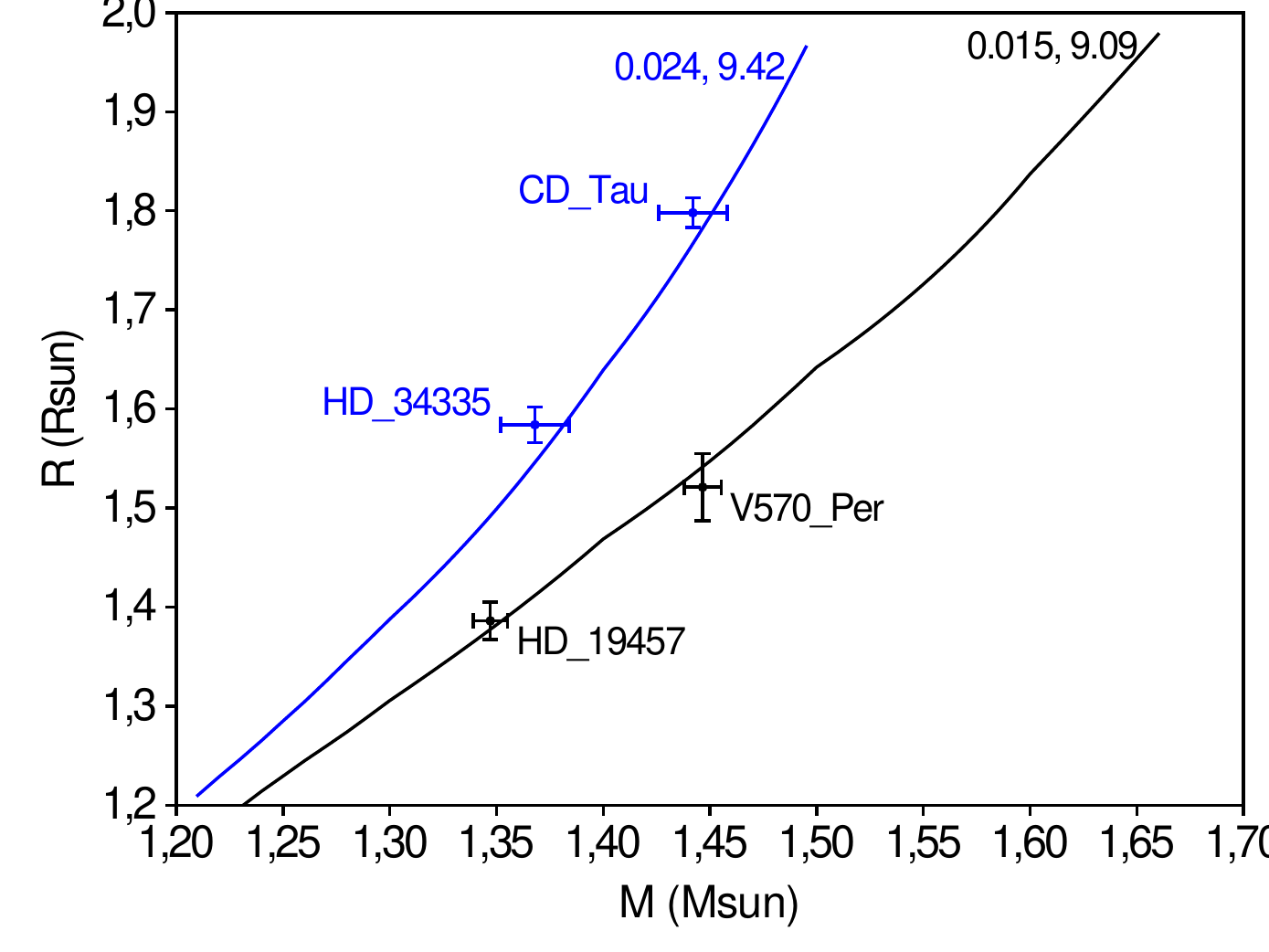}
  \includegraphics[width=0.8\columnwidth]{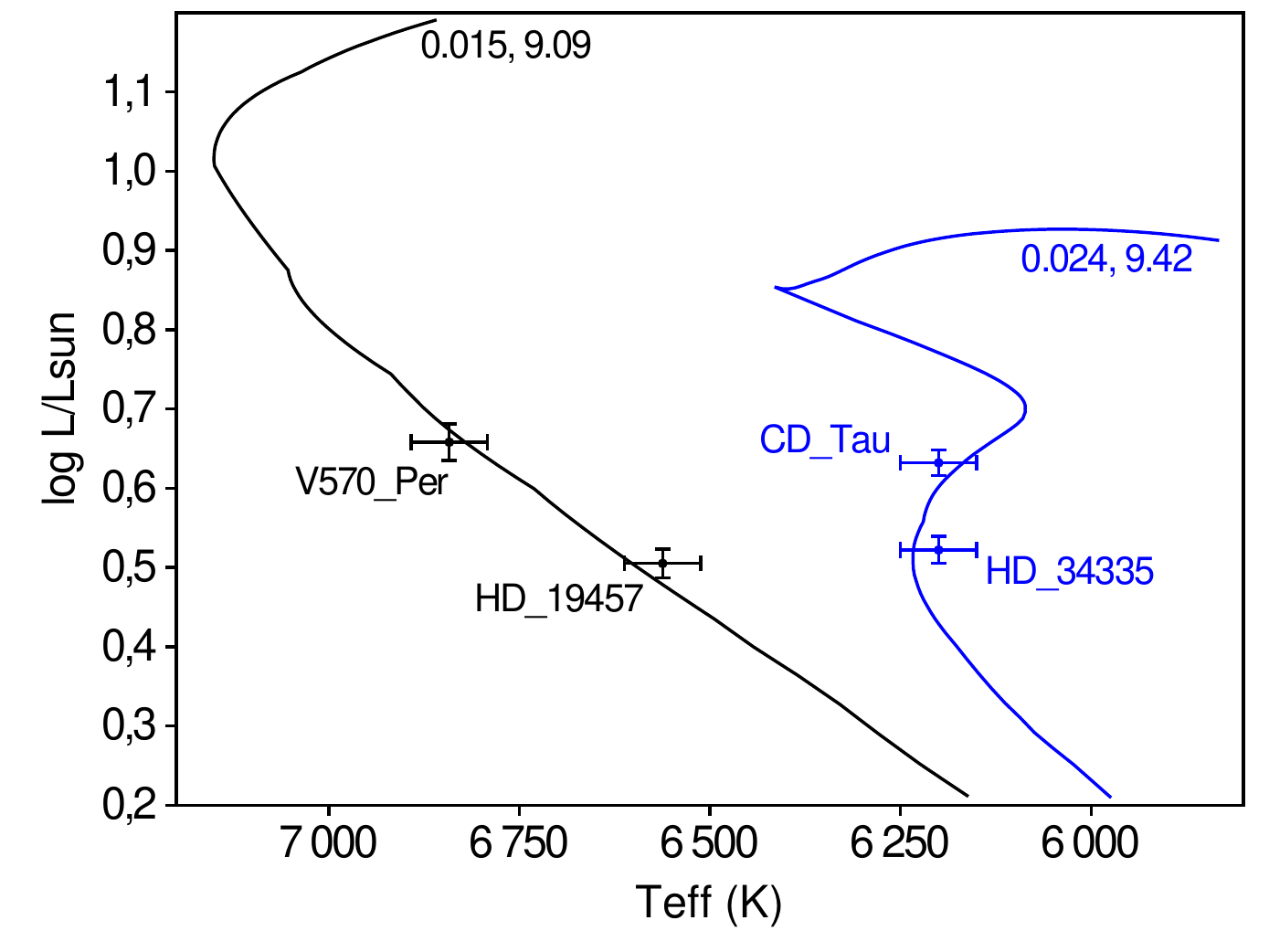}
  \caption{
  Same as Fig.~\ref{Fig:binaries1}, but for binaries 64 (V570 Per + HD 19457) and 65 (CD Tau + HD 34335) of \cite{TorresAndersenGimenez10}.
  }
\label{Fig:binaries2}
\end{figure}

Data from detached binary systems provide unique opportunities to test predictions of stellar evolution models.
We display in Fig.~\ref{Fig:massRadius} the masses and radii of binary stars from \cite{TorresAndersenGimenez10} and \cite{KrausTuckerThompson_etal11} that have masses less than \mass{1.1}.
We plot in the same figure isochrones from our grids at several ages and metallicities.
They encompass well the range of masses and radii observed in binary stars for masses above \mass{0.7}.
For lower mass stars, however, the models predict radii that are bigger than what is measured, a conclusion already stressed by \cite{KrausTuckerThompson_etal11} when comparing their measurements with predictions of low-mass star models from \cite{BaraffeChabrierAllard_etal98}.
\cite{KrausTuckerThompson_etal11} attribute this discrepancy to the possibility that short-period systems are ``inflated by the influence of the close companion, most likely because they are tidally locked into very high rotation speeds that enhance activity and inhibit convection''.

Figures~\ref{Fig:binaries1} and \ref{Fig:binaries2} display four of the binary systems listed by \cite{TorresAndersenGimenez10} in more details, two with masses below \mass{1.0} (Fig.~\ref{Fig:binaries1}) and two with masses around \mass{1.4}.
The positions of the eight stars are shown in both the mass-radius plane (upper panels) and the HR diagram (lower panels).
Visual isochrone fits to each of the four systems are also shown.

Figure~\ref{Fig:binaries1} shows the four stars of binary systems 90 (RW Lac +TYC 3629-740-1) and 91 (HS Aur + BD+47 1350) of \cite{TorresAndersenGimenez10}.
According to these authors, the metallicity of those systems is unknown and their age estimated to be around 10~Gyr for both systems.
Our isochrones confirm that both systems have a similar age, of log(age)\,$\simeq$\,10.05.
According to our models, RW~Lac and TYC 3629-740-1 would be born with a subsolar metallicity $\Zinit$\,=\,0.010, while the initial metallicity of HS~Aur and BD+47 1350 was above solar with $\Zinit$\,=\,0.018.
The current value of their surface metallicities, which must have decreased during the course of their evolution due to atomic diffusion, are predicted to be $\FeH$\,=\,$-0.16$ ($\Zsurf$\,=\,0.0086) and $\FeH$\,=\,+0.11 ($\Zsurf$\,=\,0.016) for the stars in each system, respectively.
The difference in $\FeH$ between the two components of each system is about 0.01.

Figure~\ref{Fig:binaries2} shows the four stars of binary systems 64 (V570~Per + HD~19457) and 65 (CD~Tau + HD~34335) of \cite{TorresAndersenGimenez10}.
The authors quote an almost solar metallicity for V570~Per  ($\FeH = +0.01 \pm 0.03$) and over-solar for CD~Tau ($\FeH = +0.08 \pm 0.15$), and log~ages of 8.8 and 9.5, respectively.
We get higher metallicity estimates, of $\FeH$\,=\,+0.08 ($Z$\,=\,0.015) for V570~Per and $\FeH$\,=\,+0.30 ($Z$\,=\,0.024) for CD~Tau.
We also get slightly larger ages, with log(age)\,=\,9.09 and 9.42, respectively.

A deeper analysis should, however, be done to evaluate the range of metallicities and ages predicted by our grids compatible with the error bars on the masses, radii, $\Teff$, and $\log L$ of those stars.
This is outside the scope of this article and should be the object of a separate study.
These comparisons are, however, quite encouraging and indicate that reasonable solutions can be obtained.

\subsection{Comparison with open clusters}
\label{Sect:openClusters}

\begin{figure}
  \centering
  \includegraphics[width=0.9\columnwidth]{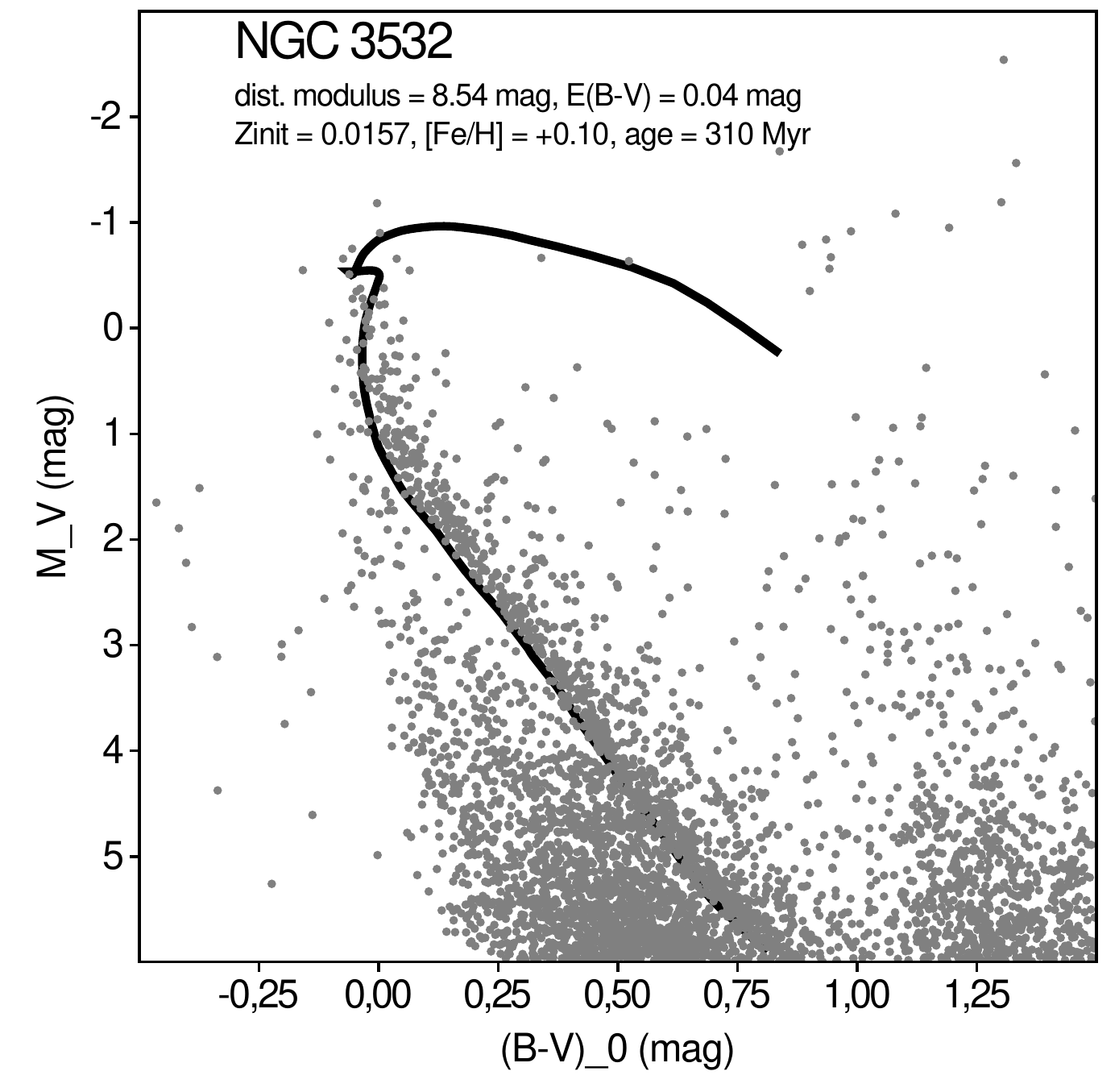}
  \caption{
  Isochrone fit to the open cluster NGC~3532.
  The data points are from \cite{ClemLandoltHoard_etal11}.
  The thick line is an isochrone at 300~My.
  }
\label{Fig:NGC3532}
\end{figure}

\begin{figure}
  \centering
  \includegraphics[width=0.9\columnwidth]{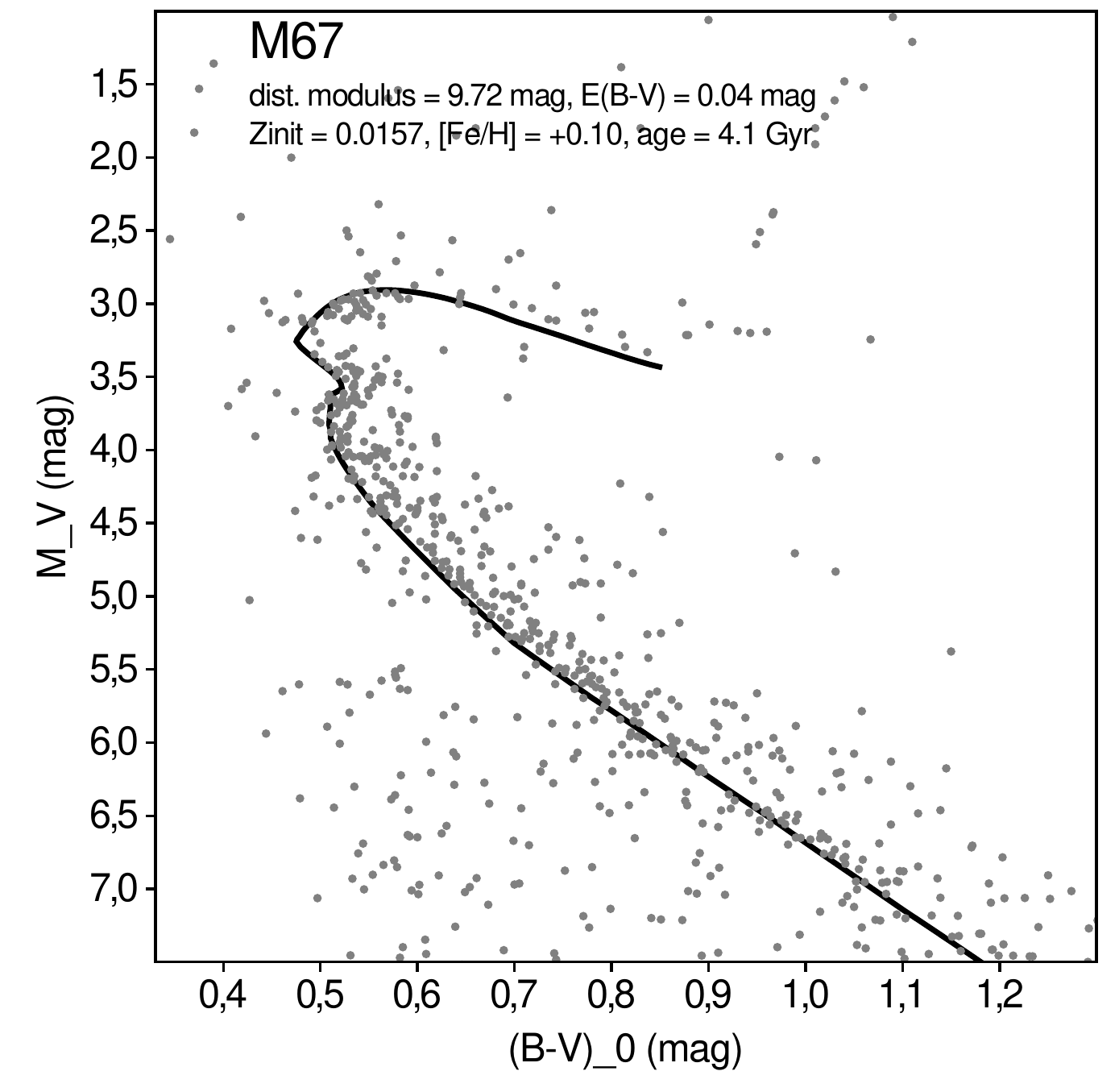}
  \caption{Isochrone fit to the open cluster M67.
  The data points are from \cite{MontgomeryMarschall93}.
  The thick line is an isochrone at 4.1~Gyr.
  }
\label{Fig:M67}
\end{figure}

We also tested our isochrones on two open clusters.
The first one is NGC~3532.
With an age estimated at 300~Myr \citep{ClemLandoltHoard_etal11}, it allows our models to be checked at the higher mass range of our grids.
The data collected by \cite{ClemLandoltHoard_etal11} are plotted in Fig.~\ref{Fig:NGC3532} for all stars with $M_\mathrm{V}$ below 6~mag.
Our 300~My isochrone adopts a distance modulus $(m-M)_\mathrm{V}$\,=\,8.54 (492~pc) and a reddening $E(B-V)$\,=\,0.04.
The conversion from $\{\Teff, \log L\}$ to $\{(B-V), M_\mathrm{V}\}$ uses the $\Teff$ to $(B-V)$ conversion relations of \cite{SekiguchiFukugita00} and \cite{Boehm-Vitense81} and the bolometric corrections of \cite{Flower96}.
A 310~My isochrone at $\FeH$\,=\,+0.10 nicely fits the MS lower boundary of NGC~3532.
This slightly over-solar metallicity agrees with the $\FeH$\,=\,+0.0$\pm$0.1 value adopted by \cite{ClemLandoltHoard_etal11}.

The second cluster is M67.
Its turn-off mass of about \mass{1.28} makes this cluster particularly attractive for comparison with isochrone predictions.
It displays a clear hook in the CMD (see Fig.~\ref{Fig:M67}).
\cite{VandenBergGustafssonEdvardsson_etal07} claims that stellar models with the \textit{new} solar composition have difficulties in predicting this hook, while \cite{MagicSerenelliWeiss_etal10} stress the sensitivity of the hook on the constitutive physics of the models.
In Fig.~\ref{Fig:M67}, we show a matching 4.1~Gyr isochrone from our models.
We take $(m-M)_\mathrm{V}$\,=\,9.72 from \cite{Sandquist04} and $E(B-V)$\,=\,0.04 from \cite{SarajediniVonHippelKozhurinaPlatais99}.
The adopted metallicity, $\FeH$\,=\,+0.10, agrees with recent abundance determinations for this cluster of +0.03$\pm$0.07 by \cite{FrielJacobsonPilachowski10} and +0.05$\pm$0.02($\pm$0.10) by \cite{PancinoCarreraRossetti10}.

As for the case of the binaries discussed in Sect.~\ref{Sect:binarySystems}, the relatively good matches obtained here are encouraging, but a deeper analysis should be done to explore the range of cluster parameters that lead to isochrones compatible with their CMDs, a task beyond the scope of this article.

\subsection{Comparison with other grids}
\label{Sect:otherGrids}

\begin{figure}
  \centering
  \includegraphics[width=0.9\columnwidth]{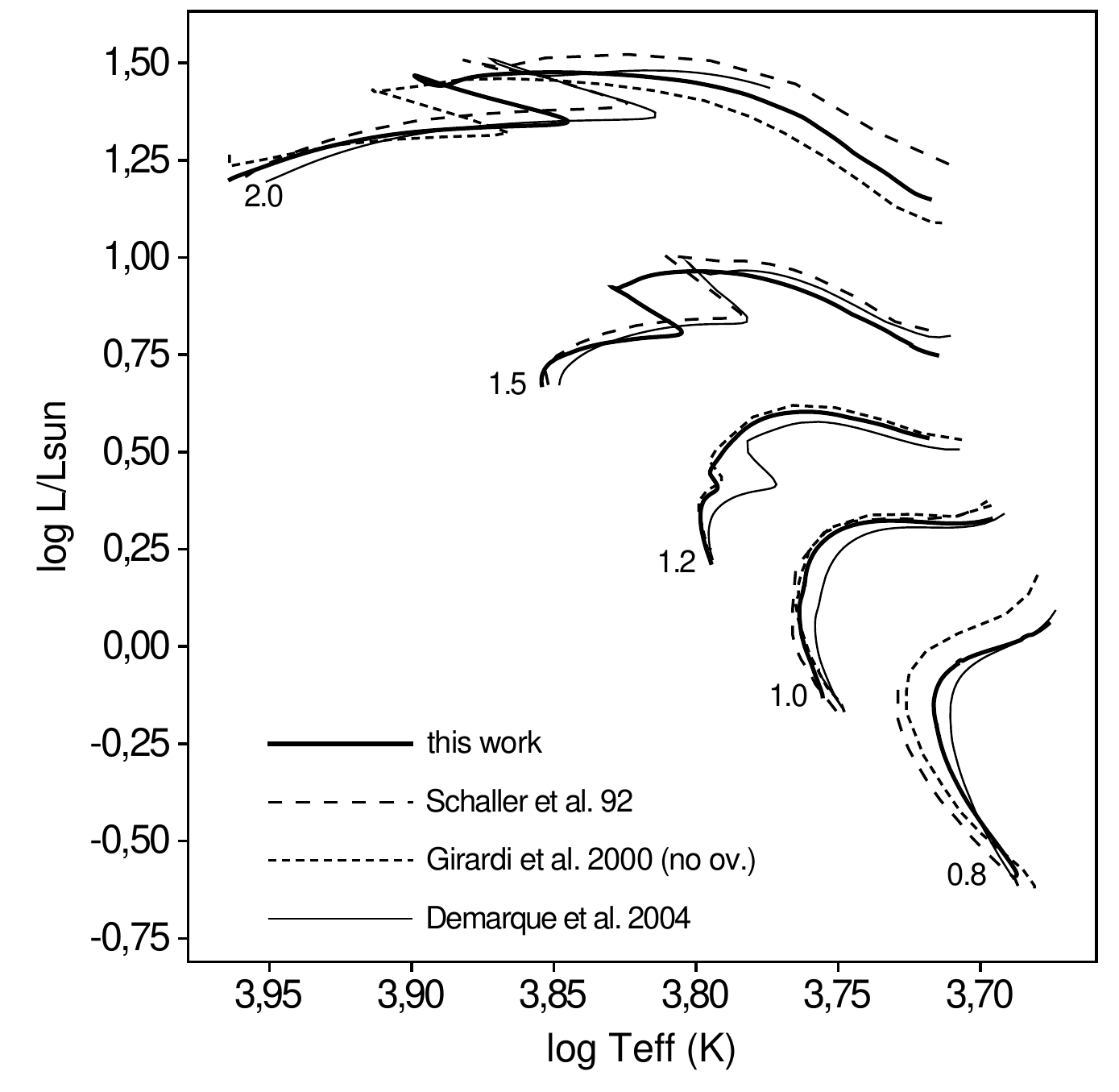}
  \caption{
  Comparison of selected tracks of this work at $\Zinit$\,=\,0.014 (solid thick lines) with the Geneva tracks of \cite{SchallerSchaererMeynet_etal92} at $Z$\,=\,0.02 (long-dashed lines), with the tracks of \cite{GirardiBressanBertelli_etal00} without overshooting at $Z$\,=\,0.019 (short-dashed lines) and with the tracks of \cite{DemarqueWooKim_etal04} at $Z$\,=\,0.02 (solid thin lines).
  The tracks from the latter three works are shown up to the base of the RGB only for clarity.
  The label next to the location of the ZAMS is the stellar mass of the tracks.
  }
\label{Fig:otherGrids}
\end{figure}

We compare in the HR diagram (Fig.~\ref{Fig:otherGrids}) a selection of our solar metallicity tracks with three sets of models from the literature:
\begin{itemize}
\item the older Geneva tracks \citep{SchallerSchaererMeynet_etal92}.
      Those used old solar abundances, had an overshooting parameter of $\alpha_{ov}$\,=\,0.20, and did not include diffusion;
\item the tracks by \cite{GirardiBressanBertelli_etal00} without overshooting.
      Those use  old solar abundances and did not include diffusion;
\item the Yonsei-Yale tracks \citep{DemarqueWooKim_etal04}.
      Those use the old solar abundances, include overshooting with $\alpha_\mathrm{ov}$\,=\,0.20 and use diffusion.
      It must be noted that their $Z$\,=\,0.02 models, displayed in Fig.~\ref{Fig:otherGrids}, have not been calibrated to the Sun.
      It is their models at $Z$\,=\,0.0181 that have been calibrated to the Sun, see \cite{YiDemarqueKim_etal01}.
      This explains their redder \mass{1.0} track in Fig.~\ref{Fig:otherGrids} compared to the three other \mass{1.0} tracks. 
\end{itemize}
It is of course difficult to compare in detail the tracks from different authors given the differences in the constitutive physics and chemistry.
We can nevertheless draw some general conclusions by comparison with our grids.

Above \mass{1.2}, the main parameter differentiating the tracks is the overshooting parameter.
The higher value of this parameter in the old Geneva and in the Yonsei-Yale models with respect to our models leads to wider MS tracks.
The 1992 Geneva models were computed with $\alpha_\mathrm{ov}$\,=\,0.20 in order to reproduce the observed width of the MS in the HR diagram.
Since then, rotation is known to also contribute to broadening the MS width \citep{TalonZahnMaeder_etal97}.
Our nonrotating models with moderate overshooting of $\alpha_\mathrm{ov}$\,=\,0.10, representative of slowly rotating stars, thus leads to narrower MS widths.
For illustration purposes, we also show in Fig.~\ref{Fig:otherGrids} tracks computed with no overshooting by \cite{GirardiBressanBertelli_etal00}.
They show a narrower MS, compatible with the explanation given here.

For masses below solar, the redder models displayed by our \mass{0.8} track compared to the older Geneva models and to the ones from \cite{GirardiBressanBertelli_etal00} mainly come from including of atomic diffusion in our models, a process neglected in the two other works.
The \mass{0.8} track of \cite{DemarqueWooKim_etal04}, who do include helium diffusion in their calculations, is also redder than the tracks of the old Geneva models and of \cite{GirardiBressanBertelli_etal00}, supporting our results with diffusion.
It is however not possible to compare our models in detail with those of \cite{DemarqueWooKim_etal04} because of the differences in both the constitutive physics and chemical composition.

\section{Data access}
\label{Sect:dataDownload}

\begin{table*}
\caption{Extract of a file content downloadable from our web site. See Sect.~\ref{Sect:dataDownload} for more details.
}
\tiny
\begin{verbatim}
# Grid of Geneva stellar models
# Number of tracks : 39
#===========================================================================================================================
# Basic track
# Zinit    :  0.0140
# Minit    :  0.5000
# numModels:  601
#   Phase 20 starting at age = 1.3477464e+09 yr, with first model at    0
#   Phase 21 starting at age = 4.4900246e+10 yr, with first model at  100
#   Phase 22 starting at age = 6.8427257e+10 yr, with first model at  200
#   Phase 23 starting at age = 7.6593657e+10 yr, with first model at  300
#   Phase 30 starting at age = 7.7696842e+10 yr, with first model at  380
#   Phase 31 starting at age = 1.1418004e+11 yr, with first model at  600
#---------------------------------------------------------------------------------------------------------------------------
#    phase      Age         Mass      LogL     LogTeff   LogRhoc    LogTc    Qcc        XcH         XcHe3       XcHe4    ...
   0  20   1.3477464e+09  0.500000  -1.357416 3.604120   1.861599 6.945235  0.0000   7.05600e-01 8.13559e-04 2.79558e-01 ...
   1  20   1.7832714e+09  0.500000  -1.356323 3.604018   1.861825 6.944023  0.0000   7.00524e-01 8.24166e-04 2.84595e-01 ...
   2  20   2.2187964e+09  0.500000  -1.355400 3.603940   1.862834 6.943293  0.0000   6.95546e-01 8.28330e-04 2.89541e-01 ...
   ...
\end{verbatim}
\label{Tab:fileExample}
\end{table*}

Basic and interpolated tracks, as well as iso-$\Zsurf$ lines, will be made available for download on our web site\footnote{%
\texttt{\tiny{http://obswww.unige.ch/Recherche/evol/-Base-de-donnees-}}
}.
Each downloaded file corresponds to either a given metallicity (for basic and interpolated tracks and for isochrones) or a given surface metallicity (for iso-$\Zsurf$ lines), and contains as many tracks as the number of stellar masses requested by the user.
An extract of the file is given in Table~\ref{Tab:fileExample}.
It contains a file header that indicates the number of tracks included in the file, and then lists the different tracks one after the other.
For each track, a track header first summarizes general information on the track.
Those are
\begin{itemize}
\item the type of track (in the example given in Table~\ref{Tab:fileExample}, it is a basic track);
\item the metallicity, either $\Zinit$ or $\Zsurf$ depending on the type of track;
\item the age or the stellar mass of the track, depending on whether it is an isochrone or not;
\item the number of models in the track;
\item the model indexes of the reference points in the track.
      The reference points define phases in the following way:
      \begin{itemize}
      \item phase 20: starts at the ZAMS,
      \item phase 21: starts at the second (after ZAMS) reference point on the MS,
      \item phase 22: starts at the third reference point, which corresponds to the occurrence of the first turn-off point on the MS or to the equivalent point if there is no hook,
      \item phase 23: starts at the fourth reference point, which corresponds to the occurrence of the second turn-off point on the MS, or to the equivalent point if there is no hook,
      \item phase 30: starts at the TAMS,
      \item phase 31: base of the red giant branch.
      \end{itemize}
\end{itemize}
The data of the models are then listed, one line per model.
For each model, we give, ordered by column:
\begin{enumerate}
\item the model index;
\item the phase of the model;
\item $t$ (yr): the age;
\item $M/M_\odot$: the stellar mass;
\item $\log L/L_\odot$: the logarithm of the surface luminosity;
\item $\log\Teff$: the logarithm of the effective temperature;
\item $\log\rho_\mathrm{c}$ (g/cm$^3$): the logarithm of the central density;
\item $\log T_\mathrm{c}$ (K): the logarithm of the central temperature;
\item $Q_\mathrm{cc}$: the mass of the convective core relative to the stellar mass;
\item $X_\mathrm{c}(\mathrm{H})$: the hydrogen mass fraction at the center;
\item $X_\mathrm{c}(\chem{He}{3})$;
\item $X_\mathrm{c}(\chem{He}{4})$;
\item $X_\mathrm{c}(\chem{C}{12})$;
\item $X_\mathrm{c}(\chem{C}{13})$;
\item $X_\mathrm{c}(\chem{N}{14})$;
\item $X_\mathrm{c}(\chem{O}{16})$;
\item $X_\mathrm{c}(\chem{O}{17})$;
\item $X_\mathrm{c}(\chem{O}{18})$;
\item $X_\mathrm{c}(\chem{Ne}{20})$;
\item $X_\mathrm{c}(\chem{Ne}{22})$;
\item $X_\mathrm{s}(\mathrm{H})$: the hydrogen mass fraction at the surface;
\item $X_\mathrm{s}(\chem{He}{3})$;
\item $X_\mathrm{s}(\chem{He}{4})$;
\item $X_\mathrm{s}(\chem{C}{12})$;
\item $X_\mathrm{s}(\chem{C}{13})$;
\item $X_\mathrm{s}(\chem{N}{14})$;
\item $X_\mathrm{s}(\chem{O}{16})$;
\item $X_\mathrm{s}(\chem{O}{17})$;
\item $X_\mathrm{s}(\chem{O}{18})$;
\item $X_\mathrm{s}(\chem{Ne}{20})$;
\item $X_\mathrm{s}(\chem{Ne}{22})$;
\item $M_\mathrm{init}/M_\odot$: the stellar mass of the initial model of the evolutionary track from which this model has been computed;
\item $X_\mathrm{init}$: the hydrogen mass fraction of the initial model of the evolutionary track from which this model has been computed;
\item $Y_\mathrm{init}$: same as $X_\mathrm{init}$, but for the initial helium mass fraction;
\item $\Zinit$: same as $X_\mathrm{init}$, but for the initial metallicity;
\item $\Zsurf$: the surface metallicity;
\item $R/R_\odot$: the stellar radius;
\item $\log \bar{\rho}/\bar{\rho}_\odot$: the logarithm of the mean stellar density relative to the solar mean density, with $\bar{\rho}_\odot$\,=\,1.411~g/cm$^3$;
\item $\nu_\mathrm{max}$: the frequency of stellar oscillation at maximum amplitude, given by Eq.~(\ref{Eq:astro_numax});
\item $A_\mathrm{max}/A_{\mathrm{max},\odot}$: the maximum oscillation amplitude relative to that of the Sun given by Eq.~(\ref{Eq:astro_Amax});
\item $\Delta \nu$: the large frequency separation, given by Eq.~(\ref{Eq:astro_DeltaNu}).

\end{enumerate}

\section{Conclusions}
\label{Sect:conclusions}

The new grids of stellar models presented in this paper cover a dense distribution of masses between 0.5 and \mass{3.5} and metallicities $\Zinit$ between 0.006 and 0.04, corresponding to \FeH\,=\,$-0.33$ to $+0.54$.
The high density of tracks in mass and metallicity allows obtaining reliable interpolated tracks for pairs of values (\Zinit, $M$) that have not been computed.
We estimated from the checks presented in Sect. 4 that the dense grids lead to interpolation accuracies lower than one percent for most of the models for $R_\mathrm{eff}$ and $\Teff$, and better than five percent for the age of most of the interpolated models (Sect.~\ref{Sect:interpolatedTracks}).
This precision is required to test them against modern highquality observational data.

Including of atomic diffusion in our models with $M$\,$<$\,\mass{1.1} leads to variations in the surface abundances that should be taken into account when comparing with observational data of stars with measured metallicities.
For that purpose, iso-$\Zsurf$ lines are computed.
Due to these variations in the surface abundances, the relations linking $Z$ and $\FeH$ (Appendix \ref{Sect:FeH2Z}) become inaccurate in low-mass stars to about 5~\%.

In the transition mass range \mass{1.15-1.25}, isochrones may display more than two hooks in the HR diagram as a result of the development of the convective core in the MS phase.
The exact morphology of the isochrone may depend on the numerical treatment of convection.
At solar metallicity, this transition mass range corresponds to ages between 3.8 and 4.2~Gyr.

The basic tracks are available for download at \texttt{\tiny{http://obswww.unige.ch/Recherche/evol/-Base-de-donnees-}}~.
An interface will also be added to compute and download interpolated tracks, iso-$\Zsurf$ lines and isochrones, on demand, within the limits of the parameters space of our grids.

\begin{acknowledgements}

We thank the referee, A. Weiss, for his careful reading of the paper and his fruitful suggestions.

\end{acknowledgements}

\bibliographystyle{aa}
\bibliography{bibTex}

\begin{thebibliography}{47}
\expandafter\ifx\csname natexlab\endcsname\relax\def\natexlab#1{#1}\fi

\bibitem[{{Asplund} {et~al.}(2005){Asplund}, {Grevesse}, \&
  {Sauval}}]{AsplundGrevesseSauval05}
{Asplund}, M., {Grevesse}, N., \& {Sauval}, A.~J. 2005, in ASPC, Vol. 336,
  Cosmic Abundances as Records of Stellar Evolution and Nucleosynthesis, ed.
  T.~G. {Barnes}, III \& F.~N. {Bash} (San Francisco: ASP), 25

\bibitem[{{Baraffe} {et~al.}(1998){Baraffe}, {Chabrier}, {Allard}, \&
  {Hauschildt}}]{BaraffeChabrierAllard_etal98}
{Baraffe}, I., {Chabrier}, G., {Allard}, F., \& {Hauschildt}, P.~H. 1998, \aap,
  337, 403

\bibitem[{{Basu} \& {Antia}(2008)}]{BasuAntia08}
{Basu}, S. \& {Antia}, H.~M. 2008, \physrep, 457, 217

\bibitem[{{Bergbusch} \& {Vandenberg}(1992)}]{BergbuschVandenberg92}
{Bergbusch}, P.~A. \& {Vandenberg}, D.~A. 1992, \apjs, 81, 163

\bibitem[{{Boehm-Vitense}(1981)}]{Boehm-Vitense81}
{Boehm-Vitense}, E. 1981, \araa, 19, 295

\bibitem[{{B{\"o}hm-Vitense}(1958)}]{Bohm58}
{B{\"o}hm-Vitense}, E. 1958, \zap, 46, 108

\bibitem[{{Brown} {et~al.}(1991){Brown}, {Gilliland}, {Noyes}, \&
  {Ramsey}}]{BrownGillilandNoyesRamsey91}
{Brown}, T.~M., {Gilliland}, R.~L., {Noyes}, R.~W., \& {Ramsey}, L.~W. 1991,
  \apj, 368, 599

\bibitem[{{Chaplin} {et~al.}(2011{\natexlab{a}}){Chaplin}, {Kjeldsen},
  {Bedding}, {Christensen-Dalsgaard}, {Gilliland}, {Kawaler}, {Appourchaux},
  {Elsworth}, {Garc{\'{\i}}a}, {Houdek}, {Karoff}, {Metcalfe},
  {Molenda-{\.Z}akowicz}, {Monteiro}, {Thompson}, {Verner}, {Batalha},
  {Borucki}, {Brown}, {Bryson}, {Christiansen}, {Clarke}, {Jenkins}, {Klaus},
  {Koch}, {An}, {Ballot}, {Basu}, {Benomar}, {Bonanno}, {Broomhall},
  {Campante}, {Corsaro}, {Creevey}, {Esch}, {Gai}, {Gaulme}, {Hale},
  {Handberg}, {Hekker}, {Huber}, {Mathur}, {Mosser}, {New}, {Pinsonneault},
  {Pricopi}, {Quirion}, {R{\'e}gulo}, {Roxburgh}, {Salabert}, {Stello}, \&
  {Suran}}]{ChaplinKjeldsenBedding_etal11}
{Chaplin}, W.~J., {Kjeldsen}, H., {Bedding}, T.~R., {et~al.}
  2011{\natexlab{a}}, \apj, 732, 54

\bibitem[{{Chaplin} {et~al.}(2011{\natexlab{b}}){Chaplin}, {Kjeldsen},
  {Christensen-Dalsgaard}, {Basu}, {Miglio}, {Appourchaux}, {Bedding},
  {Elsworth}, {Garc{\'{\i}}a}, {Gilliland}, {Girardi}, {Houdek}, {Karoff},
  {Kawaler}, {Metcalfe}, {Molenda-{\.Z}akowicz}, {Monteiro}, {Thompson},
  {Verner}, {Ballot}, {Bonanno}, {Brand{\~a}o}, {Broomhall}, {Bruntt},
  {Campante}, {Corsaro}, {Creevey}, {Do{\u g}an}, {Esch}, {Gai}, {Gaulme},
  {Hale}, {Handberg}, {Hekker}, {Huber}, {Jim{\'e}nez}, {Mathur}, {Mazumdar},
  {Mosser}, {New}, {Pinsonneault}, {Pricopi}, {Quirion}, {R{\'e}gulo},
  {Salabert}, {Serenelli}, {Aguirre}, {Sousa}, {Stello}, {Stevens}, {Suran},
  {Uytterhoeven}, {White}, {Borucki}, {Brown}, {Jenkins}, {Kinemuchi}, {Van
  Cleve}, \& {Klaus}}]{ChaplinKjeldsenChristensenDalsgaard_etal11}
{Chaplin}, W.~J., {Kjeldsen}, H., {Christensen-Dalsgaard}, J., {et~al.}
  2011{\natexlab{b}}, Science, 332, 213

\bibitem[{{Chapman} \& {Cowling}(1970)}]{ChapmanCowling70}
{Chapman}, S. \& {Cowling}, T.~G. 1970, {The mathematical theory of non-uniform
  gases. an account of the kinetic theory of viscosity, thermal conduction and
  diffusion in gases} (Cambridge: University Press, 1970, 3rd ed.)

\bibitem[{{Clem} {et~al.}(2011){Clem}, {Landolt}, {Hoard}, \&
  {Wachter}}]{ClemLandoltHoard_etal11}
{Clem}, J.~L., {Landolt}, A.~U., {Hoard}, D.~W., \& {Wachter}, S. 2011, \aj,
  141, 115

\bibitem[{{Cunha} {et~al.}(2006){Cunha}, {Hubeny}, \&
  {Lanz}}]{CunhaHubenyLanz06}
{Cunha}, K., {Hubeny}, I., \& {Lanz}, T. 2006, \apjl, 647, L143

\bibitem[{{Demarque} {et~al.}(2004){Demarque}, {Woo}, {Kim}, \&
  {Yi}}]{DemarqueWooKim_etal04}
{Demarque}, P., {Woo}, J.-H., {Kim}, Y.-C., \& {Yi}, S.~K. 2004, \apjs, 155,
  667

\bibitem[{{Eggenberger} {et~al.}(2008){Eggenberger}, {Meynet}, {Maeder},
  {Hirschi}, {Charbonnel}, {Talon}, \&
  {Ekstr{\"o}m}}]{EggenbergerMeynetMaeder08}
{Eggenberger}, P., {Meynet}, G., {Maeder}, A., {et~al.} 2008, \apss, 316, 43

\bibitem[{{Eggenberger} {et~al.}(2010{\natexlab{a}}){Eggenberger}, {Meynet},
  {Maeder}, {Miglio}, {Montalban}, {Carrier}, {Mathis}, {Charbonnel}, \&
  {Talon}}]{EggenbergerMeynetMaeder10}
{Eggenberger}, P., {Meynet}, G., {Maeder}, A., {et~al.} 2010{\natexlab{a}},
  \aap, 519, A116

\bibitem[{{Eggenberger} {et~al.}(2010{\natexlab{b}}){Eggenberger}, {Miglio},
  {Montalban}, {Moreira}, {Noels}, {Meynet}, \&
  {Maeder}}]{EggenbergerMiglioMontalban10}
{Eggenberger}, P., {Miglio}, A., {Montalban}, J., {et~al.} 2010{\natexlab{b}},
  \aap, 509, A72

\bibitem[{{Ekstr{\"o}m} {et~al.}(2011){Ekstr{\"o}m}, {Georgy}, {Eggenberger},
  {Meynet}, {Mowlavi}, {Wyttenbach}, {Granada}, {Decressin}, {Hirschi},
  {Frischknecht}, \& {Charbonnel}}]{EkstroemGeorgyEggenberger11}
{Ekstr{\"o}m}, S., {Georgy}, C., {Eggenberger}, P., {et~al.} 2011, ArXiv
  e-prints

\bibitem[{{Flower}(1996)}]{Flower96}
{Flower}, P.~J. 1996, \apj, 469, 355

\bibitem[{{Friel} {et~al.}(2010){Friel}, {Jacobson}, \&
  {Pilachowski}}]{FrielJacobsonPilachowski10}
{Friel}, E.~D., {Jacobson}, H.~R., \& {Pilachowski}, C.~A. 2010, \aj, 139, 1942

\bibitem[{{Gillon} {et~al.}(2009){Gillon}, {Smalley}, {Hebb}, {Anderson},
  {Triaud}, {Hellier}, {Maxted}, {Queloz}, \&
  {Wilson}}]{GillonSmalleyHebb_etal09}
{Gillon}, M., {Smalley}, B., {Hebb}, L., {et~al.} 2009, \aap, 496, 259

\bibitem[{{Girardi} {et~al.}(2000){Girardi}, {Bressan}, {Bertelli}, \&
  {Chiosi}}]{GirardiBressanBertelli_etal00}
{Girardi}, L., {Bressan}, A., {Bertelli}, G., \& {Chiosi}, C. 2000, \aaps, 141,
  371

\bibitem[{{Huber} {et~al.}(2011){Huber}, {Bedding}, {Stello}, {Hekker},
  {Mathur}, {Mosser}, {Verner}, {Bonanno}, {Buzasi}, {Campante}, {Elsworth},
  {Hale}, {Kallinger}, {Silva Aguirre}, {Chaplin}, {De Ridder}, {Garcia},
  {Appourchaux}, {Frandsen}, {Houdek}, {Molenda-Zakowicz}, {Monteiro},
  {Christensen-Dalsgaard}, {Gilliland}, {Kawaler}, {Kjeldsen}, {Broomhall},
  {Corsaro}, {Salabert}, {Sanderfer}, {Seader}, \&
  {Smith}}]{HuberBeddingStello_etal11}
{Huber}, D., {Bedding}, T.~R., {Stello}, D., {et~al.} 2011, ApJ, in press
  (astro-ph 1109.3460

\bibitem[{{Kjeldsen} \& {Bedding}(1995)}]{KjeldsenBedding95}
{Kjeldsen}, H. \& {Bedding}, T.~R. 1995, \aap, 293, 87

\bibitem[{{Kraus} {et~al.}(2011){Kraus}, {Tucker}, {Thompson}, {Craine}, \&
  {Hillenbrand}}]{KrausTuckerThompson_etal11}
{Kraus}, A.~L., {Tucker}, R.~A., {Thompson}, M.~I., {Craine}, E.~R., \&
  {Hillenbrand}, L.~A. 2011, \apj, 728, 48

\bibitem[{{Lagarde} {et~al.}(2011){Lagarde}, {Charbonnel}, {Decressin}, \&
  {Hagelberg}}]{LagardeCharbonnelDecressin11}
{Lagarde}, N., {Charbonnel}, C., {Decressin}, T., \& {Hagelberg}, J. 2011,
  ArXiv e-prints

\bibitem[{{Lebreton} {et~al.}(1999){Lebreton}, {Perrin}, {Cayrel}, {Baglin}, \&
  {Fernandes}}]{LebretonPerrinCayrel_etal99}
{Lebreton}, Y., {Perrin}, M.-N., {Cayrel}, R., {Baglin}, A., \& {Fernandes}, J.
  1999, \aap, 350, 587

\bibitem[{{Magic} {et~al.}(2010){Magic}, {Serenelli}, {Weiss}, \&
  {Chaboyer}}]{MagicSerenelliWeiss_etal10}
{Magic}, Z., {Serenelli}, A., {Weiss}, A., \& {Chaboyer}, B. 2010, \apj, 718,
  1378

\bibitem[{{Miglio} {et~al.}(2009){Miglio}, {Montalb{\'a}n}, {Baudin},
  {Eggenberger}, {Noels}, {Hekker}, {De Ridder}, {Weiss}, \&
  {Baglin}}]{MiglioMontalbanBaudin_etal09}
{Miglio}, A., {Montalb{\'a}n}, J., {Baudin}, F., {et~al.} 2009, \aap, 503, L21

\bibitem[{{Montgomery} {et~al.}(1993){Montgomery}, {Marschall}, \&
  {Janes}}]{MontgomeryMarschall93}
{Montgomery}, K.~A., {Marschall}, L.~A., \& {Janes}, K.~A. 1993, \aj, 106, 181

\bibitem[{{Morel} \& {Baglin}(1999)}]{MorelBaglin99}
{Morel}, P. \& {Baglin}, A. 1999, \aap, 345, 156

\bibitem[{{Morel} {et~al.}(2000){Morel}, {Provost}, \&
  {Berthomieu}}]{MorelProvostBerthomieu00}
{Morel}, P., {Provost}, J., \& {Berthomieu}, G. 2000, \aap, 353, 771

\bibitem[{{Mowlavi} {et~al.}(1998){Mowlavi}, {Meynet}, {Maeder}, {Schaerer}, \&
  {Charbonnel}}]{MowlaviMeynetMaeder_etal98}
{Mowlavi}, N., {Meynet}, G., {Maeder}, A., {Schaerer}, D., \& {Charbonnel}, C.
  1998, \aap, 335, 573

\bibitem[{{Pancino} {et~al.}(2010){Pancino}, {Carrera}, {Rossetti}, \&
  {Gallart}}]{PancinoCarreraRossetti10}
{Pancino}, E., {Carrera}, R., {Rossetti}, E., \& {Gallart}, C. 2010, \aap, 511,
  A56

\bibitem[{{Rogers} \& {Nayfonov}(2002)}]{RogersNayfonov02}
{Rogers}, F.~J. \& {Nayfonov}, A. 2002, \apj, 576, 1064

\bibitem[{{Salaris} {et~al.}(2000){Salaris}, {Groenewegen}, \&
  {Weiss}}]{SalarisGroenewegenWeiss00}
{Salaris}, M., {Groenewegen}, M.~A.~T., \& {Weiss}, A. 2000, \aap, 355, 299

\bibitem[{{Sandquist}(2004)}]{Sandquist04}
{Sandquist}, E.~L. 2004, \mnras, 347, 101

\bibitem[{{Sarajedini} {et~al.}(1999){Sarajedini}, {von Hippel},
  {Kozhurina-Platais}, \& {Demarque}}]{SarajediniVonHippelKozhurinaPlatais99}
{Sarajedini}, A., {von Hippel}, T., {Kozhurina-Platais}, V., \& {Demarque}, P.
  1999, \aj, 118, 2894

\bibitem[{{Schaller} {et~al.}(1992){Schaller}, {Schaerer}, {Meynet}, \&
  {Maeder}}]{SchallerSchaererMeynet_etal92}
{Schaller}, G., {Schaerer}, D., {Meynet}, G., \& {Maeder}, A. 1992, \aaps, 96,
  269

\bibitem[{{Sekiguchi} \& {Fukugita}(2000)}]{SekiguchiFukugita00}
{Sekiguchi}, M. \& {Fukugita}, M. 2000, \aj, 120, 1072

\bibitem[{{Sozzetti} {et~al.}(2007){Sozzetti}, {Torres}, {Charbonneau},
  {Latham}, {Holman}, {Winn}, {Laird}, \&
  {O'Donovan}}]{SozzettiTorresCharbonneau_etal07}
{Sozzetti}, A., {Torres}, G., {Charbonneau}, D., {et~al.} 2007, \apj, 664, 1190

\bibitem[{{Stello} {et~al.}(2009{\natexlab{a}}){Stello}, {Chaplin}, {Basu},
  {Elsworth}, \& {Bedding}}]{StelloChaplinBasu09}
{Stello}, D., {Chaplin}, W.~J., {Basu}, S., {Elsworth}, Y., \& {Bedding}, T.~R.
  2009{\natexlab{a}}, \mnras, 400, L80

\bibitem[{{Stello} {et~al.}(2009{\natexlab{b}}){Stello}, {Chaplin}, {Bruntt},
  {Creevey}, {Garc{\'{\i}}a-Hern{\'a}ndez}, {Monteiro}, {Moya}, {Quirion},
  {Sousa}, {Su{\'a}rez}, {Appourchaux}, {Arentoft}, {Ballot}, {Bedding},
  {Christensen-Dalsgaard}, {Elsworth}, {Fletcher}, {Garc{\'{\i}}a}, {Houdek},
  {Jim{\'e}nez-Reyes}, {Kjeldsen}, {New}, {R{\'e}gulo}, {Salabert}, \&
  {Toutain}}]{StelloChaplinBruntt_etal09}
{Stello}, D., {Chaplin}, W.~J., {Bruntt}, H., {et~al.} 2009{\natexlab{b}},
  \apj, 700, 1589

\bibitem[{{Talon} {et~al.}(1997){Talon}, {Zahn}, {Maeder}, \&
  {Meynet}}]{TalonZahnMaeder_etal97}
{Talon}, S., {Zahn}, J.-P., {Maeder}, A., \& {Meynet}, G. 1997, \aap, 322, 209

\bibitem[{{Torres} {et~al.}(2010){Torres}, {Andersen}, \&
  {Gim{\'e}nez}}]{TorresAndersenGimenez10}
{Torres}, G., {Andersen}, J., \& {Gim{\'e}nez}, A. 2010, \aapr, 18, 67

\bibitem[{{Ulrich}(1986)}]{Ulrich86}
{Ulrich}, R.~K. 1986, \apjl, 306, L37

\bibitem[{{VandenBerg} {et~al.}(2007){VandenBerg}, {Gustafsson}, {Edvardsson},
  {Eriksson}, \& {Ferguson}}]{VandenBergGustafssonEdvardsson_etal07}
{VandenBerg}, D.~A., {Gustafsson}, B., {Edvardsson}, B., {Eriksson}, K., \&
  {Ferguson}, J. 2007, \apjl, 666, L105

\bibitem[{{Yi} {et~al.}(2001){Yi}, {Demarque}, {Kim}, {Lee}, {Ree}, {Lejeune},
  \& {Barnes}}]{YiDemarqueKim_etal01}
{Yi}, S., {Demarque}, P., {Kim}, Y.-C., {et~al.} 2001, \apjs, 136, 417

\end{thebibliography}

\begin{appendix}
\section{Transformation between $Z$ and \FeH}
\label{Sect:FeH2Z}
 
A relation between the mass fraction $Z$ of all elements heavier than helium and the metallicity
$\FeH = \log[X({\mathrm{Fe}})/X] - \log[X({\mathrm{Fe}})/X]_\odot$ (where the subscript $_\odot$ refers to the photospheric abundances of the Sun today) can be established if we make two assumptions on the chemical evolution of the Galaxy.
In doing that, we assume to deal with abundances in the interstellar medium, from which stars are born.
We therefore explicitly write $Z_\mathrm{init}=1-X_\mathrm{init}-Y_\mathrm{init}$ and
$\FeH_\mathrm{init} = \log[X({\mathrm{Fe}})/X]_\mathrm{init} - \log[X({\mathrm{Fe}})/X]_\odot$
in order to distinguish initial abundances from photospheric abundances of running models, the latter having possibly been modified in the course of stellar evolution by atomic diffusion.

The first assumption concerns the galactic evolution of the helium mass fraction $Y_\mathrm{init}$ with metallicity $Z_\mathrm{init}$.
A linear relation is assumed of the form
\begin{equation}
Y_\mathrm{init} = Y_0 + \deltaYdeltaZ \times Z_\mathrm{init},
\label{Eq:YfromZ}
\end{equation}
where $Y_0$ is the primordial helium mass fraction of the universe resulting from Big Bang nucleosynthesis and \deltaYdeltaZ\ reflects the chemical enrichment of the Galaxy whereby H is globally transformed into He and heavier elements.
Equation \ref{Eq:YfromZ} gives
\begin{equation}
 X_\mathrm{init} = 1 - Y_0 - \left(1+\deltaYdeltaZ \right) \; Z_\mathrm{init}
\label{Eq:XfromZ}
\end{equation}

Second, we assume a constant $X(\mathrm{Fe})/Z$ ratio throughout the evolution of the Galaxy, i.e we neglect enhancement of $\alpha$ elements.
We then have, for all $Z$,
\begin{equation}
  \frac{X_\mathrm{init}(\mathrm{Fe})}{Z_\mathrm{init}} = \frac{X_\mathrm{\odot,init}(\mathrm{Fe})}{\Z_\mathrm{\odot,init}}.
\label{Eq:X(Fe)/Z}
\end{equation}

Transformation relations between $\FeH_\mathrm{init}$ and $Z_\mathrm{init}$ result from assumptions \ref{Eq:XfromZ} and \ref{Eq:X(Fe)/Z}:
\begin{equation}
  \left\{
  \begin{array}{rll}
  \FeH_\mathrm{init} & = & \displaystyle \log \frac{Z_\mathrm{init}}
                       {1-Y_0-\left( 1+\deltaYdeltaZ \right) \; Z_\mathrm{init}}
              - \log \frac{Z_\odot}{X_\odot} \\
       &   & \\
     Z_\mathrm{init} & = & \displaystyle \frac{1-Y_0}
               {1+\deltaYdeltaZ+\frac{X_\odot}{Z_\odot}\times 10^{-\FeH_\mathrm{init}}}
              
  \end{array}
  \right.
\label{Eq:FeHandZ}
\end{equation}

With the values adopted in this paper ($Y_0$\,=\,0.248, $\Delta Y/\Delta Z$\,=\,1.2857 and $Z_\odot/X_\odot$=0.0174), Eqs.~\ref{Eq:FeHandZ} become:
\begin{equation}
\left\{
\begin{array}{rll}
  \FeH_\mathrm{init} & = & - \log \displaystyle \left(\frac{0.013}{Z_\mathrm{init}}-0.04\right)\\
       &   & \\
     Z_\mathrm{init} & = & \displaystyle \frac{0.013}
                                {\displaystyle 0.04 + 10^{-\FeH_\mathrm{init}}}
\end{array}
\right.
\label{Eq:finalFeHZ}
\end{equation}

It is important to keep in mind that Eqs.~\ref{Eq:FeHandZ} and \ref{Eq:finalFeHZ} relate to abundances at the time of star formation.
In our grids, these relations are also valid during the MS and post-MS models for stars equal or more massive than \mass{1.1}.
\textsl{For stars less massive than \mass{1.1}, however, these relations are no more valid} because they do not take into account the alteration with time of the surface abundances due to atomic diffusion.
For example, the solar model in our grids, obtained when the $Z_\mathrm{init}$\,=\,0.014, \mass{1.0} model has reached the solar luminosity and radius at the age of 4.57~Gyr, predicts surface abundances equal to  $X_\odot$\,=\,0.7524, $Y_\odot$\,=\,0.2346, $Z_\odot$\,=\,0.0131, and $Z_\odot/X_\odot$\,=\,0.0174, in agreement with observations.
With this value for $Z_\odot$ at the surface of the Sun today, however, Eq.~\ref{Eq:finalFeHZ} predicts a value for $\FeH$ of 0.018.
This is obviously wrong as, by definition, $\FeH_\odot$\,=\,0.0.
The error comes from the use of Eq.~\ref{Eq:finalFeHZ}, which does not include the effect of atomic diffusion during the past 4.57~Gyr evolution of the solar model.
The effect is seen to amount to $\sim$~2\% on $\FeH$ for the Sun, and is expected to amount to at most $\sim$5\% for the other low-mass models (see Fig.~\ref{Fig:XsDiffusion}).
Within this precision, Eqs.~\ref{Eq:FeHandZ} and \ref{Eq:finalFeHZ} can be used for all cases.

\section{Construction of the basic tracks}
\label{Sect:normalizationProcedure}

\subsection{Main sequence}

\subsubsection{Reference points in the HR diagram}
\label{Sect:MSNormalizationInHR}

\begin{figure}
  \centering

  \includegraphics[width=0.49\columnwidth]{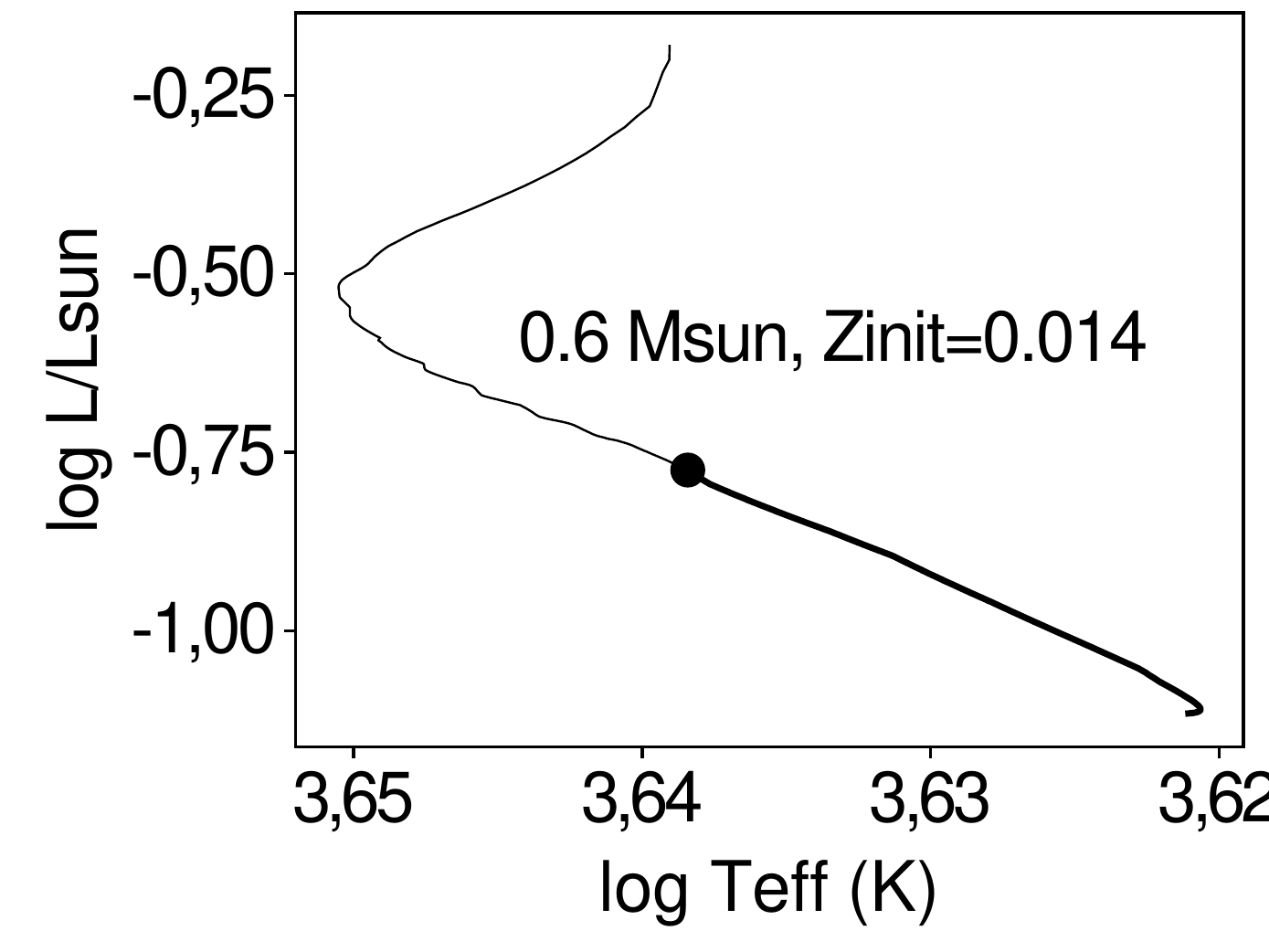}
  \includegraphics[width=0.49\columnwidth]{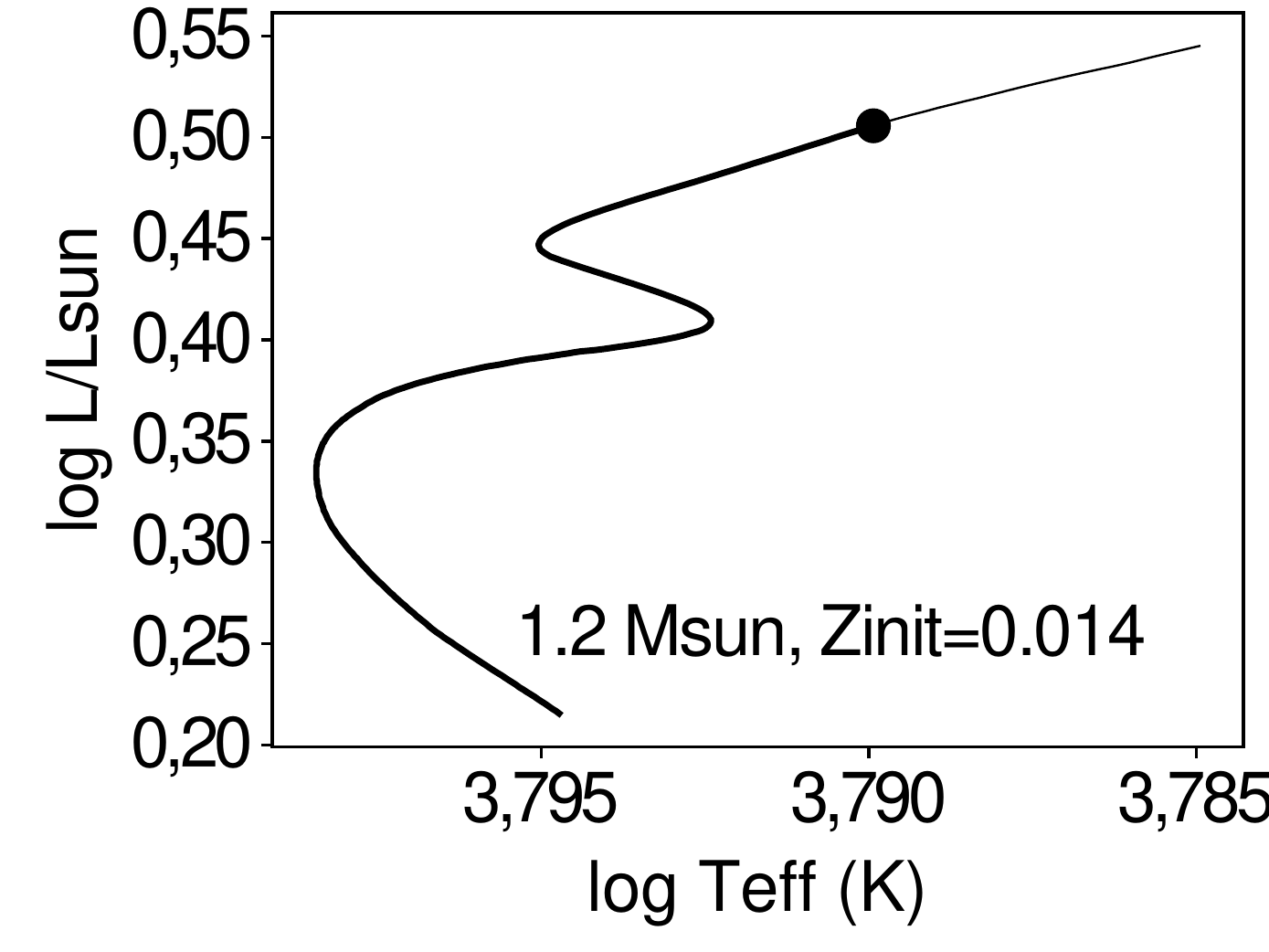}
  
  \includegraphics[width=0.49\columnwidth]{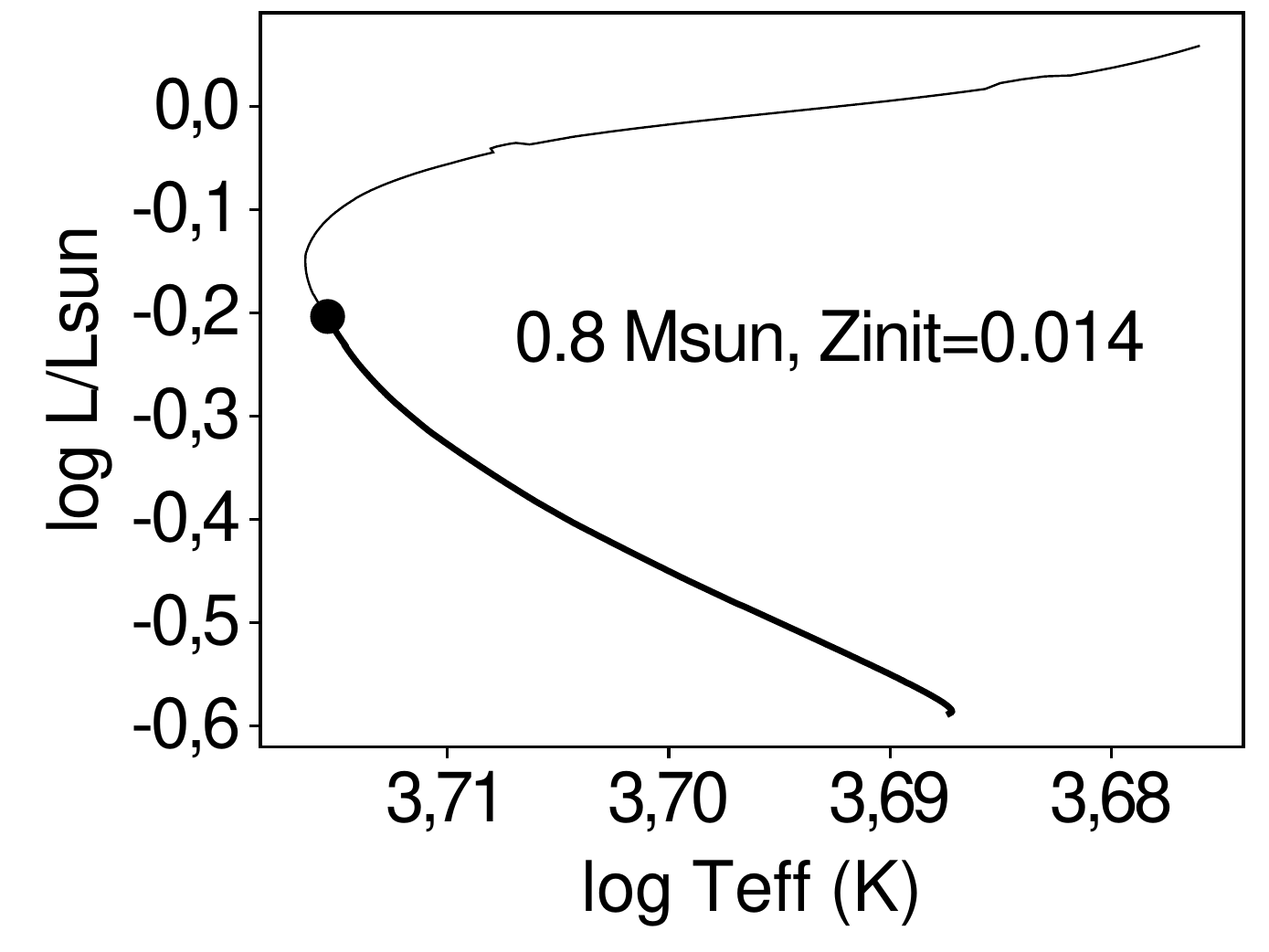}
  \includegraphics[width=0.49\columnwidth]{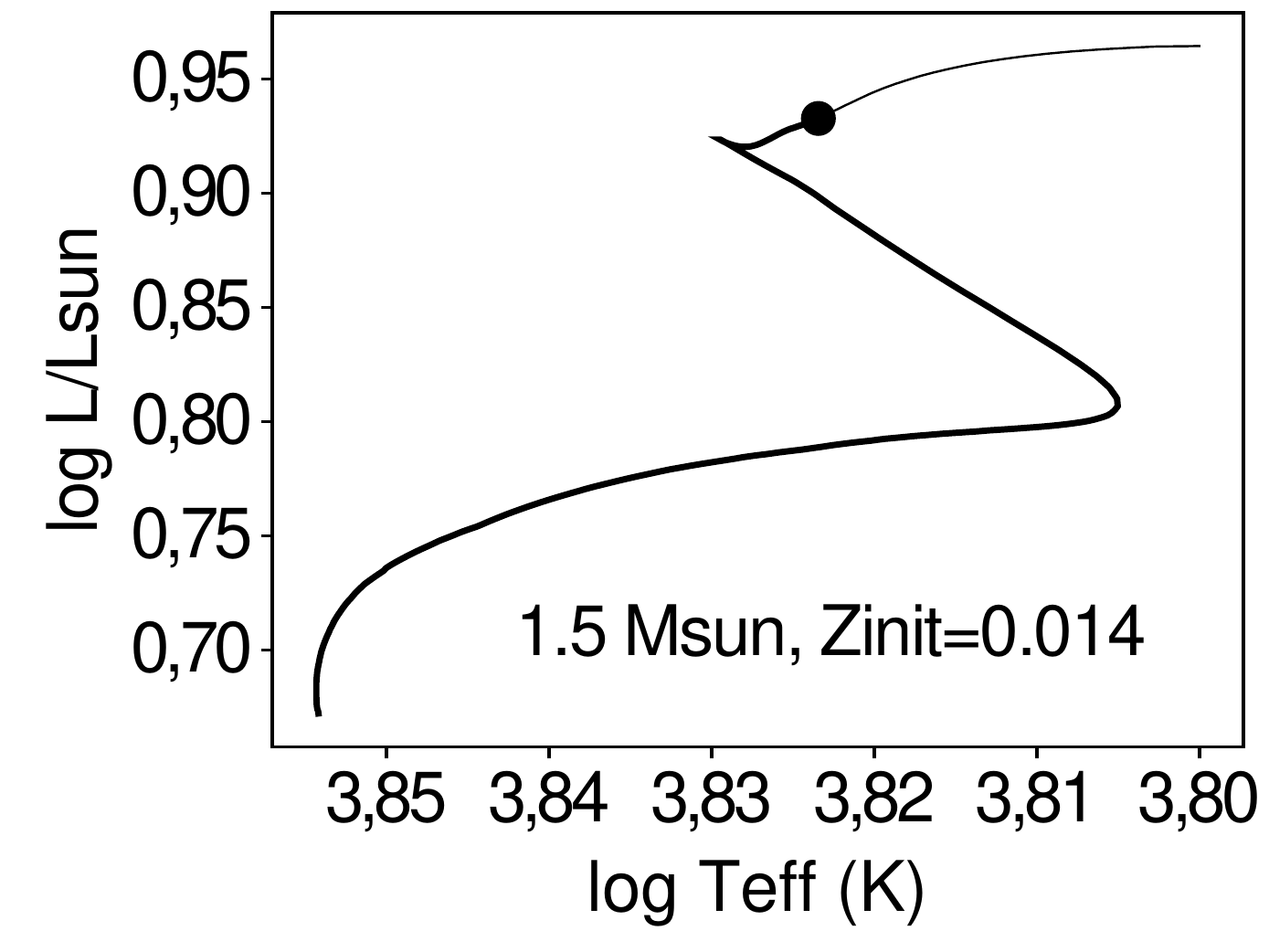}
  
  \includegraphics[width=0.49\columnwidth]{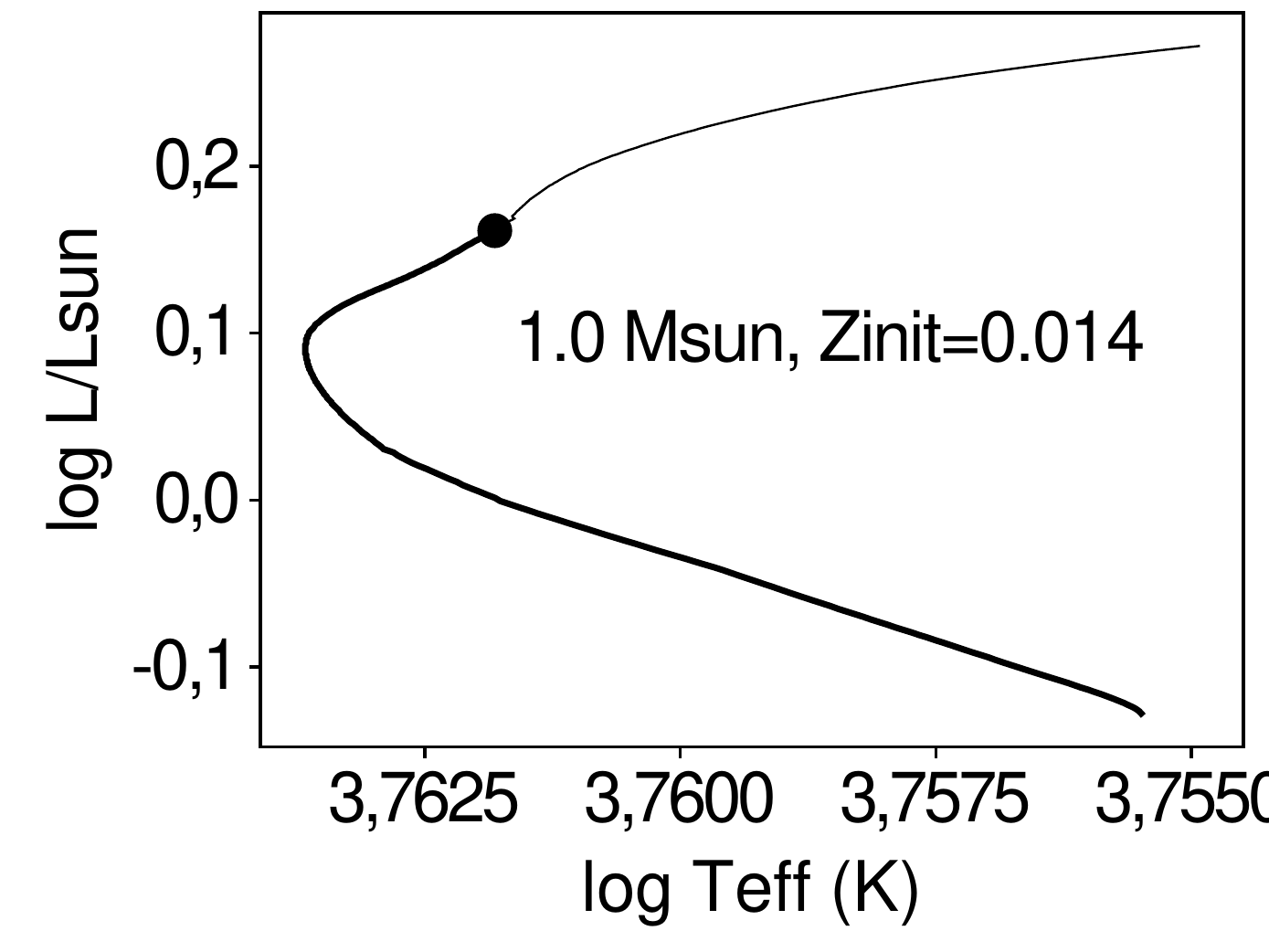}
  \includegraphics[width=0.49\columnwidth]{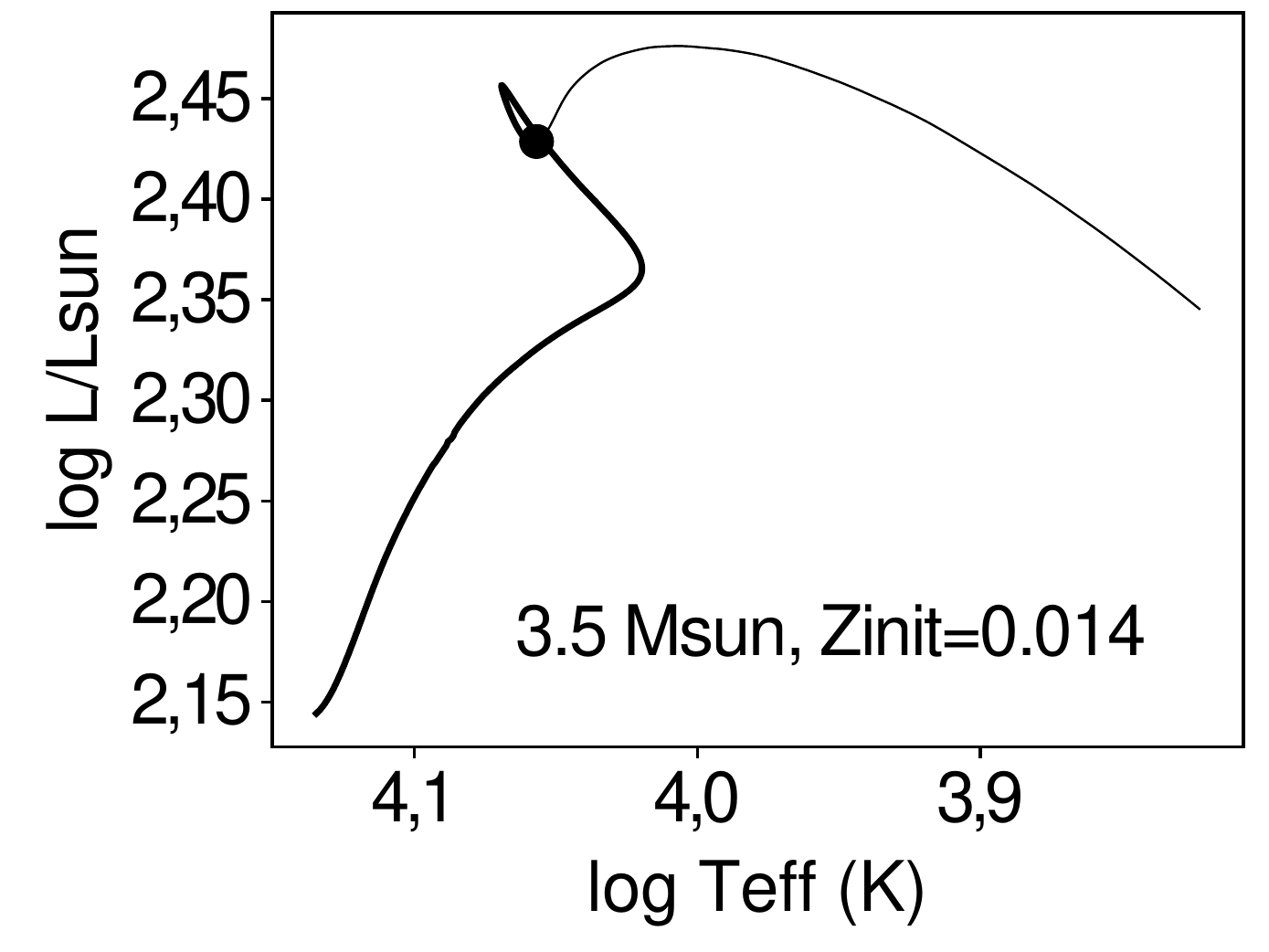}
  \caption{Selected $\Zinit$\,=\,0.014 tracks in the HR diagram covering the MS and part of the post-MS to illustrate the different morphologies.
  The main sequence is represented in thick lines, the filled circle locating its end.
  }
\label{Fig:msHRtracks}
\end{figure}

The tracks of several representative masses in the HR diagram are shown in Fig.~\ref{Fig:msHRtracks}.
They reveal a morphology of the MS that depends on the stellar mass.
In the low-mass regime ($M$\,$<$\,\mass{0.9}), the morphology is rather simple: both the effective temperature and the luminosity are monotonically increasing with time during the whole MS phase.
At higher masses, several complexities appear in the MS morphology:
\begin{enumerate}

\item For \mass{M \gtrsim 0.9}, $\Teff$ starts to decrease during the MS. 
The point where it starts to decrease is located at the end of the MS for the smaller masses, and occurs earlier as we consider more massive stars.
$\Teff$ decreases right from the ZAMS for $M$\,$\ge$\,\mass{1.6}.

\vspace{1mm}
\item Two turn off points appear in the HR diagram for $M$\,$\gtrsim$\mass{1.2}.

\vspace{1mm}
\item 
The surface luminosity is a strictly increasing function of time during all the MS for $M$\,$\lesssim$\,\mass{1.4}.
For masses $M$\,$\gtrsim$\,\mass{1.4}, however, the surface luminosity decreases after the second turn off point, reaches a minimum and increases again as the star evolves off the MS.

\end{enumerate}

To better cope with these different morphologies on the MS, we shift from the HR diagram to the $\log R - \log L$ plane.
This removes the first complexity because the stellar radius is always a strictly increasing function of time from the ZAMS up to the first loop, if existent, as illustrated in Fig.~\ref{Fig:msLogRLogLtracks}.
For constructing the basic tracks we use the morphologies in the $\log R - \log L$ plane.

\subsubsection{Reference points in the $\log R - \log L$ diagram}
\label{Sect:MSNormalizationInRL}

\begin{figure}
  \centering

  \includegraphics[width=0.49\columnwidth]{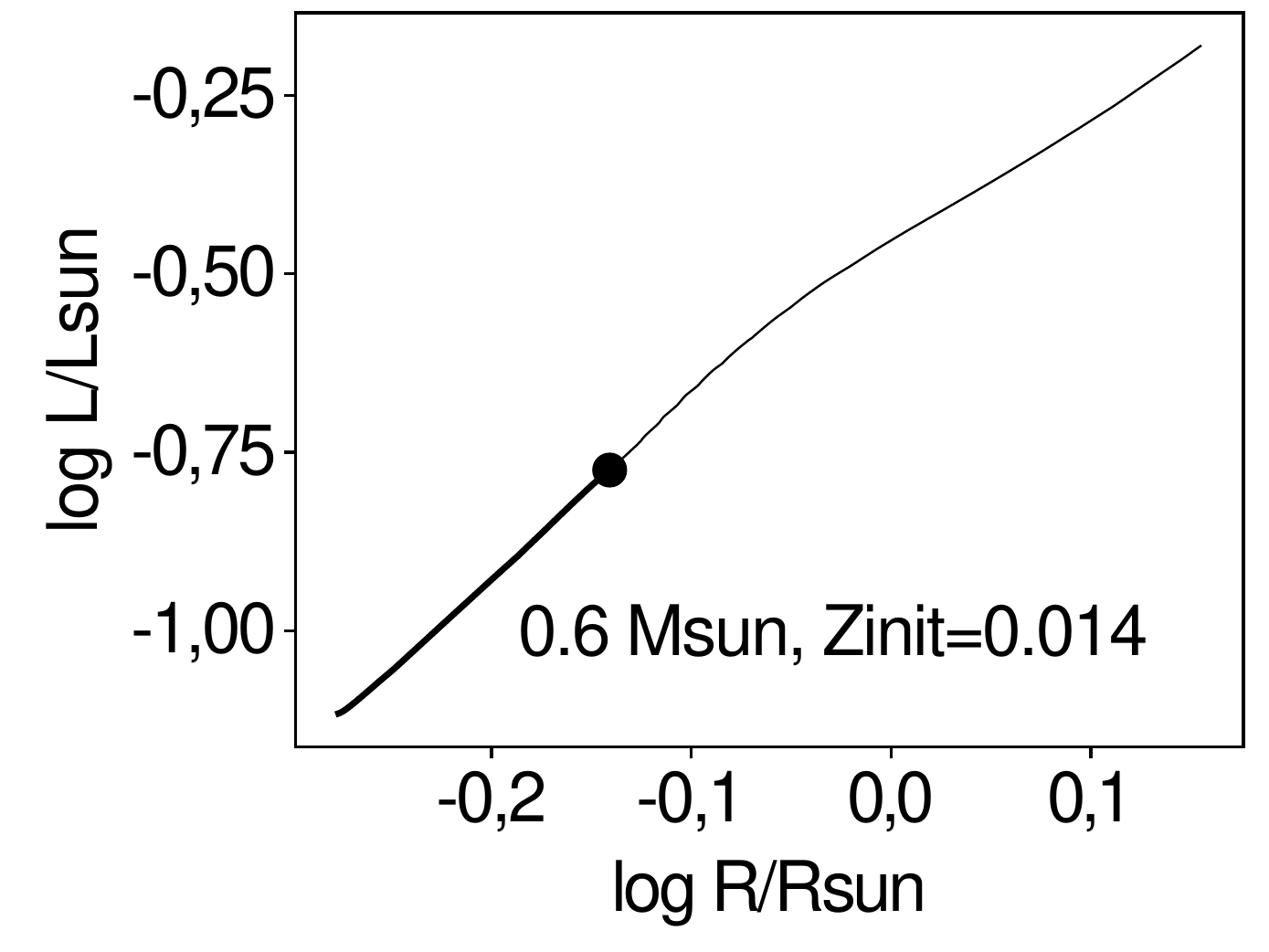}
  \includegraphics[width=0.49\columnwidth]{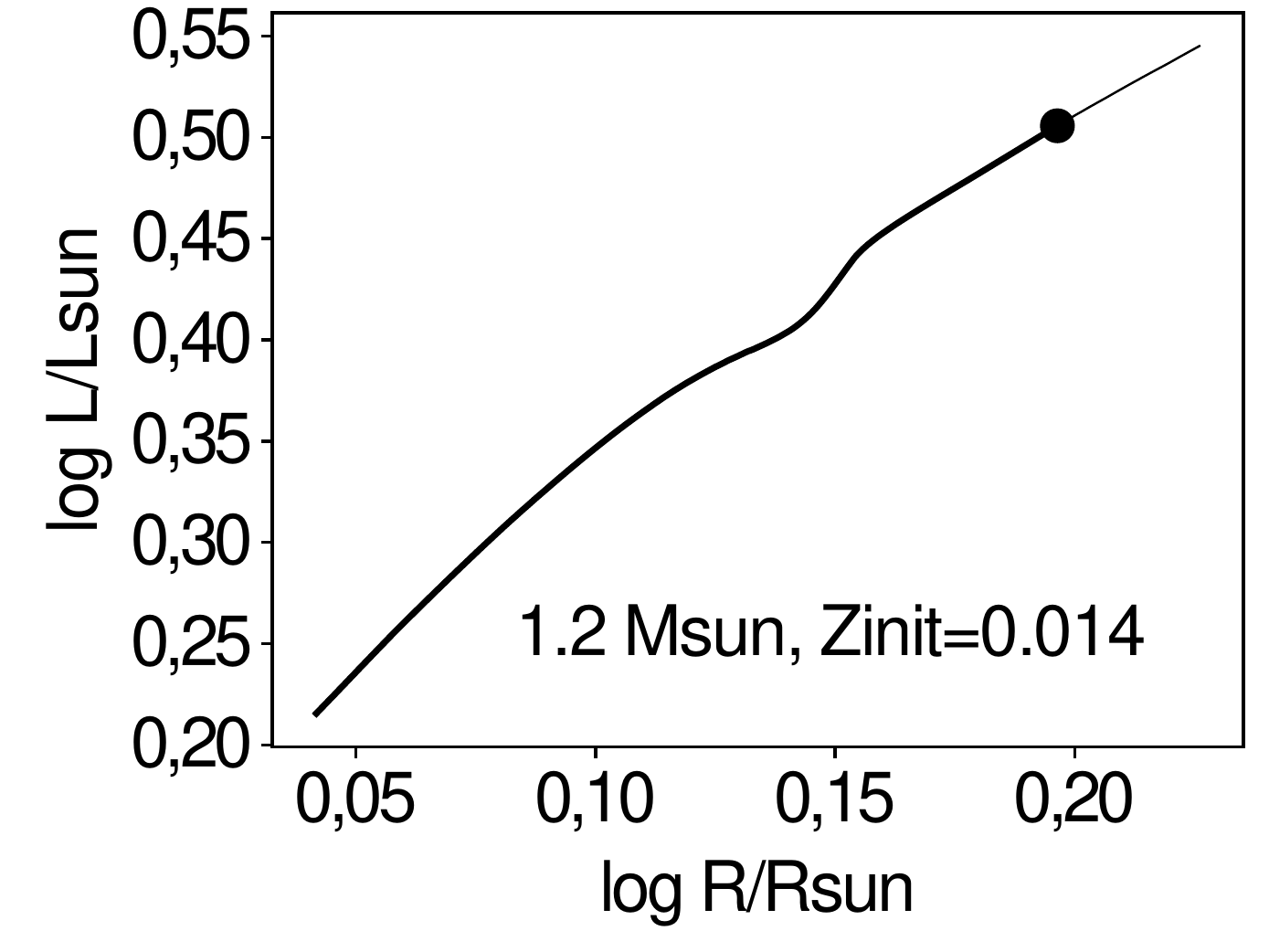}
  
  \includegraphics[width=0.49\columnwidth]{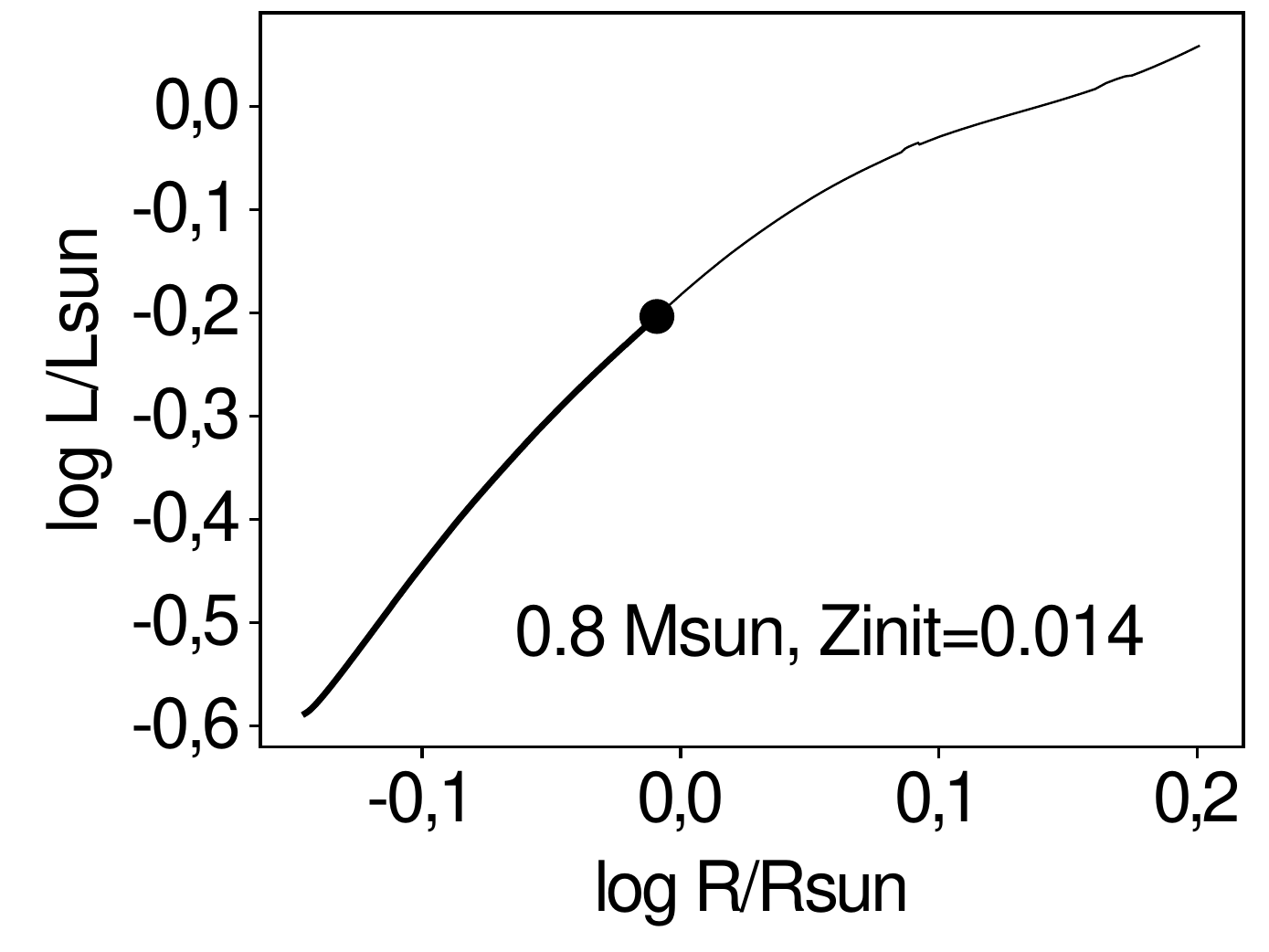}
  \includegraphics[width=0.49\columnwidth]{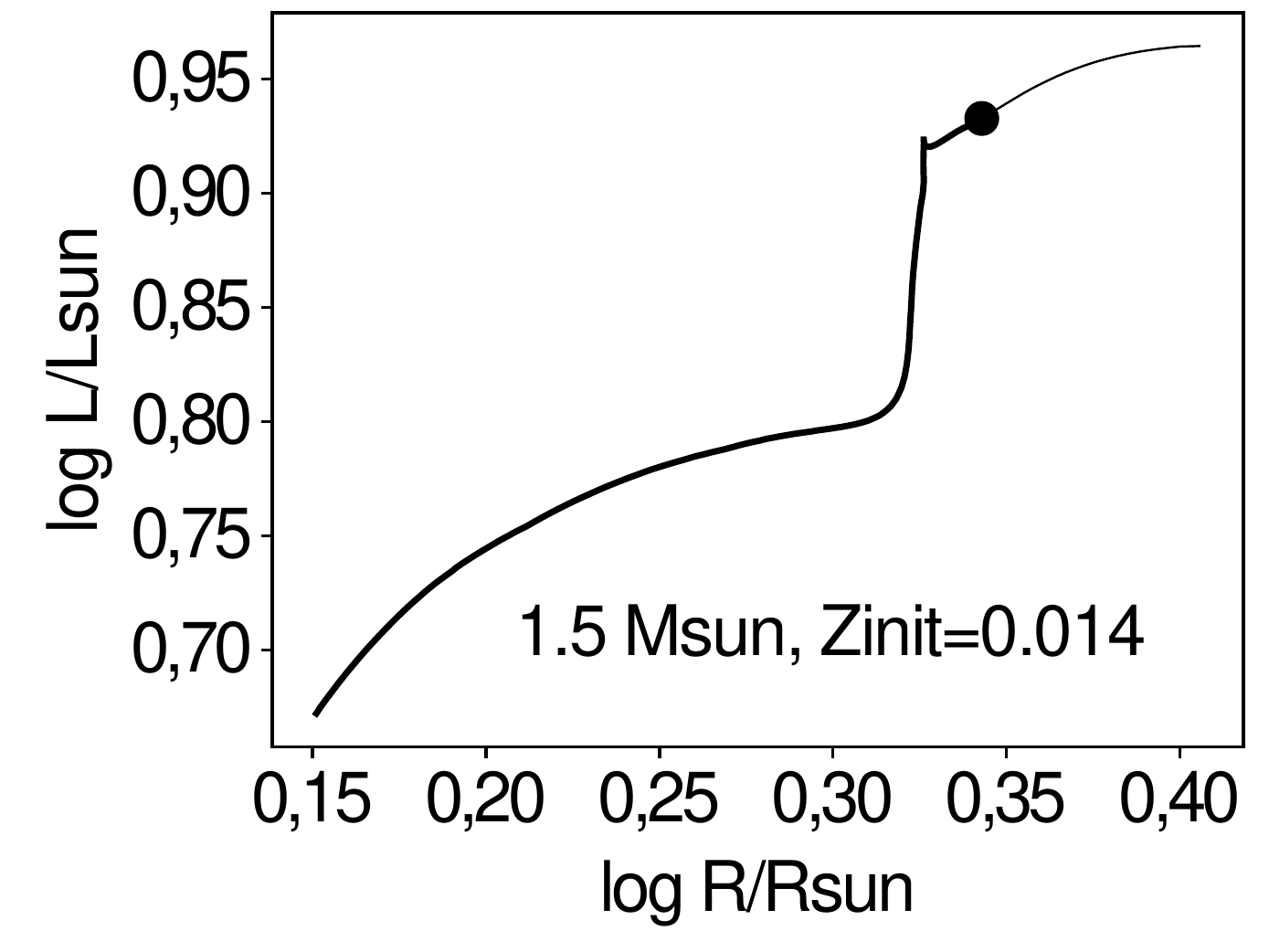}
  
  \includegraphics[width=0.49\columnwidth]{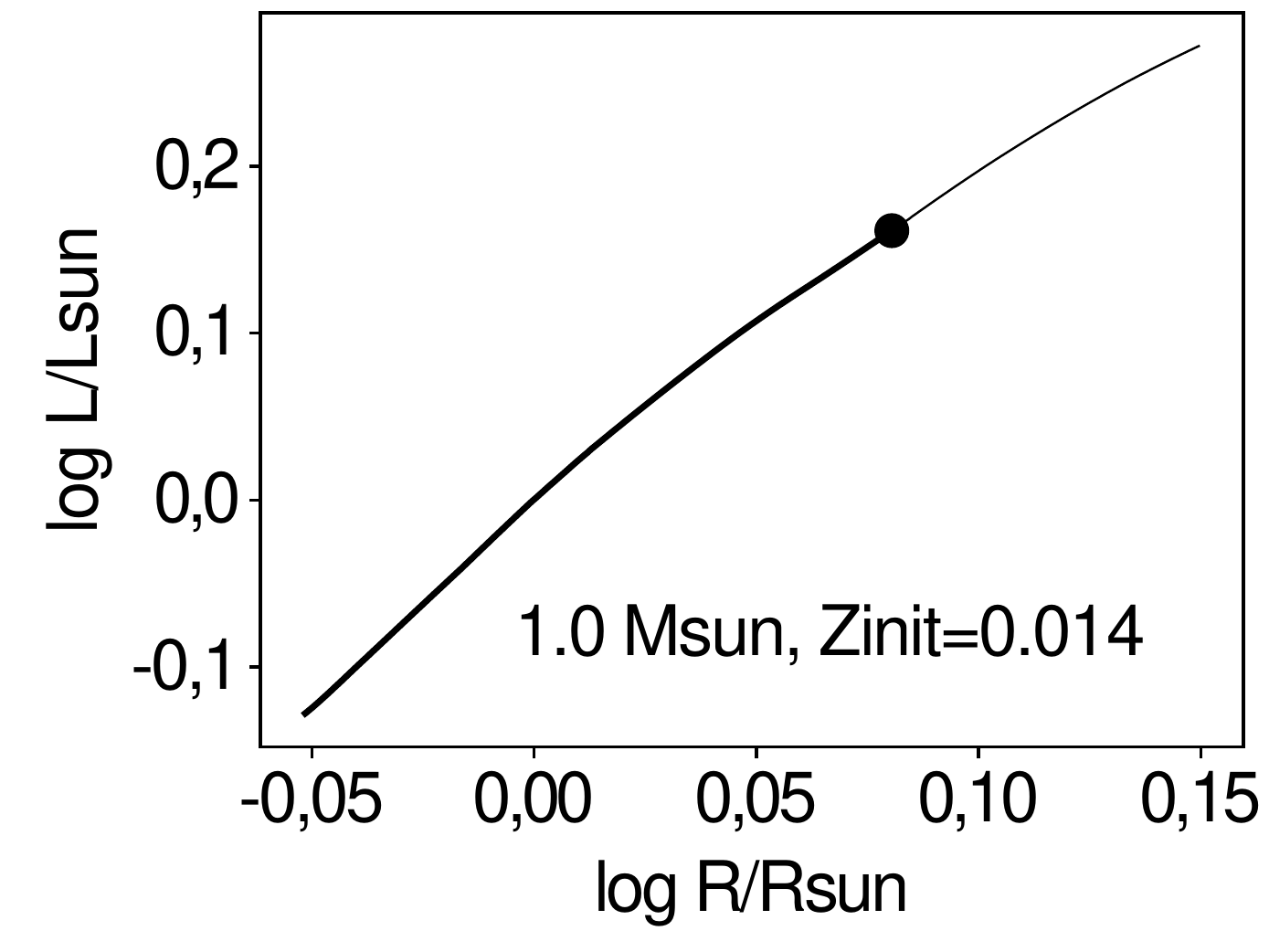}
  \includegraphics[width=0.49\columnwidth]{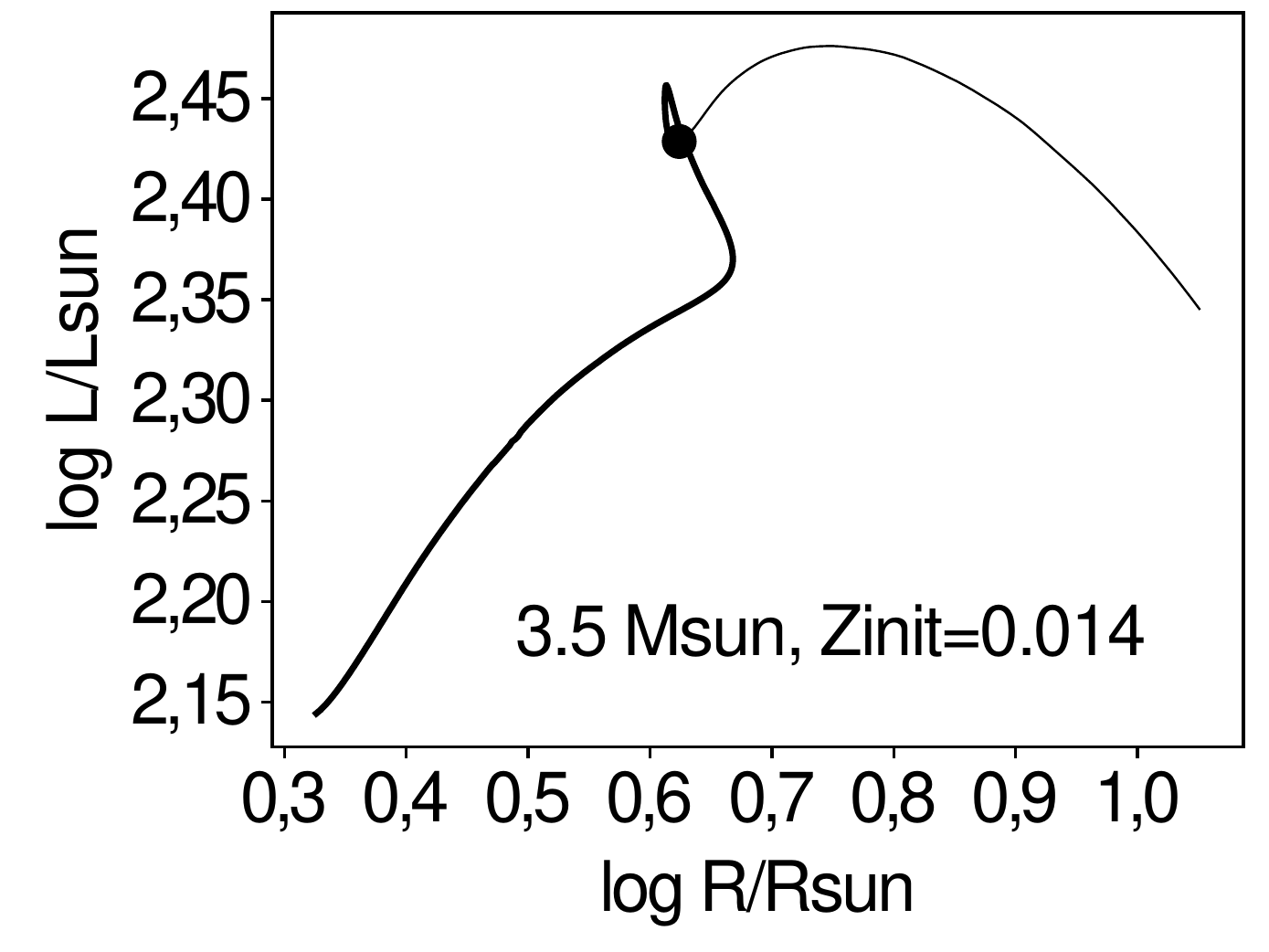}
  \caption{Same as Fig.~\ref{Fig:msHRtracks}, but in the $\log R - \log L$ diagram.
  }
\label{Fig:msLogRLogLtracks}
\end{figure}

\begin{figure}
  \centering
  \includegraphics[width=0.9\columnwidth]{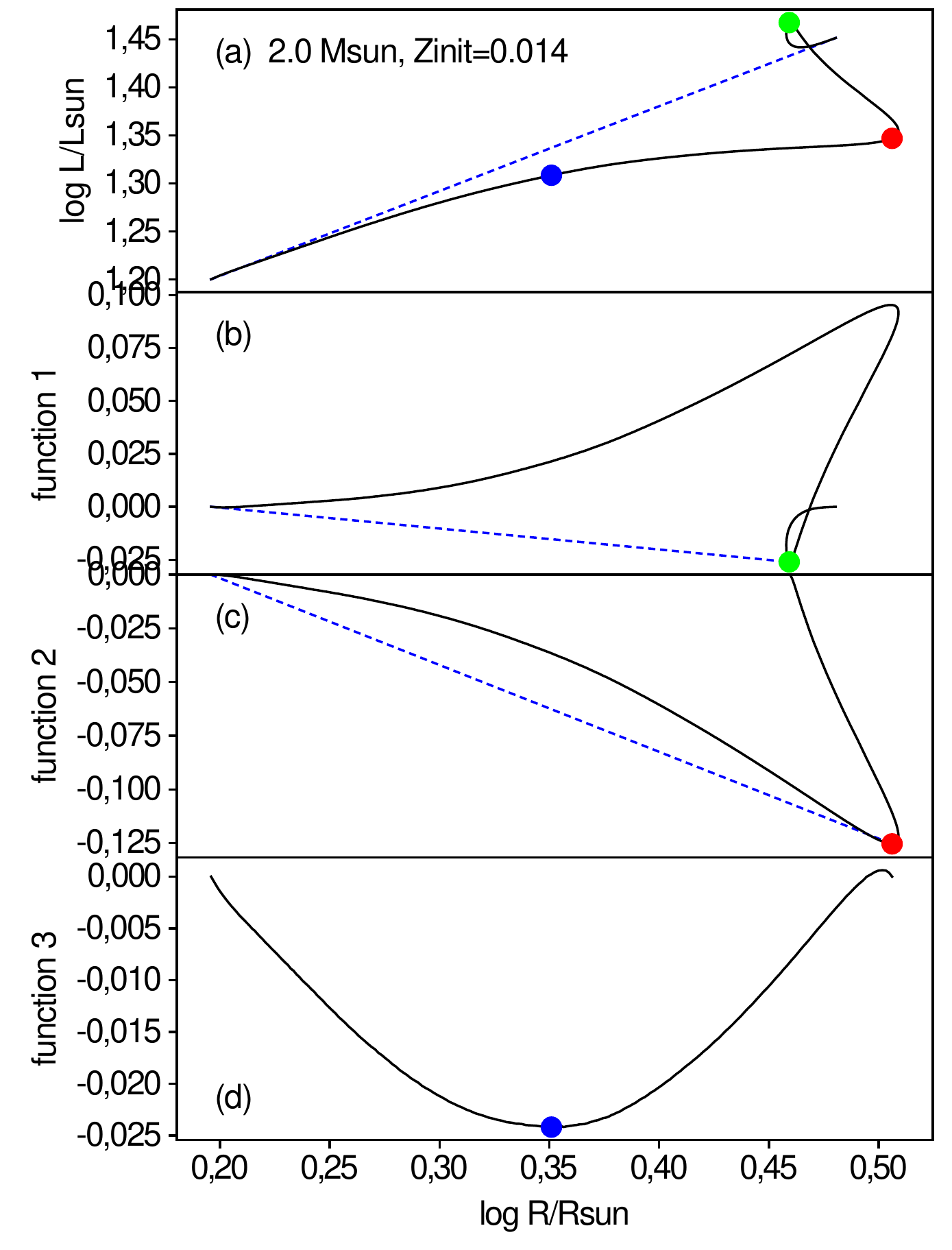}
  \caption{Upper panel: MS track of the $Z_\mathrm{init}$\,=\,0.014, \mass{2} models in the $\log R - \log L$ plane.
  Second to lowest panels: functions used to identify reference points on the MS track (see Sect.~\ref{Sect:MSNormalizationInRL}).
  }
\label{Fig:msNormFct_Z014M02p00}
\end{figure}

\begin{figure}
  \centering
  \includegraphics[width=0.9\columnwidth]{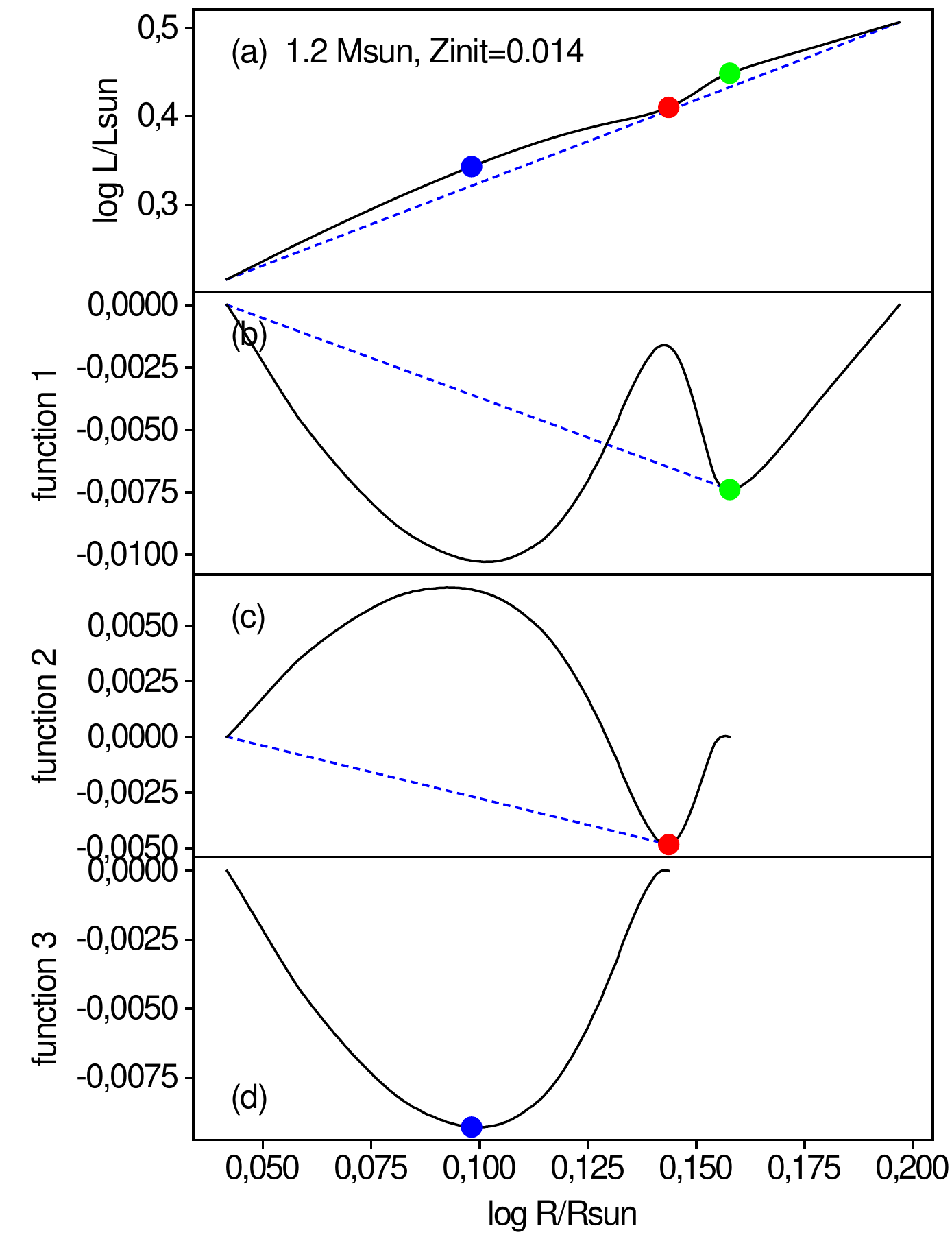}
  \caption{Same as Fig.~\ref{Fig:msNormFct_Z014M02p00}, but for the \mass{1.2} models.
  }
\label{Fig:msNormFct_Z014M01p20}
\end{figure}

\begin{figure}
  \centering
  \includegraphics[width=\columnwidth]{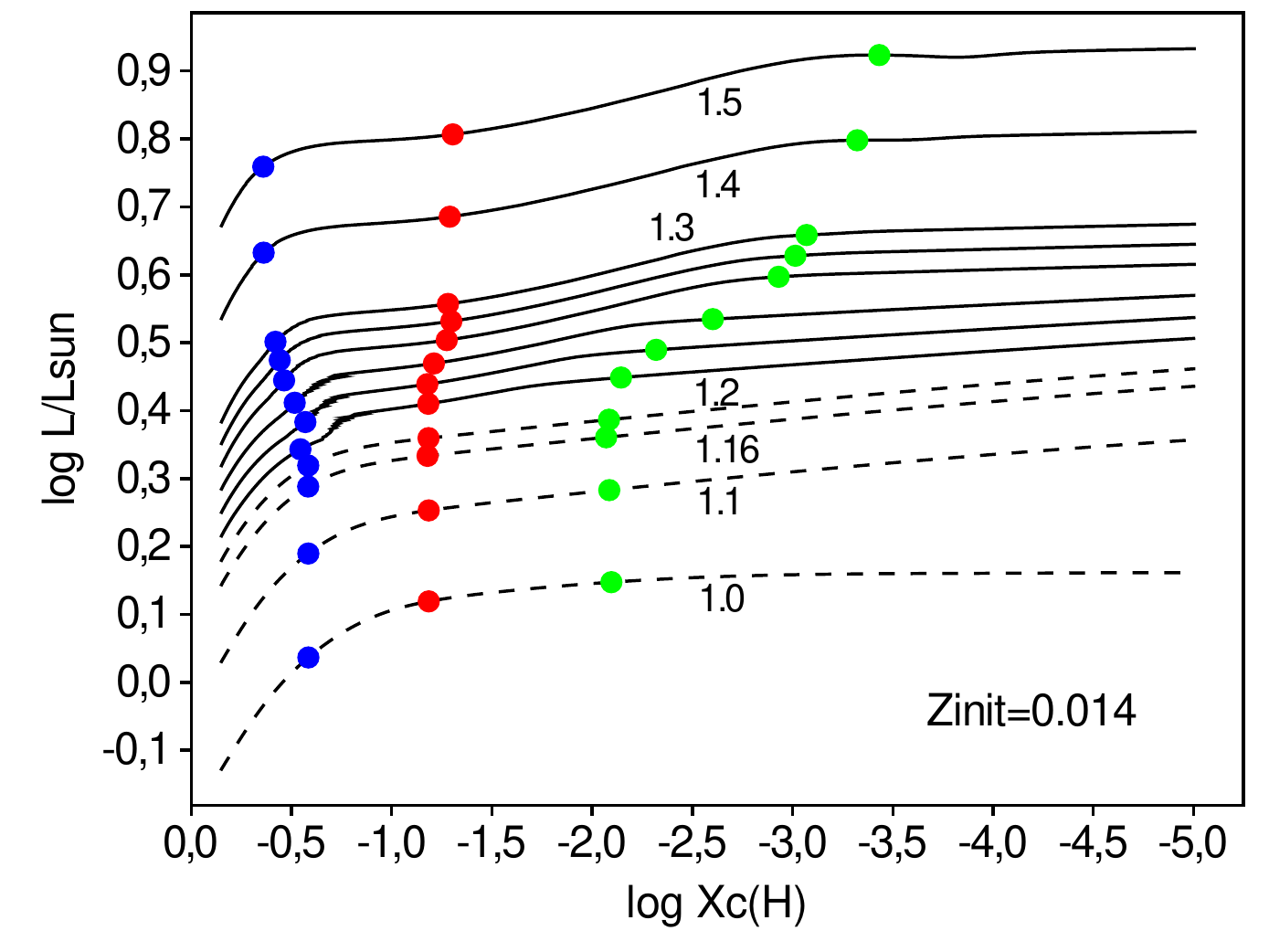}
  \caption{$\log L$ as a function of decreasing core hydrogen mass fraction, in logarithm, for \Zinit\,=\,0.014
   models of, from bottom to top, 1.0, 1.1, 1.16, 1.18, 1.2, 1.22, 1.24, 1.26, 1.28, 1.3, 1.4 and \mass{1.5}.
  The locations of the three reference points within the MS are shown by filled circles on each track.
  The dashed lines denote tracks for which the reference points are defined by fixed values of $X_\mathrm{c}(\mathrm{H})$ (see Sect.~\ref{Sect:MSNormalizationInRL}).
  }
\label{Fig:msRefPointsXcH_Z014}
\end{figure}

The tracks in the $\log R - \log L$ plane are shown in Fig.~\ref{Fig:msLogRLogLtracks}.
They display the hooks already identified in the HR diagram for stellar masses above $\sim$\mass{1.2} (Sect.~\ref{Sect:MSNormalizationInHR}).
The construction of the basic tracks is made in three steps:
\begin{enumerate}
\item We first identify the second turn off point in the $\log R - \log L$ plane.
To this aim, we define the line connecting the locations of the ZAMS and TAMS in that plane (dotted line in panel a of Fig.~\ref{Fig:msNormFct_Z014M02p00}), and compute the geometrical distance of every model on the track to that line.
This distance, called function 1, is shown in panel b of the figure.
The first minimum of function 1, searched backward from the end of the MS, defines the second turning point.
It is shown by a filled circle in panel b of the figure.
\vskip 1mm
\item We then identify the first turn off point of the track.
We proceed in a similar way as above, but work in the $\log R -$ function 1 plane rather than in the $\log R - \log L$ plane:
we define the line connecting the ZAMS to the second turning point (dotted line in panel b of Fig.~\ref{Fig:msNormFct_Z014M02p00}), and compute the geometrical distance of function 1 to that line;
this new distance, called function 2, is shown in panel c.
The location of the first turning point is defined as the minimum of that function.
It is shown by a filled circle in the panel.
\vskip 1mm
\item Finally, we define a third reference point on the MS track, between the ZAMS and the first turning point, based on the distance of function 2 to the line connecting the ZAMS and the first turning point in the $\log R -$ function 2 plane (dotted line in panel c of Fig.~\ref{Fig:msNormFct_Z014M02p00}).
That new distance is called function 3,
and the reference point is defined by the minimum of that function.
It is shown by a filled circle in panel d.
\end{enumerate}
The three reference points so defined are summarized in panel a of Fig.~\ref{Fig:msNormFct_Z014M02p00}.
The procedure is applicable to masses as low as \mass{1.2}, even if no hook is visible in the $\log R - \log L$ plane at that mass, as illustrated in Fig.~\ref{Fig:msNormFct_Z014M01p20} for the \mass{1.2} case.

For masses lower than \mass{1.2}, the technique fails because there is no deflection point on the MS tracks in the $\log R - \log L$ due to the absence of a convective core.
For those low mass stars, the three reference points are taken at fixed values of $X_\mathrm{c}(\mathrm{H})$ based on the values found at the higher mass stars.
Fig.~\ref{Fig:msRefPointsXcH_Z014} displays $\log L$ as a function of the core hydrogen mass fraction for stars in the mass range 1.2 to \mass{1.4}, and locates the three reference points on the tracks.
From these results, the reference points are fixed at $X_\mathrm{c}(\mathrm{H})=0.26$, 0.065 and 0.008 for all the stars with \mass{M<1.2} at \Zinit\,=\,0.014.
The same values of $X_\mathrm{c}(\mathrm{H})$ are adopted for the tracks at higher metallicities.
At lower metallicities, different values are adopted: $X_\mathrm{c}(\mathrm{H})$\,=\,0.29, 0.070 and 0.010 at \Zinit\,=\,0.010, and $X_\mathrm{c}(\mathrm{H})$\,=\,0.33, 0.090 and 0.015 at \Zinit\,=\,0.006.
At \Zinit\,=\,0.006, these fixed values of $X_\mathrm{c}(\mathrm{H})$ are used to define the reference points of the \mass{1.2} track as well.

\subsubsection{Basic MS tracks}

Once the reference points are defined on the MS, basic tracks are constructed by distributing a fixed number of models between the reference points.
The distribution is done regularly in age in the first two intervals (i.e. between ZAMS and the first reference point and between the first and second reference points) and regularly in $\log X_\mathrm{c}(\mathrm{H})$ between the two remaining intervals (i.e. between the second and third reference points and between the third reference point and the TAMS).
A regular distribution in $\log X_\mathrm{c}(\mathrm{H})$ cannot be done before the first turning point because the hydrogen mass fraction in the core is not a monotonically decreasing function of time for all masses.

\subsection{Post-main sequence}
\label{Sect:PostMSConstruction}

The construction of the basic tracks during the post-MS phase is easier to perform than that of the MS.
We locate the base of the RGB with a technique similar as the one used to locate the turn off points on the MS.
For that purpose, we use the angle between the X-axis in the $\log R - \log L$ plane and the line connecting the TAMS and the running point on the post-MS track.
The base of the RGB is located at the point where this angle starts to increase.
We then distribute a fixed number of models between the TAMS and that reference point at the base of the RGB, regularly spaced in age.

\section{Isochrone construction}
\label{Sect:isochroneConstruction}

\begin{figure}
  \centering
  \includegraphics[width=\columnwidth]{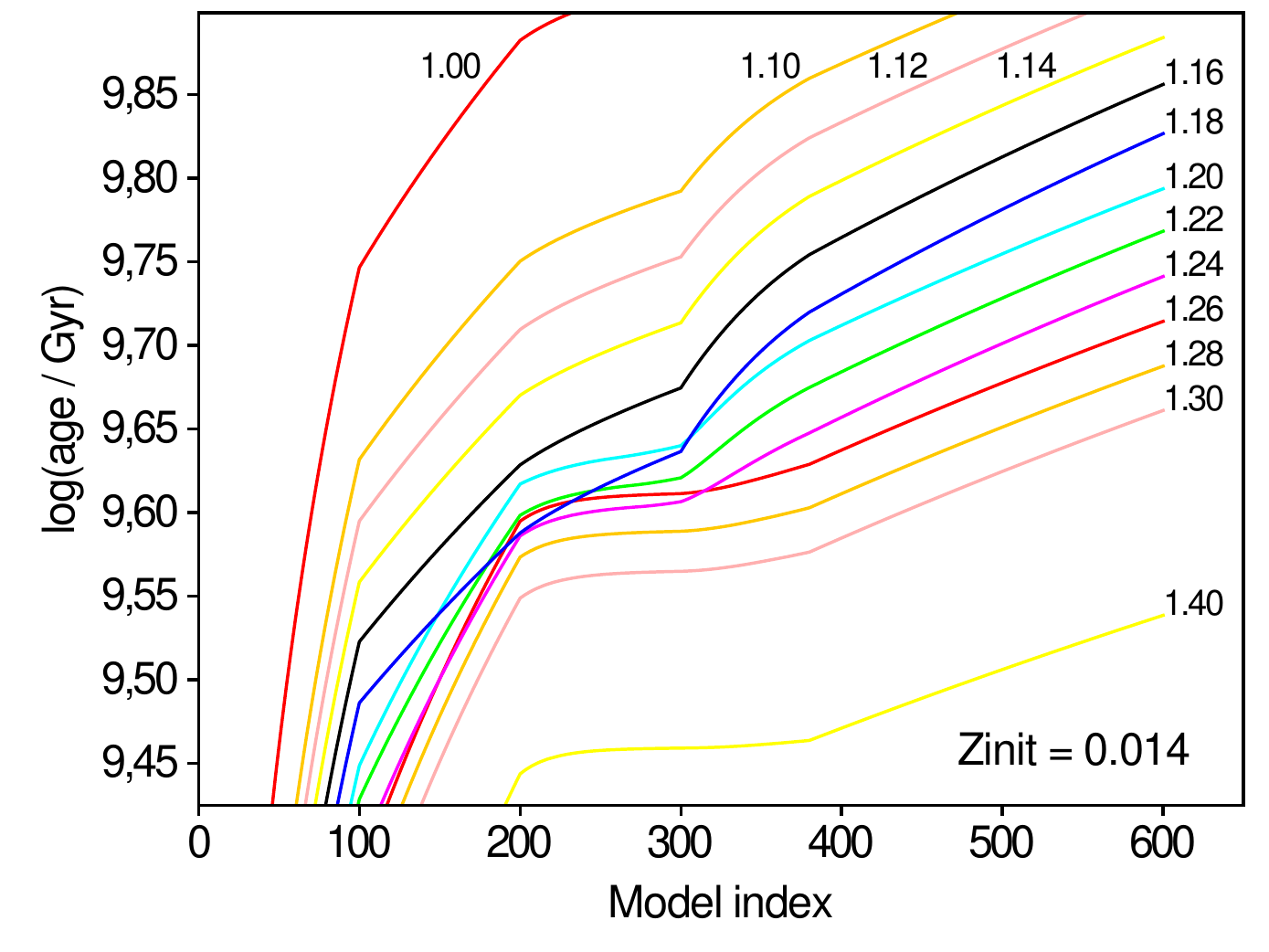}
  \caption{
  Logarithm of the age, in $10^9$~yr, of the models in the basic tracks as a function of model index for stellar masses between 1.0 and \mass{1.4} as labeled next to the curves.
  The models are taken at \Zinit\,=\,0.014.
  }
\label{Fig:isochroneTest}
\end{figure}

\begin{figure}
  \centering
  \includegraphics[width=\columnwidth]{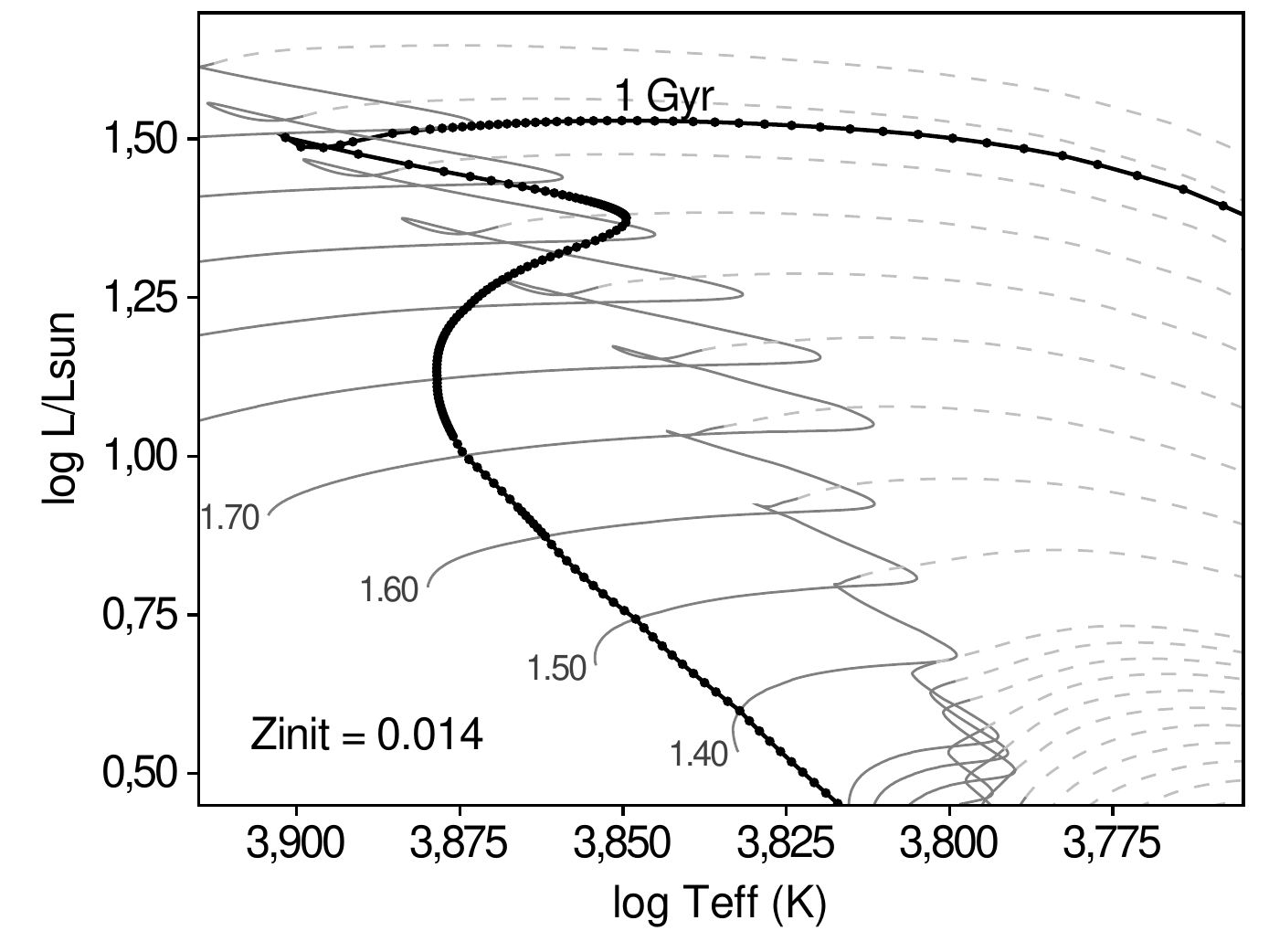}
  \caption{Isochrone at 1~Gyr, $\Zinit=0.014$, in the HR diagram, illustrating the isochrone computation procedure.
  Basic tracks are shown in grey for different stellar masses, continuous lines for the MS and dashed lines for the post-MS phase.
  The isochrone is plotted in black, each filled circle representing a stellar model.
  }
\label{Fig:isochroneProcedure}
\end{figure}

\begin{figure}
  \centering
  \includegraphics[width=\columnwidth]{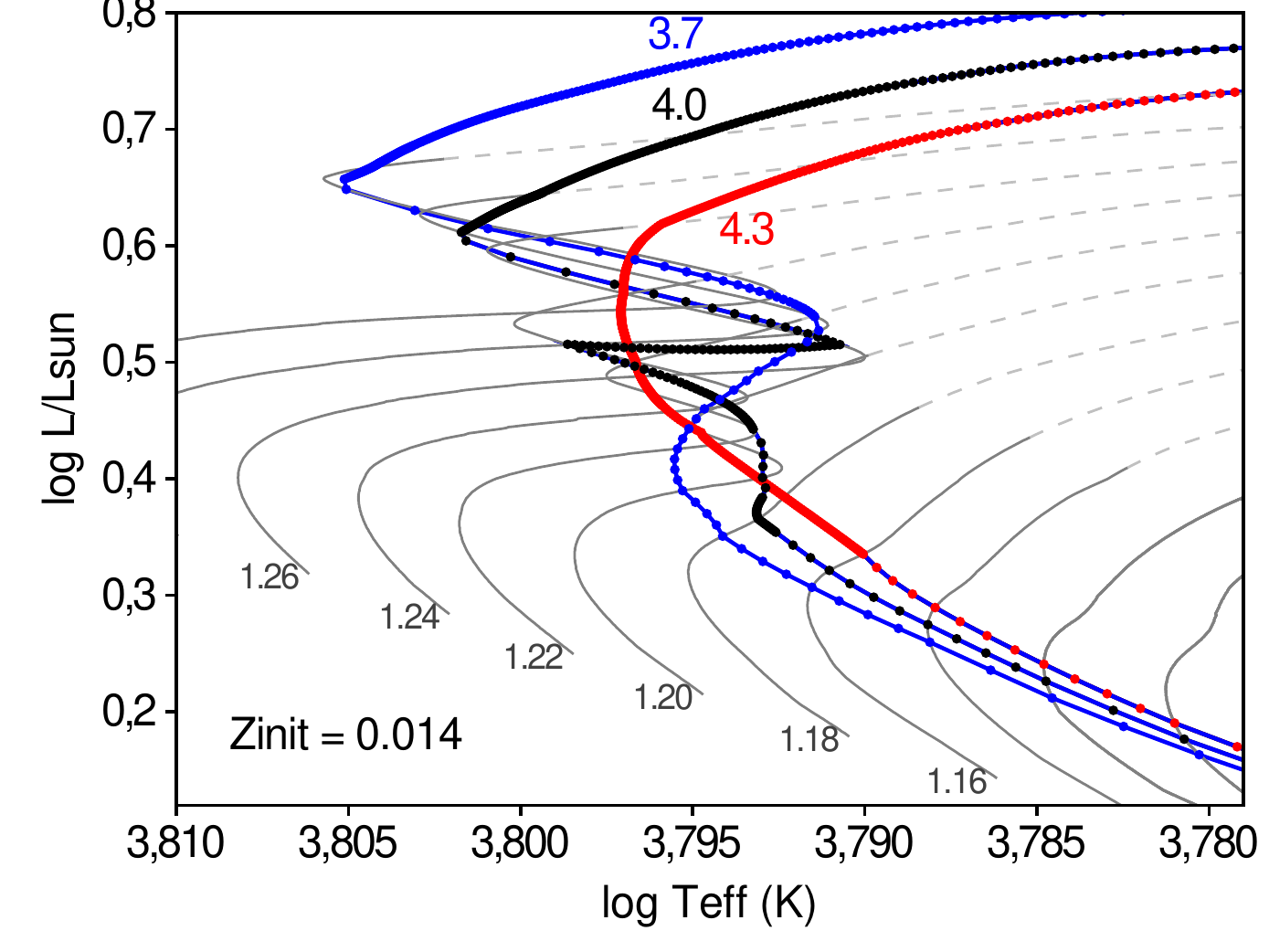}
  \caption{Same as Fig.~\ref{Fig:isochroneProcedure}, but for isochrones at 3.7, 4.0 and 4.3~Gyr as labeled next to the isochrones.
  }
\label{Fig:isochroneProcedure4Gy}
\end{figure}

Isochrones are computed from the basic tracks by taking care to reproduce adequately the hooks at the end of the MS.
The simplest procedure would consist in considering, in turn, each evolutionary stage available in the basic tracks (identified by the model indexes in those tracks as described in Sect.~\ref{Sect:basicTracks}), and interpolating in the set of models with the same index from the different tracks to get a model at the requested age.
The resulting isochrone would then consist of stellar models distributed according to the evolutionary stages defined in the basic tracks, which are, by construction, well sampled around the hooks in the HR diagram.

This procedure is, however, not applicable due to the appearance of a convective core during the MS at stellar masses around \mass{1.20}.
Figure~\ref{Fig:isochroneTest} plots the age of the models in the $Z$\,=\,0.014 basic tracks as a function of model index for masses between 1.0 and \mass{1.4}.
The age at the end of the MS (indexes above 400) is seen to monotonically decrease with increasing stellar masses, as expected.
But this is not the case in the middle of the MS.
The age of models at index 200, for example, is decreasing with increasing mass for all masses except when passing from 1.18 to \mass{1.20} and again when passing from 1.24 to \mass{1.26}.
The former case corresponds to the mass range at which a convective core appears in the middle-to-end of the MS, and the latter case to the mass range at which the convective core settles during the whole MS phase (see Fig.~\ref{Fig:convCore}).

We therefore proceed differently.
At the given metallicity, we compute a very dense grid of models by interpolating in the basic tracks with
a mass step that depends on the evolutionary stage.
We take mass steps of \mass{0.01} between the ZAMS and the second MS reference point, \mass{0.005} between the second and third reference points, \mass{0.0005} between the third and fourth, \mass{0.0001} between the fourth reference point and the TAMS, and \mass{0.0005} in the post-MS phase.
We then interpolate in each track of this dense grid to obtain a model at the requested age, and add it to the isochrone.
The resulting isochrone obtained at 1~Gyr is displayed in Fig.~\ref{Fig:isochroneProcedure}.

The issue mentioned above related to the development of the convective core during the MS does impact the morphology of isochrones that have their MS hooks at stellar masses between 1.18 and \mass{1.26}.
At \Zinit\,=\,0.014, it occurs for isochrones around 4~Gyr, shown in Fig.~\ref{Fig:isochroneProcedure4Gy}.
The 4~Gyr isochrone clearly displays the signature of the appearance of the convective core in the middle of the MS at stellar masses between 1.18 and \mass{1.20} and of the propagation of the convective core to the whole MS at stellar masses between 1.24 and \mass{1.26}.
These effect of the development of the convective core on the morphology of the isochrones is limited to ages within 200 million years around 4.0~Gyr.
The isochrones at 3.7 and 4.3~Gyr, for example, are seen in Fig.~\ref{Fig:isochroneProcedure4Gy} to display a normal behavior, the former with the two hooks characteristic of more massive stars and the latter with no hook characteristic of lower mass stars.

\end{appendix}

\end{document}